\documentclass[]{aa}

\usepackage{graphicx}
\usepackage{multirow}
\usepackage{txfonts}
\usepackage{hyperref}
\begin{document}

   \title{The kinematics of local thick discs do not support an accretion origin\thanks{Based on observations made at the European Southern Observatory using the VLT under programmes 096.B-0054(A) and 097.B-0041(A).}\fnmsep \thanks{The reduced datacubes and the data required to recreate the kinematical maps shown in Appendix~\ref{appen} are available to the community. The data for each galaxy in our sample can be accessed through the following links: \href{https://etsin.avointiede.fi/dataset/urn-nbn-fi-csc-kata20190129144516677076}{ESO~157-49}, \href{https://etsin.avointiede.fi/dataset/urn-nbn-fi-csc-kata20190129145112815373}{ESO~443-21}, \href{https://etsin.avointiede.fi/dataset/urn-nbn-fi-csc-kata20190129145329227945}{ESO~469-15}, \href{https://etsin.avointiede.fi/dataset/urn-nbn-fi-csc-kata20190129145511188303}{ESO~544-27}, \href{https://etsin.avointiede.fi/dataset/urn-nbn-fi-csc-kata20190129145647429085}{IC~217}, \href{https://etsin.avointiede.fi/dataset/urn-nbn-fi-csc-kata20190129145832850935}{IC~1553}, \href{https://etsin.avointiede.fi/dataset/urn-nbn-fi-csc-kata20190129150027164097}{PGC~28308}, and \href{https://etsin.avointiede.fi/dataset/urn-nbn-fi-csc-kata20190129150158580983}{PGC~30591}}}

   \author{S.~Comer\'on\inst{1},
          H.~Salo\inst{1},
          J.~H.~Knapen\inst{2,3,4},
          and R.~F.~Peletier\inst{5}}

   \institute{University of Oulu, Astronomy Research Unit, P.O.~Box 3000, FI-90014 Oulu, Finland\\
              \email{seb.comeron@gmail.com}
              \and Instituto de Astrof\'isica de Canarias E-38205, La Laguna, Tenerife, Spain
              \and Departamento de Astrof\'isica, Universidad de La Laguna, E-38205, La Laguna, Tenerife, Spain
              \and Astrophysics Research Institute, Liverpool John Moores University, IC2, Liverpool Science Park, 146 Brownlow Hill, Liverpool, L3 5RF, UK
              \and Kapteyn Astronomical Institute, University of Groningen, P.O.~Box 800, NL-9700 AV Groningen, the Netherlands}

 
  \abstract{Thick discs are nearly ubiquitous components of the discs of present-day galaxies. It has been proposed that a fraction of their stars have been accreted. Here, we aim to find whether accretion of satellites is the main formation mechanism of thick discs. To do so, we observed a sample of eight nearby edge-on galaxies with the Multi-Unit Spectroscopic Explorer (MUSE) integral field unit at the Very Large Telescope (VLT). Six of the galaxies have a distinct thick disc. We derived thick-disc velocities and velocity dispersions for the galaxies in our sample. We devise a formalism to estimate the fractions of retrograde material in the thick discs by using kinematical maps and thin/thick disc decompositions. None of the galaxies in our sample show strong evidence for retrograde material at large distances from the centre. Including those found in the literature, there are seventeen thick discs with studied kinematics, with only one showing unambiguous signatures of retrograde material. Literature numerical studies of dynamical friction allow us to estimate that at the current cosmic time about one in six mergers for which the stars of the accreted galaxy ended in a thick disc were retrograde. This is in tension with the observed fraction of galaxies with a partly retrograde thick disc (one in seventeen). We conclude that satellite accretion is not favoured by observations to be the main formation mechanism of thick discs.}


   \maketitle
%

\section{Introduction}

Thick discs are the dim and vertically extended counterparts of canonical thin discs. They were discovered in external galaxies by \citet{BURS79} and \citet{TSI79}. In the Milky Way, the thick disc was found by \citet{YO82} -- although he identified it as the stellar halo -- and \citet{GIL83}. In spite of their low surface brightness at the heights where they dominate the disc luminosity, thick discs are fairly massive; their stellar masses typically come second only to that of the thin disc, as shown by \citet{YOA06} and \citet{CO11, CO12, CO18} for external galaxies and by \citet{FUHR08}, \citet{FUHR12}, and \citet{SNAITH14} for the Milky Way.

The growing consensus is that the bulk of the Milky Way thick disc has an internal origin. \citet{RUCH14} found no evidence for a significant accreted component in the Galaxy disc. By studying the stellar velocity dispersions in the solar neighbourhood, \citet{KROU02} deduced that, in the past, star formation episodes in clusters with embedded masses up to $\sim10^{6}\,\mathcal{M}_{\bigodot}$ have occurred in the Milky Way. Stars in such massive clusters must have a large velocity dispersion and the cluster dissolution would have resulted in a thick component. \citet{LEHN14} and \citet{HAY15} proposed that the thick disc in the Milky Way was born thick out of a turbulent and well-mixed interstellar medium during the first gigayears of the Galaxy. The Milky Way thick disc might however contain a small fraction of accreted stars in streams such as those detected by \citet{GIL02} and \citet{MAR04}. 

While several studies of local galaxies have been interpreted to favour an internal origin for thick discs \citep{CO14, CO15, CO16}, we are aware of the existence of at least one retrograde thick disc \citep[in FGC~227;][]{YOA05, YOA08}, which strongly indicates that it was accreted. The retrograde nature of the thick disc of FGC~227 was found by studying its rotation curve. The possibility of thick discs being, at least partly, accreted was already pointed out by \citet{A03}. A recent study of FCC~170, an S0 galaxy in the Fornax Cluster, shows clear signs of an accreted thick disc component, although the bulk of the the thick disc is believed to have formed in situ \citep{PIN19}.

The number of published thick-disc rotation curves is small due to the difficulties in obtaining spectra of regions with low surface brightness. Furthermore, it is hard to pinpoint the heights at which the thick disc dominates \citep{CO18}. Hence, several rotation curves thought to characterise the thick disc might actually sample the thin disc. To our knowledge, rotation curves thought to belong to the thick disc are available for seven galaxies in \citet{YOA05, YOA08}\footnote{\citet{YOA08} discuss nine galaxies, but they did not obtain a thick disc rotation curve for FGC~1440 and they failed to find coherent rotation in FGC~1948.}, two in \citet{CO15, CO16}, one in \citet{GUE16}, and one more in \citet{KAS16}. This last paper studies single-slit spectroscopy of three edge-on galaxies. The spectra were obtained for two slits: one was placed at the midplane and the second one was parallel to the first one and was placed at a height of $5^{\prime\prime}-7^{\prime\prime}$, depending on the galaxy. However, according to the decompositions in \citet{CO18}, only in NGC~5422 was the slit placed high enough so as to include a significant thick-disc contribution. Based on the above references, we estimate the total number of galaxies with available thick-disc rotation curves to be eleven, of which ten have prograde thick discs. Increasing the number of available rotation curves for thick discs is necessary to check whether a population of FGC~227-like thick discs exists or whether FGC~227 is a peculiar object with an exceptionally troubled youth.

In this paper we study the rotation curves of a small sample of galaxies at different heights above the midplanes. The data were obtained for an ongoing Multi-Unit Spectroscopic Explorer \citep[MUSE;][]{BA10} survey of edge-on galaxies aimed at studying the kinematics and the stellar populations of a statistical sample of thick discs. Six out of eight galaxies obtained so far in our survey have a distinct thick disc \citep[according to the decompositions in][]{CO18}. As a consequence, we are increasing the number of available thick-disc rotation curves by $\sim50\%$.

We note that the reported properties of thick discs depend on how thick discs are defined. Indeed, in the Galaxy the thick disc is defined by selecting individual stars based on their kinematics and/or composition \citep[e.g.~][]{CA10, BENS11, BO12, CHENG12} or using star counts \citep[e.g.~][]{REY01, JU08, JAY13, LO14, WANG18}. In external galaxies, the thick disc is most often defined based on photometric decompositions. The different definitions might explain why the thick disc of the Milky Way is often -- mostly in studies where the thick-disc stars are kinematically- and/or abundance-selected -- found to have a shorter radial scale length than that of the thin disc, whereas thick discs characterised in extragalactic objects are typically as extended as the thin discs, or more so. The not-quite-overlapping thick-disc definitions have caused some authors to demand that the expression ``thick disc'' be used for the geometrically defined features only \citep{HAY17}. The different thick-disc definitions are tentatively reconciled in models such as those by \citet{MIN15}, where what is defined as a thick disc in photometric decompositions is actually the sum of the flares of mono-age populations. In these models, the metal-poor, $\alpha$-enhanced, centrally concentrated thick disc found in the Galaxy would correspond to the inner portion of the thick discs in external galaxies, whereas the outer portion would be made of much younger flared populations. We note however that these models only account for internal secular evolution and do not include the possible accretion of external material.

This paper is organised as follows. We describe our sample in Sect.~\ref{ssample}. In Sect.~\ref{aquis} we explain how the data were collected and processed. In Sect.~\ref{results} we show our main observational results and in Sect.~\ref{numerical} we present a method to estimate the fraction of retrograde material in the thick discs. We present our discussion in Sect.~\ref{discussion} and summarise our findings and conclusions in Sect.~\ref{conclusions}.

\section{The sample}

\label{ssample}

\begin{table*}
 \caption{The sample and some of its properties.}
 \label{sample}
 \centering
 \begin{tabular}{l c c c c c c  c l}
 \hline\hline
  ID & $r_{25}$ & PA  & $cz$ &$d$ &\multicolumn{2}{c}{$v_{\rm c}$} & $z_{\rm c1}$ & Disc structure\\
     &  & & & & H\,{\sc i} data& This paper & &\\
     & ($\arcsec$) & ($\degr$) & (km\,s$^{-1}$) & (Mpc) & (km\,s$^{-1}$) & (km\,s$^{-1}$)& ($\arcsec$)  &\\
  \hline
  ESO~157-49 & 52.1 & 30.4  & 1656 & 17.3 & 99  & 107  & 6.1 &Well defined thin and thick disc\\
  ESO~443-21 & 36.9 & 160.8 & 2819 & 52.2 & 164 & 196  & $-$ &No distinct thin and thick disc\\
  ESO~469-15 & 55.9 & 149.2 & 1636 & 28.3 & 104 & 83   & 4.8 &Well defined thin and thick disc\\
  ESO~544-27 & 46.5 & 153.3 & 2454 & 45.9 & 125 & 129  & 4.0 &Well defined thin and thick disc\\
  IC~217     & 59.9 & 35.7  & 1890 & 21.1 & 104 & 115  & 5.4 &Well defined thin and thick disc\\
  IC~1553    & 40.5 & 15.0  & 2905 & 36.5 & 128 & 142  & 5.0 &Well defined thin and thick disc\\
  PGC~28308  & 59.9 & 125.2 & 2719 & 43.1 & 180 & 130  & $-$ &Well defined thin and thick disc; the\\
             &      &       &      &       &     &     &     & thin disc dominates at all heights\\
  PGC~30591  & 45.4 & 169.2 & 2028 & 35.5 & 122 & 97   & $-$ & No distinct thin and thick disc\\
  \hline
 \end{tabular}
 \tablefoot{Position angle (PA) values from \citet{SA15}. Heliocentric recession velocities ($cz$) from the NED except for ESO~157-49 and IC~1553 where we estimated $cz$ by making the amplitude of the stellar rotation curve in the midplane roughly similar for the approaching and the receding sides. Distance estimates ($d$) from \citet{TU08, TU16}. H\,{\sc i} circular velocity values ($v_{\rm c}$) from \citet{MAT96}, \citet{THEU06}, and \citet{COUR09}. The $v_{\rm c}$ values as estimated in this paper come from fitting Eq.~\ref{equation} to the midplane line emission curves extracted from the maps in Figs.~\ref{ESO157-49}-\ref{PGC30591} (see also Fig.~\ref{vcs}). The heights above which the thick disc dominates in S$^4$G images ($z_{\rm c1}$) are from \citet{CO18} and so are the disc structure descriptions.}

\end{table*}

The sample of eight galaxies studied in this paper was selected from the sample of 70 galaxies in \citet{CO12} where we studied edge-on galaxies in the {\it Spitzer} Survey of Stellar Structure of Galaxies \citep[S$^4$G;][]{SHETH10}. The S$^4$G was made using the IRAC camera \citep{FA04} on-board the {\it Spitzer Space Telescope}. The galaxies were decomposed assuming two vertically isothermal discs in hydrostatic equilibrium. The eight galaxies that we study here were selected to be observable from the southern hemisphere and to have an $r_{25}$ radius smaller than $60\arcsec$ according to HyperLeda\footnote{http://leda.univ-lyon1.fr/} \citep{MAK14}. In that way, it was guaranteed that the region between the centre of a target and its edge would be covered in a single MUSE pointing (this is necessary in order to be able to build a rotation curve out of the observations). Some of the properties of the galaxies in the sample are summarised in Table~\ref{sample}. The sample is made of galaxies less massive than the Milky Way, with maximum circular velocities obtained from H\,{\sc i} kinematics in the range $99\,{\rm km\,s^{-1}}\leq v_{\rm c}\leq180\,{\rm km\,s^{-1}}$.

Our original decompositions in \citet{CO12} had an over-simplistic point spread function (PSF) treatment and are superseded by those in \citet{CO18}, where the extended wings of the IRAC PSF are accounted for \citep[the `correct' IRAC PSF is described in][]{HO12}\footnote{A copy of this document is available at \url{http://irsa.ipac.caltech.edu/data/SPITZER/docs/irac/calibrationfiles/psfprf/}}. With the updated PSF at hand we found that two of the galaxies in the sample that had been observed with MUSE -- ESO~443-21 and PGC~30591 -- have no distinct thin and thick-disc components. Another galaxy -- PGC~28308 -- has a thick-disc component that is dimmer than the thin disc component at all observed heights once the scattering of the thin-disc light in IRAC is accounted for.

As the IRAC and the MUSE PSFs are different, the range of heights at which the thick-disc light dominates in \citet{CO18} is unlikely to be the same as in the MUSE datacubes\footnote{How the mass-to-light ratio of the discs changes as a function of wavelength also plays a role here.}. The MUSE PSF has a smaller full width at half maximum (FWHM) than that of IRAC. However, when studying thick discs, the extended PSF wings play a role that is more important than that of the PSF core\footnote{On the importance of a proper PSF treatment see, e.g.~\cite{DEJONG08}, \citet{SAN14, SAN15}, and \citet{CO18}}. These wings are not described by the FWHM and are an unknown for MUSE. This is not critical in our case, as integral field data cover a large region, meaning that different heights of a galaxy can be studied in a single exposure. Hence, even if the range of heights where the thick-disc light dominates in MUSE is different to that in IRAC, we  still expect to see a significant thick-disc contribution at some heights. We would  only lose the thick-disc kinematic signatures if the discs were vertically unresolved. This is clearly not the case for our observations because the rotation curves change with height, as shown in Sect.~\ref{results}.

\section{Data acquisition and processing}

\label{aquis}

\begin{table}
 \caption{Best-fit parameters of the rotation curves fitted with Eq.~\ref{equation}}
 \label{fittable}
 \centering
 \begin{tabular}{l c c c}
 \hline\hline
 ID & $v_{\rm c}$ & $x_{\rm f}$ &$\gamma$\\
 &(${\rm km\,s^{-1}}$)&$(^{\prime\prime})$&\\
 \hline
 ESO~157-49 & 107 & 18.0 & 6.72\\
 ESO~443-21 & 196 & 13.7 & 1.76\\
 ESO~469-15 &  83 & 10.4 &12.15\\
 ESO~544-27 & 129 & 18.5 & 1.80\\
 IC~217     & 115 & 15.0 & 0.91\\
 IC~1553    & 142 & 13.5 & 2.02\\
 PGC~28308  & 130 & 15.4 & 4.20\\
 PGC~30591  &  97 &  8.5 & 2.08\\
 \hline
 \end{tabular}

\end{table}

The data were obtained with the Integral Field Unit MUSE at Unit 4 of the VLT. The instrument has a $1^{\prime}\times1^{\prime}$ field of view (FOV), a $0\farcs2$ spaxel size, and a spectral range that stretches between $4750\,\AA$ and $9351\,\AA$ with a $1.25\,\AA$ spectral pixel size. The spectral resolution is $R\approx1700$ in the blue side and $R\approx3700$ in the red side. This translates into a $\sim2.5\,\AA$ spectral resolution that varies little -- but noticeably -- with the wavelength, as discussed below. The sample galaxies were observed in service mode between December 2015 and August 2016. Each of the eight galaxies had four 2624\,s on-target exposures, except for IC~217, for which the number of exposures was three. The exposures for a given galaxy were all centred at the same position and taken with different rotation angles separated by $90\degr$. Thus, the total exposure time per galaxy was almost three hours, except for IC~217. Off-target sky frames with a 240\,s exposure were also taken, but were not used in the reduction (see below).

The data were reduced using version 1.6.2 of the MUSE pipeline \citep{WEI12} within the \texttt{Reflex} environment \citep{FREUD13}. Because of our relatively long on-target exposures, the sky variations between the on- and the off-target exposures were large and the sky frames were unsuitable for sky correction. Instead, the sky was modelled from regions in the on-target exposures that were distant from the galaxy midplane (\texttt{skymethod=model} in the pipeline). This is possible because the galaxies are edge-on, and thus fill only a portion of the field of view (FOV).

In our first data-reduction attempts the sky model was obtained from selecting the spaxels with the least light in a white-light image, that is, an image obtained from collapsing the full datacube along its spectral direction. We however later found that several of our galaxies have regions with significant off-plane ionised gas emission that would cause genuine galaxy emission to be included in a sky defined from white-light images. We thus ran the pipeline twice for each galaxy. The first time, we did it for a spectral window 20\,\AA\ in width centred in the galaxy rest-frame H$\alpha$. The pipeline created sky masks based on H$\alpha$ images obtained from these datacubes. They were built by selecting the 10\% least luminous spaxels (\texttt{skyfr\_2}=0.10). Additionally, the pipeline, by default, excludes the 5\% dimmest spaxels as possible artefacts. The sky masks based on H$\alpha$ were then fed into a second data reduction, this time run over the whole MUSE spectral range.

Because the observations were scheduled as a filler, most of the exposures had relatively poor seeing ($\sim0\farcs8$ to $\sim1\farcs5$ in the combined datacubes). Hence, the auto-alignment recipe of the pipeline, \texttt{muse\_exp\_align}, often failed. We manually aligned the cubes for each of the on-target exposures by blinking white-light images built from the datacubes. The small angular shifts found with this method were then fed into the \texttt{muse\_exp\_combine} recipe to obtain a single combined datacube.

The combined datacubes were cleaned from small sky residuals with version 2.1 of \texttt{ZAP} \citep{SO16}. \texttt{ZAP} creates a sky mask similar to that made in the main MUSE pipeline. It then performs a principal component analysis of the sky residuals and corrects the datacube for them. In this case the sky was defined to be the 7\% of spaxels with the least emission in a H$\alpha$ image obtained from the combined datacube \footnote{This image was not continuum-subtracted and was made with a 20\,\AA\ window centered in the redshifted H$\alpha$ line. The dimmest 3\% spaxels were ignored as possible artefacts}. \texttt{ZAP} worked well across most of the FOV, but spurious features appeared for the regions with the largest line emission. Specifically, a significant noise was introduced. To circumvent this problem, we found the spectra where \texttt{ZAP} introduced noise by measuring the standard deviation of the fluxes in the range between spectral pixels 600 and 1000 for both the cleaned and the uncleaned cubes\footnote{This corresponds to the region between $\lambda\approx5500\,\AA$ and $\lambda\approx6000\,\AA$, which is dominated by the continuum with little contribution from sky and emission lines.}. We found those spaxels for which this standard deviation was increased by \texttt{ZAP}. We then replaced the noisy spectra in the cleaned datacube with those from the uncleaned datacube.

We produced white-light images out of the datacubes to obtain an optical-near-infrared image of the area under study (top-left panels in Figs.~\ref{ESO157-49}--\ref{PGC30591}). The images were used to create a mask of foreground stars (or globular clusters around the target) and background galaxies. We also masked the regions where the median signal-to-noise ratio (S/N) is smaller than $0.5$ in the spectral range between 5490\,\AA\ and 5510\,{\AA}.

The unmasked spaxels were Voronoi binned or tessellated to produce spacial bins with a median S/N of 25 in the spectral range between 5490\,\AA\ and 5510\,{\AA} with version 2.6 of the Voronoi binning code by \citet{CA03}. This kind of binning is adaptive, meaning that regions with a high surface brightness have smaller spacial bins than those with a low surface brightness. In many cases, individual spaxels close to the midplanes exceeded the S/N requirement and were left unbinned by the code. The number of spacial bins created per galaxy was typically a few thousand. The galaxy with the fewest bins is IC~217, with 847, and that with the most bins is IC~1553, with 7008. A spectrum was created for each bin by coadding all those in all spaxels within a bin. We call this binning based on the stellar emission the ``stellar binning''.

For each spacial bin we obtained a velocity -- $V$ -- and a velocity dispersion -- $\sigma$ -- with version 6.7.0 of the \texttt{python} version of the Penalized Pixel-Fitting code \citep[\texttt{pPXF};][]{CA04}. \texttt{pPXF} is a full-spectral-fitting code that fits spectra with a combination of stellar population spectral energy distribution (SED) templates. The output is a parametrised line-of-sight velocity distribution (LOSVD). We masked all the regions with possible strong emission lines and all the regions with sky lines with a peak flux larger than $6\times10^{-19}\,{\rm W\,m^{-2}\,\AA^{-1}\,arcsec^{-2}}$ according to \citet{HAN03}. This number was selected so that the three lines in the calcium triplet remain unmasked (we verified that a three-times-smaller masking threshold does not change our kinematical maps significantly). The width of the masking window was $1000\,{\rm km\,s^{-1}}$. We also masked the O$_2$-A telluric feature at around 7610\,\AA\ with a masking window width of $4400\,{\rm km\,s^{-1}}$.

\begin{figure*}
\begin{center}
  \includegraphics[width=0.98\textwidth]{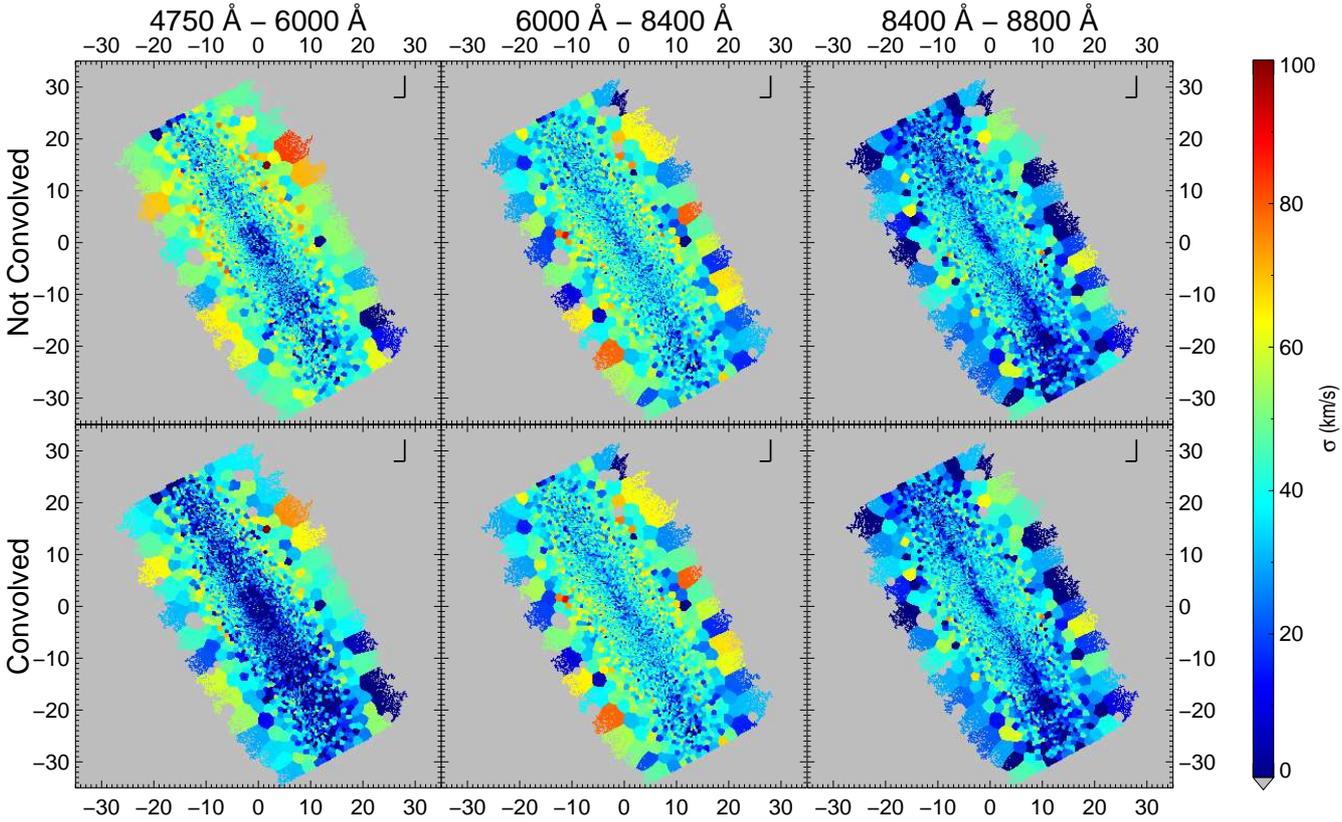}\\ 
  \end{center}
  \caption{\label{convol} Velocity dispersion maps of ESO~157-49 for different wavelength ranges (``blue'' for the {\it left} column, ``red'' for the {\it middle} column, and ``infrared'' for the {\it right} column). Maps in the {\it top} row were computed from spectra directly obtained from the processed datacubes. Maps in the {\it lower} row were made using spectra and SED templates that have been convolved to a common spectral FWHM as explained in the text. The axes are in arcseconds. The north and the east are indicated by the ``backwards-L'' symbol.}
\end{figure*}

The SED templates that we used for the fits come from the \texttt{MIUSCAT} library \citep{VAZ12}. We used the models with a Kroupa universal initial mass function \citep{KROU01} and with Padova isochrones \citep{GI00}. They have a FWHM of $2.51\,\AA$ which is similar to that of MUSE. However, some of the objects under study here have small masses and hence small velocity dispersions whose determination can be significantly affected if the FWHM mismatch between the MUSE and the template is not accounted for. We also tested in one of our galaxies -- ESO157-49 -- whether fitting different spectral ranges would affect the results. The selected spectral ranges were ``blue'' ($4750\,\AA-6000\,\AA$), ``red'' ($6000\,\AA-8400\,\AA$), and ``infrared'' ($8400\,\AA-8800\,\AA$; this last spectral window covers the calcium triplet)\footnote{When fitting the whole spectral range we used an eight-order additive Legendre polynomial to model the continuum, as detailed later. Such a high-order polynomial does not work well when fitting small spectral ranges, as its wiggles may have widths similar to those of the spectral lines. This is why we used polynomials of order four, six, and two to fit the blue, red, and infrared spectral ranges, respectively.}. In the maps in the top row in Fig.~\ref{convol} we can see that in general $\sigma$ is larger in blue than it is in infrared. This potentially indicates that the spectral resolution of MUSE at the calcium triplet is better than that in blue\footnote{Other possibilities are that the stellar populations sampled by the two wavelength ranges are not the same and dust obscuration effects.}. This is a well-know effect. For example \citet{BA17} modelled the FWHM of the line spread function as
\begin{equation}
\label{bacon}
 {\rm FWHM}(\lambda)=5.866\times10^{-8}\,\lambda^2-9.187\times10^{-4}\,\lambda+6.040
\end{equation}
with $\lambda$ in \AA\ based on a MUSE survey of the {\it Hubble} Deep Field.

To account for the changes in resolution as a function of wavelength and also for the difference in resolution between the spectra and the templates we did the following: for regions where the instrumental resolution is better than that of the template, we convolved the spectra with a Gaussian kernel of varying width so all the wavelengths had the same resolution as the template. For regions with a worse resolution than the templates, the templates were convolved to the instrumental resolution. This was again done with a varying-width Gaussian kernel to account for changes in the resolution with wavelength. We assumed the FWHM of the line spread function to be described by Eq.~\ref{bacon}. Applying such a convolution significantly lowers the $\sigma$ in the blue map (lower-left panel in Fig.~\ref{convol}). The blue FWHM might be slightly overestimated which might result in a slight underestimation of $\sigma$. Indeed, whereas for the unconvolved maps the median difference between the blue and infrared $\sigma$ values is $6\,{\rm km\,s^{-1}}$, for the convolved ones it is $-8\,{\rm km\,s^{-1}}$ (the median values are calculated over all bins and not over the surface)\footnote{Alternatively, this might be due to dust obscuring the light from stars deep inside the disc at blue wavelengths.}.

The final \texttt{pPXF} fits were done for the spectral range between 4750\,\AA\ and 8800\,{\AA}. The continuum was modelled with an eight-order additive Legendre polynomial. The initial values that we used for the fits were the galaxy radial velocity as reported in the NED and a velocity dispersion of $100\,{\rm km\,s^{-1}}$. The final velocity and velocity dispersion maps are shown in the second and third panels {in the left column in} Figs.~\ref{ESO157-49}--\ref{PGC30591}.

We stress that all the velocities and velocity dispersions obtained with the fits are line-of-sight (LOS) integrated quantities and are thus not representative of any particular radius in the galaxy. This point is important for understanding the analysis in Sect.~\ref{numerical}.

We also produced emission line velocity maps. The procedure that we followed is that used in the MaNGA survey (M.~Cappellari, private communication). We used the stellar binning to produce new \texttt{pPXF} fits. This time we did not mask the emission lines and included gas line templates to account for the gas emission. We kept the stellar velocities fixed and equal to those in the star-only \texttt{pPXF} fit. The line emission templates were chosen to be Gaussian profiles with a FWHM corresponding to that to which the spectra were degraded at the specific wavelength of the line. The line templates were centred at the wavelengths of the most prominent lines (H$\beta$, [O\,{\sc iii}]\,$\lambda4959$, [O\,{\sc iii}]\,$\lambda5007$, [O\,{\sc i}]\,$\lambda6300$, [O\,{\sc i}]\,$\lambda6364$, [N\,{\sc ii}]\,$\lambda6548$, H$\alpha$, [N\,{\sc ii}]\,$\lambda6583$, [S\,{\sc ii}]\,$\lambda6716$, and [S\,{\sc ii}]\,$\lambda6731$). All the emission lines were assumed to share the same kinematics. The resulting fitted stellar spectrum was then stored.

We then created a Voronoi tessellation of the spectra based on the H$\alpha$ emission (we refer to this second tessellation as the ``line binning''). We started by creating an H$\alpha$ line image of the galaxy from the datacubes by integrating in a window of 22\,\AA\ centred in H$\alpha$ and we subtracted the continuum emission calculated in a window of the same width 50\,\AA\ redwards. We ran the Voronoi binning code with the condition of $S/N=50$ per bin in H$\alpha$. We excluded the spaxels where the H$\alpha$ line had $S/N<0.5$. The extracted spectra were again fitted with \texttt{pPXF}. For each bin we used as a stellar template the result of the \texttt{pPXF} fits in the previous step (that where gas and stars were fitted using the stellar binning) whose bin centre was the closest to that of the line bin under study. We kept the velocity of this template fixed and we only allowed it to change by a normalisation factor. This is the fit from which we obtained the final gas kinematics. The line velocity maps are shown in the lower-left panels in Figs.~\ref{ESO157-49}--\ref{PGC30591}.

\begin{figure*}
\begin{center}

  \includegraphics[width=0.98\textwidth]{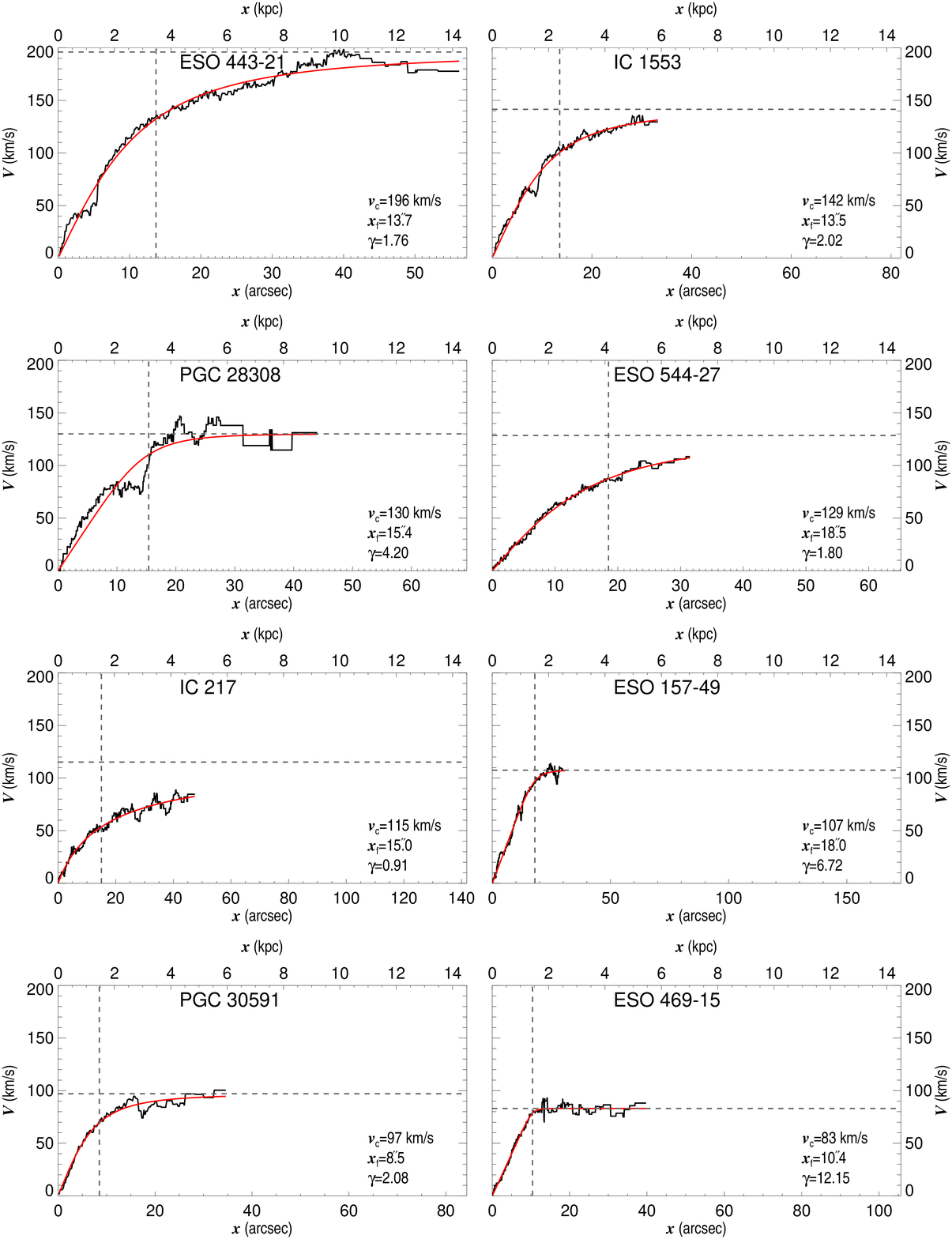}\\

  \end{center}
  \caption{\label{vcs} {Symmetrised midplane rotation curves extracted from the line emission velocity maps in Figs.~\ref{ESO157-49}-\ref{PGC30591} (black line) and their fit with Eq.~\ref{equation} (red line). All panels cover the same physical scale (horizontal axis) and the same velocity scale (vertical axis). The lower-right corner of each panel shows the best fit parameters where $v_{\rm c}$ is the circular velocity, $x_{\rm f}$ is the axial coordinate of the flattening of the rotation curve, and $\gamma$ describes the sharpness of the transition between a rising and the flat sections. The horizontal dashed lines indicate $v_{\rm c}$ and the vertical dashed lines indicate $x_{\rm f}$. The panels are ordered by descending $v_{\rm c}$.}}
\end{figure*}

We used the line emission velocity maps to estimate the circular velocity of galaxies as well as the position at which the rotation curve starts to flatten, $x_{\rm f}$. To do so we took the midplane gas rotation curve and folded it through the galaxy centre. Prior to the folding, the rotation curves were resampled to a regular sampling. Subsequently, as in \citet{YOA05}, we fitted the rotation curve with the following expression \citep[the same as in][when setting their $\beta=0$]{COUR97}
\begin{equation}
\label{equation}
 v(\chi)=v_{\rm c}\frac{1}{\left(1+\chi^\gamma\right)^{1/\gamma}},
\end{equation}
where $v_{\rm c}$ is the circular velocity, $\chi$ is the axial coordinate normalised by the position of the flattening $x_{\rm f}$ ($\chi\equiv x_{\rm f}/x$), and $\gamma$ parametrises the sharpness of the transition from a rising to a flat rotation curve. The larger $\gamma$, the sharper the transition. The fits are shown in Fig.~\ref{vcs}. The resulting $v_{\rm c}$ and $x_{\rm f}$ values are reported in Table~\ref{fittable}.

We compared our $v_{\rm c}$ values to those obtained from H\,{\sc i} surveys in the literature (Table~\ref{sample}). We find that for some galaxies both values have significant discrepancies. This might be attributed to the fact that we are only sampling the inner parts of rotation curves, which in turn might cause us to poorly constrain $v_{\rm c}$. However, this might not be the case for the galaxy with the largest discrepancy, PGC~28308 ($v_{\rm c}=130\,{\rm km\,s^{-1}}$ for us and $v_{\rm c}=180\,{\rm km\,s^{-1}}$ in the literature). The Extragalactic Distance Database \citep[EDD;][]{TU09} H\,{\sc i} line profile for this galaxy is not symmetric, which might indicate that the galaxy is disturbed in its outskirts or that the line comes from the superposition of two objects. However, within the radius that we have studied with MUSE, our maps indicate undisturbed kinematics (Fig.~\ref{PGC28308}). At larger radii, the S$^4$G images show that the galaxy is asymmetric and has a warp on its Eastern side. PGC~28308 has been identified to form a pair with MCG-02-25-019 in the HIPASS catalogue \citep{ME04}. The angular separation between the two galaxies is $4\farcm7$, which is equivalent to 59\,kpc in projection at an estimated distance of $43.1$\,Mpc.

To fully analyse the velocity maps, we produced velocity curves at different heights. Initially the curves were made at the midplane and at the heights $\pm z_{\rm c1}$ (above and below the midplane) at which the thick disc starts to dominate the surface brightness budget in \citet{CO18}. We then created additional velocity curves at equidistant heights between the midplane and $z_{\rm c1}$. The distance between cuts was set to be about $\Delta z=1\farcs5$. We created more rotation curves at heights $z>z_{\rm c1}$ with the same vertical spacing as above. Where no well-defined $z_{\rm c1}$ height existed we produced rotation curves at heights separated by $\Delta z=1\farcs5$. The adopted position angles (PA) came from \citet{SA15}. We did not use the PA values in \citet{CO18} because we found that for ESO~157-49 the midplane dust lane was obviously misaligned with the \citet{CO18} PA by $2\fdg3$. For the other galaxies the differences are smaller, ranging from $0\degr$ to $1\fdg5$. The rotation curves are shown in the top-right panels in Figs.~\ref{ESO157-49}--\ref{PGC30591}.

The panels below those displaying the rotation curves in Figs.~\ref{ESO157-49}--\ref{PGC30591} show the differences in circular velocity between the different vertical cuts and that at the midplane. The curves were produced by measuring the mean velocity at a given height for a given axial range. The size of the axial bins was set to be between $\Delta x=3\farcs5$ and $\Delta x=4\farcs0$. The exact bin size was set so that an odd number of bins would fit within the region where the rotation is not flat (marked by the vertical dashed lines in the plots).

The third row of panels in the right column in Figs.~\ref{ESO157-49}--\ref{PGC30591} shows the mean velocity of the rotation curve as a function of height at the regions beyond 1.3 times the flattening radius $x_{\rm f}$. This has been computed for both the blue- and the red-shifted halves of the galaxies when available. We could have chosen to compute these curves beyond $x_{\rm f}$, but for rotation curves where the transition between the rising and the flat sections is not sharp, regions at $x\approx x_{\rm f}$ show a substantial slope. Ideally, we would have liked to put the limit at $1.5\,x_{\rm f}$ or more, but the MUSE FOV did not cover this region in all our galaxies.

In the lower-right panels in Figs.~\ref{ESO157-49}--\ref{PGC30591} we compare the stellar rotation curves at the midplane and at the maximum heights under study with those of the ionised gas.

\section{Results}

\label{results}

The regions studied in this paper are largely the inner parts of galaxies. The maximum radius at which rotation curves have been obtained is almost always confined within $r_{25}$. The one exception is ESO~443-21. The fact that for ESO~443-21 we can obtain kinematics out to $1.5\,r_{25}$ with the same exposure time as for the other galaxies suggests some inaccuracy in the HyperLeda $r_{25}$ measurement for this galaxy.

Since we are looking at the inner parts of galaxies, we are studying a region where the different definitions of thick disc do not diverge significantly.

\begin{figure*}
 \begin{center}

   \includegraphics[width=0.98\textwidth]{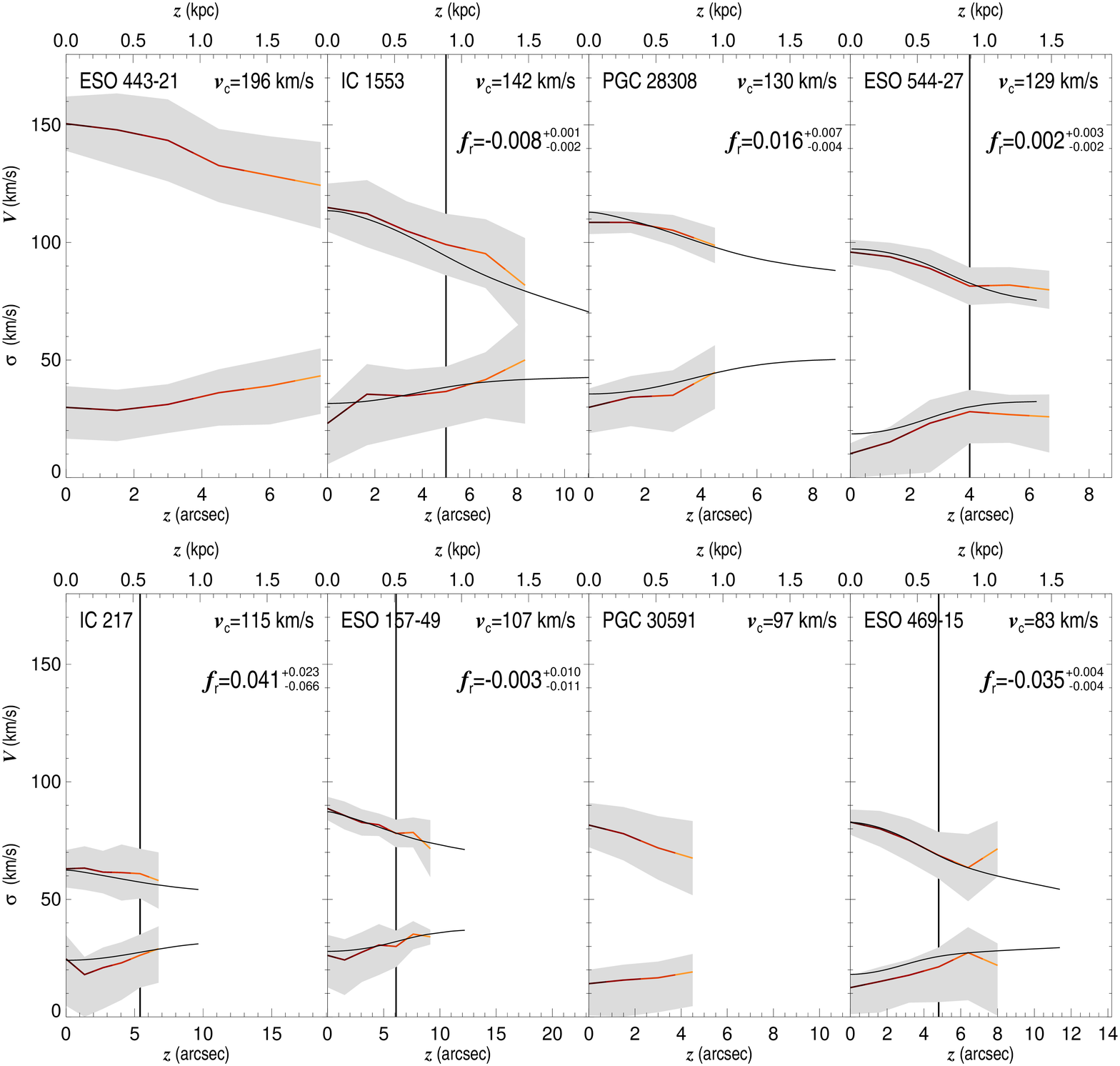}

 \end{center}
 \caption{\label{drift} Symmetrised velocity and velocity dispersion as a function of height (red-orange lines) for the regions at $x>1.3\,x_{\rm f}$. The plots are ordered by descending maximum circular velocity at a height $z=0$ as calculated from the parametrisation of the midplane gas rotation curves (Fig.~\ref{vcs} and Table~\ref{sample}). The grey bands correspond to the standard deviations of velocities and velocity dispersions at the bins used to obtain each point in the curves. The vertical solid lines show the height where the thick disc starts to dominate, if applicable to the galaxy. The black curves indicate fits to the velocity and velocity dispersion maps obtained with the formalism presented in Sect.~\ref{numerical3} for the galaxies that have two discs. For these fits the non-convolved vertical profiles from photometric decompositions of \citet{CO18} were adopted and then convolved with the MUSE FWHM (Eqs.~\ref{int1}--\ref{int2} and Table~\ref{sample2}). The horizontal axes are set so they stretch the same vertical span for all galaxies (1.85\,kpc). The vertical axis is also the same in all the panels.}
\end{figure*}

For all galaxies in the sample we are able to reach the region where the rotation curve starts to flatten (we can study regions at $x\gtrsim x_{\rm f}$). However, our rotation curves are never traced far beyond the flattening. In Fig.~\ref{drift} we present the velocity for $x>1.3\,x_{\rm f}$ as a function of height averaged over the two or four quadrants covered by pointings. The data points are computed at the heights of the rotation curves in the top-right panels in Figs.~\ref{ESO157-49}-\ref{PGC30591}. We did so by averaging the velocity values obtained for all the relevant bins (we gave an equal weight to each bin). The grey bands indicate the standard deviation in the velocity measurements for the bins corresponding to a given data point. Fig.~\ref{drift} also shows the velocity dispersion $\sigma$  calculated as
\begin{equation}
 \sigma=\sqrt{\frac{\sum_i\sigma_i^2}{N}},
\end{equation}
where $\sigma_i$ are the velocity dispersion values in the individual bins and $N$ is the number of bins. The grey bands indicate the standard deviation in the velocity dispersion fitted values.

One way to detect a retrograde thick-disc component would be to find a large velocity dispersion in a thick disc. If a significant retrograde component were present, we would expect to find velocity dispersions of the order of the galaxy circular velocity. This only occurs in PGC~28308, where high $\sigma$ values at large heights and small radii can be seen (Fig.~\ref{PGC28308}). This is probably due to the central spheroid and not to a retrograde thick disc. Indeed, PGC~28308 has a central mass concentration with a mass of 10\% of the baryonic galaxy mass, which makes it the largest one in the sample \citep{CO18}. The other galaxies fail to show a coherent cluster of bins with a large velocity dispersion.

Small fractions of retrograde material are hard to detect. The presence of such material would slightly increase the velocity dispersion and would cause some lag in the rotation curves. However, due to asymmetric drift, both the velocity dispersion and the lag increase with height even for a fully prograde disc. For example, in \citet{CO15} we made $N$-body models of ESO~533-4 and found that lags of a few tens of kilometres per second are to be expected at the height of the thick disc due to asymmetric drift and a non-perfectly-edge-on orientation. ESO~533-4 has a maximum circular velocity of $\sim150\,{\rm km\,s^{-1}}$, comparable to that of the most massive galaxies investigated here. In all our galaxies the lag in the rotation curve at $x>1.3\,x_{\rm f}$ is of a few tens of kilometres per second at the height of the thick disc. We discuss how to disentangle the effects of asymmetric drift from those of retrograde motions in Sect.~\ref{numerical}.

\begin{figure}
 \begin{center}

   \includegraphics[width=0.48\textwidth]{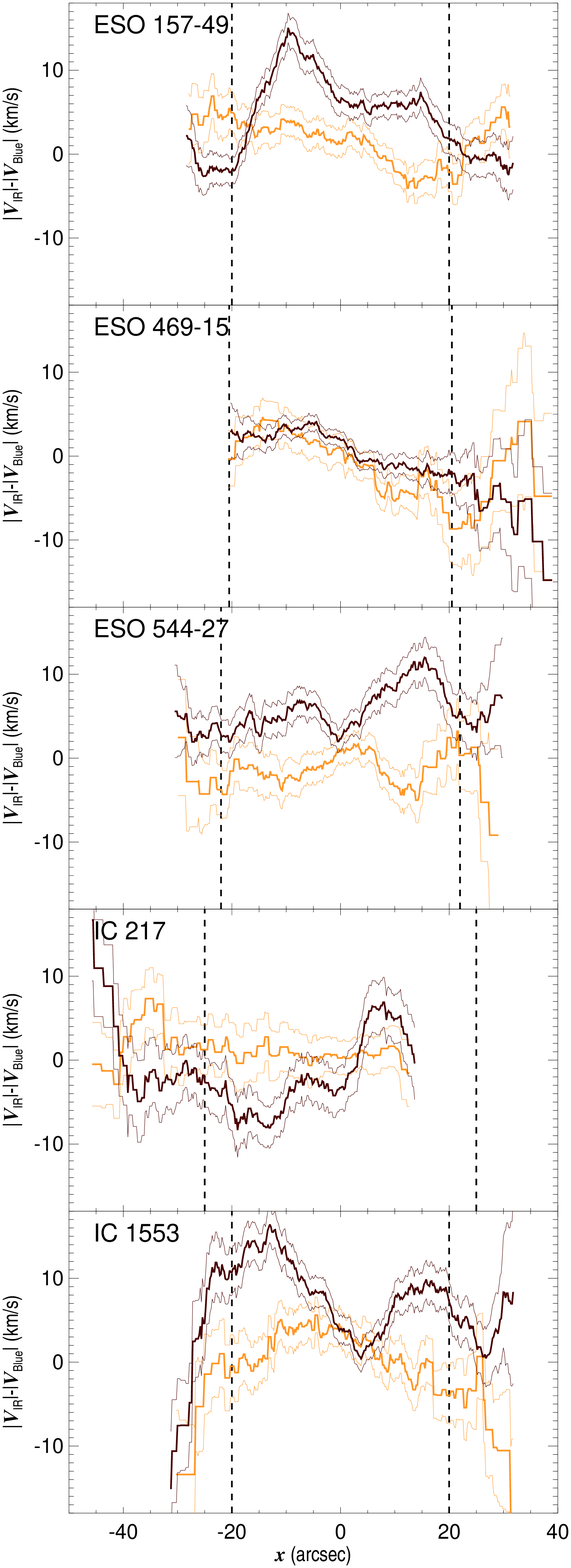}

 \end{center}
 \caption{\label{colours} Difference between the absolute infrared and blue rotation velocities at the midplane (thick dark red line) and the height where the thick-disc light starts to dominate (thick yellow line). Uncertainty intervals are indicated by the thin lines. The position at which $x=1.3\,{\rm x_{\rm f}}$ is indicated with dashed lines. Only the galaxies for which the thick-disc light dominates a height range are shown. The curves have been smoothed with an $8^{\prime\prime}$ window.}
\end{figure}

It is possible that the vertical velocity gradients shown in Fig.~\ref{drift} could be affected by dust extinction \citep[see e.g.~the effect of different extinction levels in rotation curves in][]{BOS92, ZA03}. We checked this possibility by comparing the rotation curves as obtained from the blue and the infrared spectral ranges (see Sect.~\ref{aquis}) at different heights. The results are shown in Fig.~\ref{colours} where the dark red lines represent the difference between the absolute velocities determined in blue an infrared at the midplane and the yellow lines represent the same at the height where the thick disc starts to dominate the surface brightness in \citet{CO18}. For visualisation purposes, the curves were smoothed with an $8^{\prime\prime}$ window. Uncertainties were computed as the error of the mean over the smoothing window. We use the velocity errors as computed by \texttt{pPXF} and assumed that the errors in blue and infrared are not independent. As expected, at large heights dust has little to no effect and the difference between the two velocities is close to zero. At the midplane, this is not always the case. In ESO~157-49, ESO~544-27, and IC~1553 the plot shows two humps around galactocentric projected distance $x=0$. These indicate that the rotation curve grows faster in the infrared than in blue, which in turn indicates a large dust extinction in the central parts of the galaxy.

However, our interest here is to look at the behaviour of the rotation curve at $x>x_{\rm f}$. There, large differences between the infrared and the blue velocities (larger than $\left|V_{\rm IR}\right|-\left|V_{\rm Blue}\right|=10\,{\rm km\,s^{-1}}$) are always associated with large error bars that make those velocity differences compatible with $\left|V_{\rm IR}\right|-\left|V_{\rm Blue}\right|<10\,{\rm km\,s^{-1}}$. We conclude that the midplane velocities -- and hence the asymmetric drift -- are underestimated at a level of $\sim10\,{\rm km\,s^{-1}}$ at most in the flat regions of the rotation curves. The effect would however be small, in accordance with previous findings that the outer parts of the discs are optically thin \citep{BOS92}.

In areas of the galaxies with radii smaller than that of the flat rotation curve region, some galaxies show large lags. In one, PGC~28308 (Fig.~\ref{PGC28308}), this is because of the presence of the spheroid \citep[just as can be seen in NGC~3115;][]{GUE16}. ESO~469-15 also has a similar peak in the lag curve at $x\approx10\arcsec$ (Fig.~\ref{ESO469-15}). Although this could also be attributed to a small spheroid, none were fitted in \citet{CO18}. A large lag of up to $60\,{\rm km\,s^{-1}}$ is observed in some regions of IC~1553 at an axial coordinate $x\approx10\arcsec-20\arcsec$ and at a height of $z\approx8\arcsec$ (Fig.~\ref{IC1553}). This lag appears well above the limit where the thick disc starts to dominate the luminosity. It might correspond to a small fraction of retrograde material, but no significant increase in the velocity dispersion is found in the lagging bins. IC~1553 also shows a significant asymmetry and a prominent warp (Fig.~\ref{IC1553}). It is tempting to link this warp to a galaxy covered by the MUSE FOV and found to the southeast of IC~1553. However we found it to be a background galaxy, at a redshift $z\approx0.05$ (vs.~$z\approx0.01$ for IC~1553). According to the NED, IC~1553 has no large nearby galaxy with a comparable recession velocity.

\section{Estimating the contribution of retrograde thick-disc stars}

\label{numerical}

We assume that the galaxies consist of a thin- and thick-disc component, and denote by $f_{\rm T}$ the relative contribution of the thick-disc stars at a given height, and by $\overline{V}_{\rm t}$, $\overline{V}_{\rm T}$ and $\sigma_{\rm t}$, $\sigma_{\rm T}$, the mean LOS velocities and velocity dispersions of the thin and thick components, respectively.  Moreover, we assume that a certain fraction $f_{\rm r}$ of the thick-disc stars are on retrograde orbits: these stars are assumed to have the same velocity dispersion as the prograde thick-disc component and a mean velocity $-\overline{V}_{\rm T}$. This implies that the effective mean value of the thick-disc component is $\overline{V}_{\rm T}^{\prime}=(1-2f_{\rm r}) \overline{V}_{\rm T}$. The combined thin+thick disc LOS velocities $V$ and velocity dispersions $\sigma$ can then be written as
\begin{equation}
\label{eq_vtot}
\overline{V} = \left(1-f_{\rm T}\right) \overline{V}_{\rm t} + f_{\rm T} \left(1-2f_{\rm r}\right)\overline{V}_{\rm T}, \\
\end{equation}
\begin{equation}
\overline{V^2} =\left(1-f_{\rm T}\right) \left(\overline{V}_{\rm t}^2+\sigma_{\rm t}^2\right)+ f_{\rm T} \left(\overline{V}_{\rm T}^2+\sigma_{\rm T}^2\right),\\
\end{equation}
\begin{equation}
\sigma^2 = \overline{V^2}-\overline{V}^2.
\label{eq_sigmatot}
\end{equation}
In what follows we use the observed 2D velocity and velocity dispersions maps to estimate $f_{\rm r}$. To do that we approximate the circular velocities with the gas rotation curve and apply Jeans' equations to account for the velocity dispersion support which reduces the mean tangential velocity needed to balance the central gravity (``asymmetric drift''). The derivation of the asymmetric drift equation is summarised in Sect.~\ref{numerical1}, following \citet{BI08}. We also need to deal with the difference between the actual quantities and their LOS-integrated observed counterparts: we use a self-consistent $N$-body model to check how to apply the asymmetric drift equation to LOS velocities (Sect.~\ref{numerical2}). We then devise a method to fit the fraction of retrograde thick-disc stars, and use  the $N$-body data to estimate the different sources of uncertainty, the main contribution coming from the uncertainty in $f_{\rm T}(z)$. Finally, in Sect.~\ref{numerical3}, we apply this method to MUSE data, using the photometric thin/thick-disc decompositions made in \citet{CO18}, taking into account the effect of the PSF on the observed vertical profiles.

\subsection{The effect of asymmetric drift and retrograde stars}
\label{numerical1}

For an axisymmetric steady-state galaxy the radial component of Jeans' velocity-moment equation reads \citep[see Eq.~4.222a in][]{BI08}
\begin{equation}
\label{jeans}
\frac{\partial\left(n\left<v_R^2\right>\right)}{\partial R}+\frac{\partial\left(n\left<v_Rv_z\right>\right)}{\partial z} + n\left(\frac{\left<v_R^2\right>-\left<v_\phi^2\right>}{R} +\frac{\partial\Phi}{\partial R}\right) = 0,
\end{equation}
where $R,z,$ and $\phi$ are the radial, vertical, and azimuthal coordinates, and $v_R, v_z, v_\phi$ the corresponding velocity components; brackets indicate averaging over a small volume around $R,z,\phi$. Here $n=n(R,z)$ stands for the number density of stars and $\Phi(R,z)$ denotes the gravitational potential. In a galaxy with thin and thick-disc components, stars of both components separately obey this equation. Following \citet{BI08}, we assume a zero mean radial velocity so that
\begin{equation}
\left<v_R^2\right> = \sigma_R^2,
\end{equation}
separate the mean azimuthal velocity and the azimuthal velocity dispersion,
\begin{equation}
\left<v_\phi^2\right> = \left<v_\phi\right>^2 + \sigma_\phi^2,
\end{equation}
and write the derivative of the potential in terms of circular velocity $v_{\rm c}$,
\begin{equation}
R \frac{\partial\Phi}{\partial R} = v_{\rm c}^2.
\end{equation}
Eq.~\ref{jeans} can then be recast to the form
\begin{equation}
\label{jeans2}
\left<v_\phi\right>^2 =  v_{\rm c}^2 - \sigma_R^2
\left[ 
\frac{\sigma_\phi^2}{\sigma_R^2}-1 -\frac{R}{n \sigma_R^2} 
\left(\frac{\partial\left(n\left<v_R^2\right>\right)}{\partial R}+\frac{\partial\left(n\left<v_Rv_z\right>\right)}{\partial z}
\right) 
\right].
\end{equation}
To estimate the radial and vertical gradient terms we assume exponential profiles for the number density and the radial velocity dispersion
\begin{equation}
n \propto e^{-R/h_R} e^{-\left|z\right|/h_z}
\end{equation}
and
\begin{equation}
\sigma_R^2 \propto e^{-R/h_v},
\end{equation}
where $h_R$ and $h_z$ are the disc scale length and scale height and $h_v$ is the radial scale length of the velocity dispersion. We then approximate
\begin{equation}
\left<v_Rv_z\right> = k_0 \frac{\left|z\right|}{h_z} \left(\sigma_R^2-\sigma_z^2\right),
\end{equation}
\noindent where the multiplier $k_0$ describes how the orientation of the velocity ellipsoid depends on vertical height; $k_0=0$ corresponds to cylindrical rotation, while $k_0=1$ corresponds to a situation where the principal axis of the velocity ellipsoid points to the galaxy centre. Inserting these expressions into Eq.~\ref{jeans2} finally leads to
\begin{equation}
\label{eq_asym}
\left<v_\phi\right>^2 =  v_{\rm c}^2 - \sigma_R^2
\left[ 
\frac{\sigma_\phi^2}{\sigma_R^2}-1 +\frac{R}{h_R}+\frac{R}{h_v}
+ k_0 \left(\frac{\left|z\right|}{h_z}-1\right) \frac{\sigma_R^2-\sigma_z^2}{\sigma_R^2}
\right].
\end{equation}
This deviation of the mean tangential velocity from the circular velocity due to velocity dispersion is the so-called asymmetric drift. The lag is larger for the thick disc, which has a larger velocity dispersion than the thin disc. The thick-disc contribution $f_{\rm T}$ increases with height, thereby making the combined velocity and velocity dispersion profiles depend on height. However, this is not the only factor that can cause a $z$-dependence: other contributions may come from the possible dependence of circular velocity and velocity dispersion on height. Moreover, Eq.~\ref{eq_asym} contains a term with an explicit $z$-dependence.

\begin{figure}
\begin{center}
\includegraphics[width=0.48\textwidth]{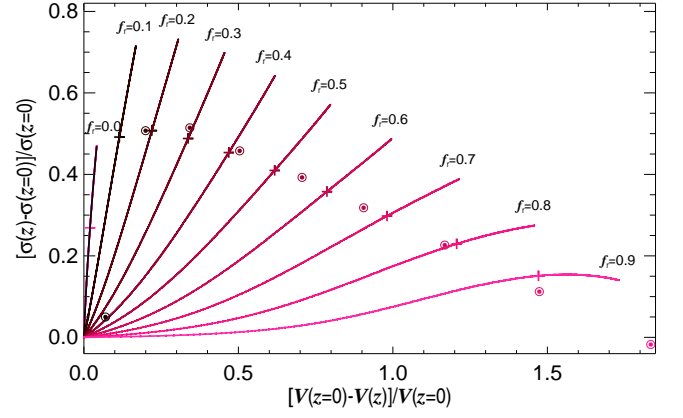}
\end{center}
\caption{Toy models for the influence of asymmetric drift and retrograde thick-disc stars on the vertical velocity and velocity dispersion gradients at $\left|z\right|=3h_{z{\rm t}}$. We use Eqs.~\ref{eq_toy1} -- \ref{eq_toy2} with $v_{\rm c}=1$, $k_i=1$, $\sigma_{\rm t}=0.2$, assuming $h_{z{\rm T}}/h_{z{\rm t}}=1-4$ and a fraction of retrograde stars $f_{\rm r}=0-0.9$. The crosses indicate $h_{z{\rm T}}/h_{z{\rm t}}=3$. Circles stand for the $N$-body simulation fits of Fig.~\ref{fig_nbody_profiles} below. Relatively small retrograde fractions of $f_{\rm r}=0.1-0.2$ cause lags of a few tens of per cent if the thin and the thick discs are well differenciated (i.e.~$h_{z{\rm T}}/h_{z{\rm t}}\leq3$, as is the case for the galaxies in our sample).}
\label{fig_toy}
\end{figure}  

Figure~\ref{fig_toy} shows a simple toy model illustrating how the vertical velocity and velocity dispersion profiles can help to indicate the presence of a retrograde thick-disc component.  We assume that the thin- and thick-disc components have the same total flux, and that they follow exponential vertical distributions with scale heights $h_{z\rm{t}}$ and $h_{z\rm T}$; this specifies the relative contribution $f_{\rm T}(z)$.  If we ignore for a moment the difference between actual and LOS velocities, we approximate Eq.~\ref{eq_asym} with
\begin{equation}
\overline{V}_i^2 =v_{\rm c}^2 -k_i\sigma_i^2, \ \ i={\rm t},{\rm T},
\label{eq_toy1}
\end{equation}
and assume that velocity dispersions depend on scale height as
\begin{equation}
{\sigma_{\rm T}^2}/{\sigma_{\rm t}^2} = {h_{z{\rm T}}}/{h_{z{\rm t}}}.
\label{eq_toy2}
\end{equation}
Fixing $v_{\rm c}=1$, $\sigma_{\rm t}=0.2$, and for simplicity using $k_i=1$, we use Eqs.~\ref{eq_vtot}--\ref{eq_sigmatot} to calculate the expected vertical gradients in combined mean velocity and velocity dispersion, for a set of $h_{z{\rm T}}/h_{z{\rm t}}$ and $f_{\rm r}$ values. In Fig.~\ref{fig_toy} the gradients are measured by the relative change of $V$ and $\sigma$ between $z=0$ and $\left|z\right|=3h_{z{\rm t}}$. In the absence of a retrograde component, the velocity drop in the vertical direction is due to the stronger asymmetric drift of the thick-disc component which becomes dominant at large height. In this case a  drop in $\overline{V}$ with height is achieved only in combination with an increase in $\sigma$. On the other hand, a larger relative decrease in $\overline{V}$ is expected in this toy model if the reduction of the mean velocity with height is due to the retrograde contribution to the effective mean velocity of the thick disc.

\begin{figure}
  \centering
   \includegraphics[width=0.48\textwidth]{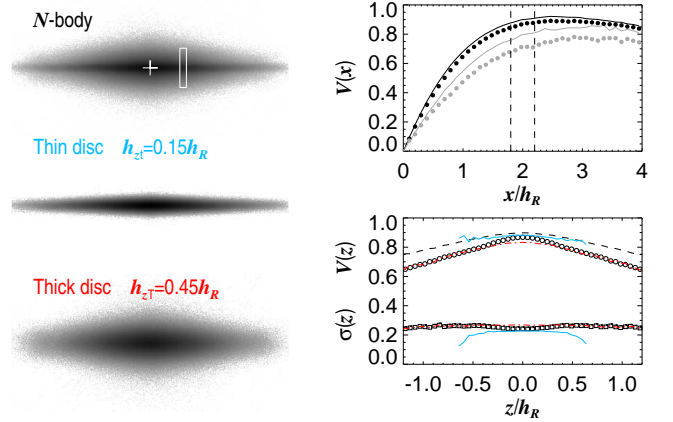}
\caption{Illustration of the $N$-body model used for checking the method for extracting the fraction of retrograde thick-disc stars. The {\it upper left} frame shows an edge-on snapshot of the simulated disc, while the {\it other two} frames display the thin and thick components separately (each with $10^6$ particles). The {\it upper right} frame displays the combined LOS-averaged rotation curves: black and grey lines correspond to the circular velocities at $z/h_R=0$ and $0.9$, respectively, while the circles indicate the same for stars. The {\it lower right} frame shows the combined vertical velocity and velocity dispersion profiles, averaged over the radial range $1.8-2.2\,h_R$, indicated by the white box on the $N$-body snapshot. The dashed line stands for the circular velocity, open circles for $\overline{V}$, and squares for $\sigma$. The continuous blue and dash-dot red curves show $\overline{V}$ and $\sigma$ for the thin and thick discs, respectively.}
\label{fig_nbody}
\end{figure}  

\subsection{$N$-body modelling}
\label{numerical2}

We next turn to an $N$-body model to assess, in a more quantitative manner, how to apply this approach to actual data. Figure~\ref{fig_nbody} displays a simulation model with superposed thin- and thick-disc components, with equal masses and with scale heights differing by a factor of three. As in \citet{SA17}, the initial values have been constructed with the GalactICS software \citep{KUIJ95}. This code was used to create two separate, self-consistent axisymmetric bulge-disc-halo models, with similar disc-to-total mass ratio and total rotation curve but different vertical thickness of the disc. The disc radial velocity dispersions were also different for the two models and were set by the ratio $\sigma_z/\sigma_R \approx 0.8$; the Toomre $Q$ at two radial scale lengths was about 1.4 and 2.8 for the thinner and the thicker disc models, respectively. The two disc models were then superposed together with the bulge and the halo of one of the single-disc models. The combined system was evolved with GADGET-2 \citep{SPRIN05} for 200 time units (a time unit is equivalent to the dynamical timescale at one exponential scale length, or about 10\,Myrs) to ensure that the two discs are mutually relaxed and also with respect to the halo and bulge. No significant non-axisymmetric perturbations developed during this time, meaning that the simulation is expected to fulfill the axisymmetric steady-state Jeans' equations. The parameters of the $N$-body model were chosen in a manner that should be relevant to the observed, fairly low-mass galaxies. In the simulations the total disc-to-halo mass ratio is approximately 0.25 within five scale lengths, with the thick and thin components contributing equally to the disc mass. The vertical thicknesses are 0.15 and 0.45 times the radial scale length (we use the initial radial scale length as a unit in all simulations). The model also has a small classical bulge component, but since this amounts to only 0.5\% of the disc mass its influence is insignificant.  Both thin and thick disc are realised with $10^6$ particles. This number is sufficiently large to allow the construction of mean stellar velocities and velocity dispersions, both locally at any point $(R,z)$, and along a line of sight with a given axial distance $x$ at height $z$. Using the accelerations calculated with GADGET-2, we can also reconstruct the circular velocities at the locations of particles; we use this to create a ``gas-particle'' population following circular orbits. To reduce the noise, five simulation snapshots are superposed when calculating kinematic quantities.

\begin{figure}
  \centering
   \includegraphics[width=0.48\textwidth]{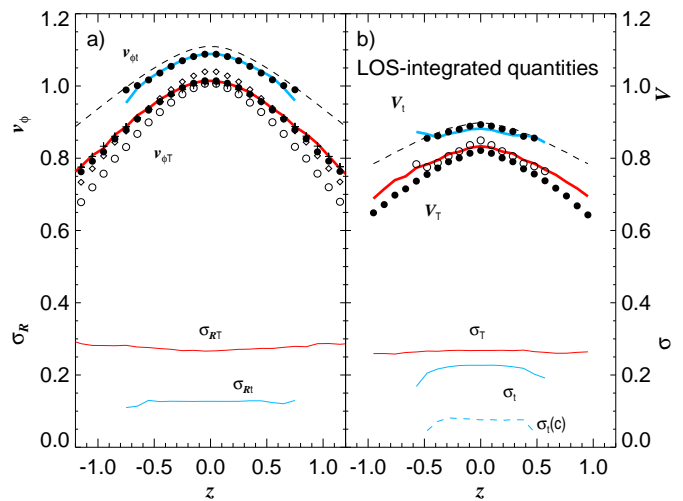}
\caption{Example of application of Eq.~\ref{eq_asym} to $N$-body data. Frame ($a$) displays velocities and velocity dispersions measured in a radial annulus at $R=1.8-2.2$, while in ($b$) LOS-integrated quantities at axial distance $x=1.8-2.2$ are shown. In ($a$) the dashed line indicates the circular velocity as a function of $z$, and the thick blue (above) and red curves (below) are the mean tangential velocities for the thin and thick discs, respectively. Thin curves are the radial-velocity dispersions. The symbols refer to variations in the application of Eq.~\ref{eq_asym}: the filled black circles include all terms, evaluated numerically at the radial annulus ($h_{R\rm t}=1.1$, $h_{R\rm T}=1.2$, $h_{v\rm t}=1.4$, $h_{v{\rm T}}=1.5$, $k_0=1$; measured $\sigma_R(z)$, $\sigma_{\phi}(z)$, and $\sigma_z(z)$ values are used). For the thick disc, other approximations are also shown. The crosses correspond to the case where $\sigma_R(z)$, $\sigma_{\phi}(z)$, and $\sigma_z(z)$ are replaced with their mean values for $\left|z\right| < 2h_{z{\rm t}}$ (0.27, 0.23, and 0.28, respectively; this improves agreement, probably because the values are less noisy). Diamonds indicate $\sigma_z/\sigma_R =0.7$ and $\sigma_\phi/\sigma_R=0.7$. Finally, open circles correspond to $h_R=1$, $h_v=1$. Frame ($b$) shows LOS integrated quantities separately for thin (blue curves; below) and thick discs (red; above). For the thin disc, the solid circles are calculated using the corrected $\sigma_{\rm t}(c)$ (see the text), while open circles use the non-corrected velocity dispersion, dominated by the large range of mean velocities along the LOS. In both cases, Eq.~\ref{eq_asym} is applied with the $h_R$ and $h_v$ values measured from the LOS-integrated radial profiles, while $\sigma_z/\sigma_R =0.7$, $\sigma_\phi/\sigma_R=0.7$, and $k_0=1$ are assumed.}
\label{fig_nbody_point_los}
\end{figure}  

With this $N$-body model we can numerically check several factors affecting the application of Jeans' equations.  First of all, the asymmetric drift equation is indeed well satisfied at any point for the two components. In Fig.~\ref{fig_nbody_point_los} the thick blue and red curves indicate the vertical $v_\phi$ profiles obtained directly from velocities of simulation particles, while solid circles indicate Eq.~\ref{eq_asym}, when numerically evaluating all the different terms contained within it. To obtain this agreement, the drop of circular velocity with height is important to include, indicating that the same effect must be accounted for when modelling the vertical profiles of observed LOS velocities. The precise shape of the velocity ellipsoid has a smaller but still noticeable effect (compare diamonds with filled circles; for clarity, only shown for the thick disc), and the same concerns the radial gradient terms (compare open and filled circles). The comparison in Fig.~\ref{fig_nbody_point_los} is made at the region where the rotation curve is turning flat: in this region $k_0 \approx 1$; however, ignoring this term has little effect.

The right-hand frame of Fig.~\ref{fig_nbody_point_los} shows a similar comparison for the LOS-integrated velocities, substituting $\overline{V}_i$ and $\sigma_i$ for $v_{\phi}$ and $\sigma_R$ in Eq.~\ref{eq_asym}. Though the trends are similar to those seen when using actual velocities, the amount of predicted asymmetric drift is clearly too large.  In particular, for the thin disc it is several times too large. This follows from the fact that since the intrinsic velocity dispersion of the thin disc is small, $\sigma_{\rm t}$ is dominated by the large range of mean velocities traversed in the LOS integration; for the thick disc this effect is smaller (compare the intrinsic and LOS velocity dispersions in Fig.~\ref{fig_nbody_point_los}a and b). To some degree this extra dispersion can be accounted for by using a correction, where the gas-velocity dispersion is quadratically subtracted from the thin-disc velocity dispersion,
\begin{equation}
\sigma_{\rm t}^2(c) = \sigma_{\rm t}^2 -\sigma_{\rm g}^2.
\end{equation}
This correction (dashed curve in Fig.~\ref{fig_nbody_point_los} b) assumes that gas has zero intrinsic dispersion and that its LOS dispersion arises solely from the mixture of mean velocities (this applies exactly to our simulated ``gas'' particles). Clearly this correction improves the applicability of Eq.~\ref{eq_asym} to the thin disc.

Our $N$-body model allows us to mimic a variable degree of retrograde rotation. To achieve this we reverse the velocities of a fraction $f_{\rm r}$ of randomly selected thick-disc particles. We also checked that the Eqs.~\ref{eq_vtot}--\ref{eq_sigmatot} are satisfied by such modified particle realisations, basically indicating that the first two moments are sufficient to describe the LOS velocity distributions with good accuracy.

Armed with this experience we next devise a method for extracting the fraction of stars in retrograde orbits $f_{\rm r}$ from synthetic $N$-body data, using only quantities which can be obtained also from real data. We write 
\begin{eqnarray}
\overline{V}_{\rm t}^2(z) &=& a_0 \left[ 
v_{\rm c}^2(z) -\sigma_{\rm t}^2 \left( f(x) +\frac{1}{2} \frac{\left|z\right|}{ h_{z{\rm t}}} \right)
\label{eq_asym_model1}
\right], \\
\overline{V}_{\rm T}^2(z) &=& a_0 \left[ 
v_{\rm c}^2(z) -\sigma_{\rm T}^2 \left( f(x) +\frac{1}{2} \frac{\left|z\right|}{h_{z{\rm T}}} \right)
\right].
\label{eq_asym_model2}
\end{eqnarray}
The difference with the toy model is that we now take into account the vertical drop of $v_{\rm c}$, and include $\sigma_{\rm t}$ and $\sigma_{\rm T}$ as unknowns; $a_0$ is an unknown scale parameter between the LOS and the actual velocities. The function $f(x)$ collects the velocity ellipsoid and radius-dependent terms of Eq.~\ref{eq_asym}: replacing $R$ with $x$, and assuming $\sigma_{\phi}^2 = \sigma_{z}^2 = 0.5 \sigma_{R}^2$ and $k_0=1$ this function becomes $f(x) = -1 + x/h_R+ x/h_v$. Combining these expressions with Eqs.~\ref{eq_vtot}--\ref{eq_sigmatot} we form models for $V(x,z)$ and $\sigma^2(x,z)$, and use a $\chi^2$ minimisation to estimate the parameters $a_0, \sigma_{\rm t},\sigma_{\rm T},f_{\rm r}$ which best match the synthetic profiles constructed from $N$-body realisations with different $f_{\rm r}$. The fraction $f_{\rm T}$ is measured from the simulation, and the scale heights are fitted to the simulated surface density profiles (in practice they are close to the input values used in GalactICS, even after the relaxation of the discs).

\begin{figure*}
  \centering
   \includegraphics[width=0.98\textwidth]{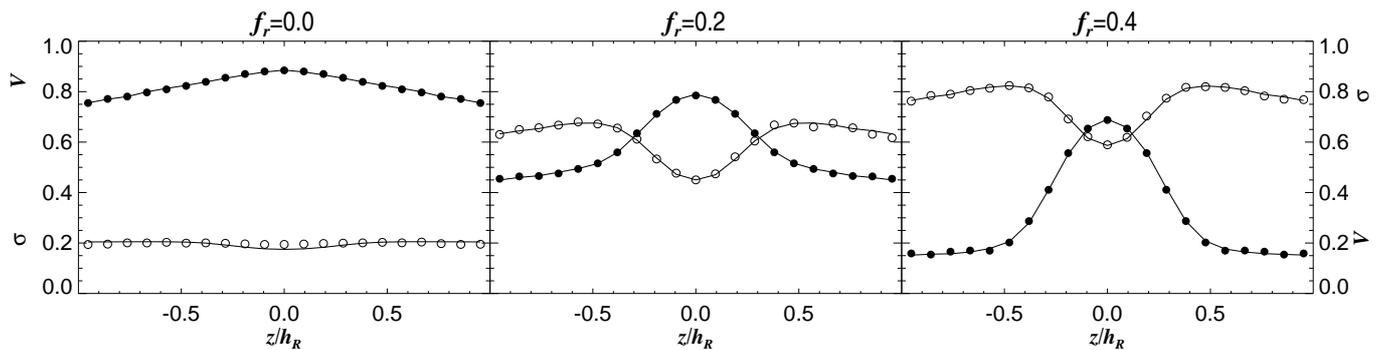}
\caption{Examples of fitting the simulated LOS velocity (filled circles) and velocity dispersion profiles (open circles) when 0, 20, and 40\% of thick-disc stars are in a retrograde component. The curves are the fitted values using the $f_{\rm T}(z)$ ratio from simulated vertical profiles.}
\label{fig_nbody_profiles}
\end{figure*}  

\begin{figure}
  \centering
   \includegraphics[width=0.48\textwidth]{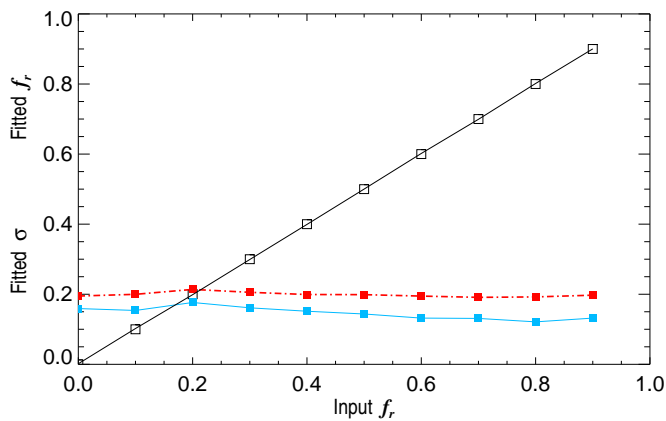}
\caption{In black, fitted retrograde fraction for a range of input fractions; boxes indicate the input values. Continuous blue and dash-dot red curves are the fitted `effective' velocity dispersions for the thin- and thick-disc components, respectively.}
\label{fig_nbody_profiles_2}
\end{figure} 

Figure~\ref{fig_nbody_profiles} shows examples of fitted profiles overplotted on synthetic data, for various amounts of retrograde thick-disc stars, while Fig.~\ref{fig_nbody_profiles_2} collects the fit results to a full range $f_{\rm r} =0.0, 0.1, ..., 0.9, 1.0$.  In the example shown, the synthetic profile is constructed at axial range $x/h_R = 2.0-2.5$, and the above model is used with $f(x) = 3.5$. Figure~\ref{fig_nbody_profiles_2} shows that altogether the fit recovers retrograde fractions remarkably well: the deviation $\left|f_{\rm r}({\rm fit})-f_{\rm r}({\rm data})\right|$ stays well below 0.01. Also the fitted values of $\sigma_{\rm t}$ and $\sigma_{\rm T}$ are practically the same for all values of $f_{\rm r}$ and within $10\%$ of the input $N$-body values.

The values obtained for this $N$-body model are also marked in Fig.~\ref{fig_toy}, overplotted on the toy-model curves. Although there is a systematic offset compared to the toy model, the qualitative trends are quite close to those in the $N$-body model. The only exception is the behaviour for $f_{\rm r}$ close to zero, where the actual vertical gradient in $\sigma$ is much smaller than what the toy model suggests. This is related to the above-discussed effect of LOS integration on the thin-disc velocity dispersion, which hides the actual intrinsic velocity dispersion. In this respect, observed galaxies can be expected to be close to $N$-body data. 

\begin{figure}
  \centering
   \includegraphics[width=0.48\textwidth]{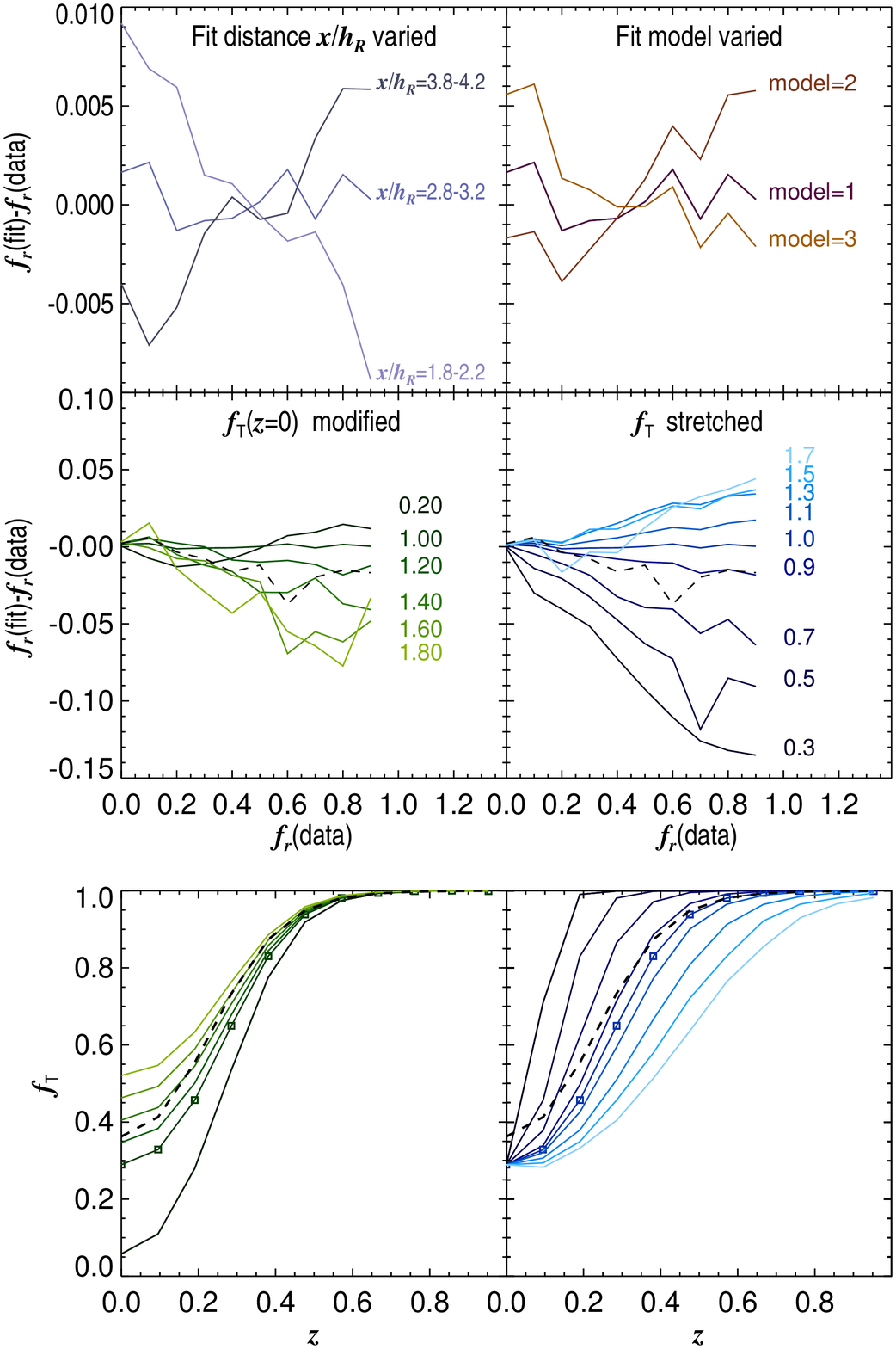}
\caption{Various factors affecting the error between the actual input fraction of retrograde thick-disc stars in the $N$-body model, and the fraction returned by the fit. The {\it upper left} panel shows fits with Eq.~\ref{eq_asym_model2}, using data from various axial distance intervals as indicated by the labels. In the {\it upper right} panel, different variants of Eq.~\ref{eq_asym_model2} are tested, using the data from the axial interval $x/h_R=2.8-3.2$: compared to the nominal case (``model=1''), in ``model=2'' the latter term in Eq.~\ref{eq_asym_model2} has been omitted, while for ``model=3'' the $\sigma_{\rm t}^2$ has been replaced with $\sigma_{\rm t}^2-\sigma_{\rm g}^2$, to correct for the contamination of LOS-velocity dispersions by systematic velocities. The {\it middle} panels (notice the change in the $y$-axis range) show the effect of using an incorrect fraction for thick disc, illustrated in the corresponding panel in the {\it lowermost} row. In the {\it left} frame the shape of $f_{\rm T}$ is retained, but the contribution of the thick disc in the equatorial plane has been multiplied with a factor indicated by the labels. In the {\it right} frame, the $f_{\rm T}$ profile has been stretched/dilated in the vertical direction by the amount indicated by the labels. The dashed lines indicated the fit obtained using the $f_{\rm T}$ estimated from a photometric thin-thick disc decomposition to the simulation data.}
\label{fig_nbody_errors}
\end{figure}  

The above success in retrieving $f_{\rm r}$ from $N$-body data is not fully representative of the application to true data, in which many uncertainties are present. To see how robust the fitting is, we experiment by varying the axial range over which the fit is done and also assess the effects of using inappropriate values for the term $f(x)$. Moreover, we introduce an uncertainty into the $f_{\rm T}$ factor because it is the most uncertain in real data. In Fig.~\ref{fig_nbody_errors} various factors are explored. In the upper left, synthetic data from different axial distances are used, while in the upper right, the details of the fitted model are varied; both have negligible effect on the results. These and similar tests with varying $f(x)$ (e.g.~dropping the  explicit $z$ dependence, or omitting the $a_0$ parameters) indicate that the precise form of the model function for $V_i^2(z)$ is not important for recovering the $f_{\rm r}$ fraction.

On the other hand, the middle row of Fig.~\ref{fig_nbody_errors} explores uncertainties due to the adopted thin/thick decomposition, quantified by using  different $f_{\rm T}$ profiles, where either the fractional contribution of thick-disc stars at the equatorial plane is multiplied by some factor (left), or the $f_{\rm T}$ profile is stretched/dilated in the vertical direction by some factor. Also shown is a photometric decomposition made on $N$-body data, using the same method that was used in \citet{CO18} for observed galaxies (see below). In this photometric decomposition, the central plane contribution of the thick disc was overestimated by a factor of approximately 1.25 (36\% instead of 29\%) and the $f_{\rm T}$ profile was additionally stretched by a factor of about 1.1 compared to the actual $N$-body profile. As expected, the errors in the fitted $f_{\rm r}$ fraction are considerably larger than those caused by the details of the fitting model. Nevertheless, even assuming relatively large errors, the ratios $f_{\rm r}$ are retrieved with a better than $10\%$ accuracy.

\subsection{Application of the formalism to the MUSE data}

\label{numerical3}

\begin{table}
 \caption{Disc scale lengths and scale heights and FWHMs of the MUSE observations}
 \label{sample2}
 \centering
 \begin{tabular}{l c c c c}
 \hline\hline
  ID & $h_R$& $h_{z{\rm t}}$&$h_{z{\rm T}}$& FWHM\\
     & ($^{\prime\prime}$)& ($^{\prime\prime}$)& ($^{\prime\prime}$)& ($^{\prime\prime}$)\\
  \hline
  ESO~157-49 & 11.5& 1.2 & 5.3 & 0.8\\
  ESO~469-15 & 7.1 & 1.0 & 4.9 & 0.9\\
  ESO~544-27 & 8.1 & 0.7 & 3.0 & 1.2\\
  IC~217     & 19.2& 1.5 & 5.3 & 1.0\\
  IC~1553    & 7.7 & 0.9 & 2.7 & 1.3\\
  PGC~28308  & 9.7 & 1.0 & 5.2 & 0.7\\
  \hline
 \end{tabular}
 \tablefoot{The scale heights are from \citet{CO18}. The scale lengths were estimated by assuming an exponential disc seen edge-on. The fit to obtain the scale lengths was performed for the axial range between the galaxy centre and the outermost position covered by the MUSE FOV.}
\end{table}

We are now ready to estimate retrograde fractions from MUSE datacubes. In the application to MUSE data we use the density profiles from the thin-thick disc decompositions made in \citet{CO18} for the $3.6\,\mu{\rm m}$ S$^4$G data. These decompositions were based on the assumption that the two components can be represented with isothermal models in mutual hydrodynamical equilibrium. The effective scale heights and the common radial scale lengths from that paper are given in Table~\ref{sample2}. We start the fitting the stellar velocity and $\sigma$ maps at $x>1.3x_{\rm f}$ using Eqs.~\ref{eq_asym_model1}--\ref{eq_asym_model2} to obtain $V_{\rm t}$ and $V_{\rm T}$; we use the observed gas velocities as a proxy for circular velocities. To obtain the combined $V$ and $\sigma$ we use the non-convolved model profiles $\rho_{\rm t}(z)$ and $\rho_{\rm T}(z)$ from \citet{CO18}, and convolve the first and second velocity moments with the 1D MUSE PSF,
\begin{equation}
\label{int1}
\overline{V}(z)=\frac{\int {\rm PSF} (\left|z-z^{\prime}\right|) \left[ \rho_{\rm t}(z^{\prime}) \overline{V}_{\rm t}(z^{\prime}) +\rho_{\rm T}(z^{\prime}) \overline{V}_{\rm T}(z^{\prime}) (1-2f_{\rm r}) \right] dz^{\prime}}{\int {\rm PSF} (\left|z-z^{\prime}\right|) \left[ \rho_{\rm t}(z^{\prime}) +\rho_{\rm T}(z^{\prime})  \right] dz^{\prime}},
\end{equation}
\begin{equation}
\label{int2}
\overline{V^2}(z)=\frac{\int {\rm PSF} (\left|z-z^{\prime}\right|) \left[ \rho_{\rm t}(z^{\prime}) \overline{V_{\rm t}^2}(z^{\prime}) +\rho_{\rm T}(z^{\prime}) \overline{V_{\rm T}^2}(z^{\prime}) (1-2f_{\rm r}) \right] dz^{\prime}}{\int {\rm PSF} (\left|z-z^{\prime}\right|) \left[ \rho_{\rm t}(z^{\prime}) +\rho_{\rm T}(z^{\prime})  \right] dz^{\prime}},
\end{equation}
yielding $\sigma^2 = \overline{V^2}-\overline{V}^2$. We assumed the MUSE PSF to be a Gaussian with the FWHM determined using {\sc iraf}'s {\sc imexamine}\footnote{The 1D integration in Eqs.~\ref{int1} and \ref{int2} is only valid for a Gaussian PSF. Other PSF shapes need to account for a full 2D integration \citep[see Fig.~2 in][]{CO18}.}. The adopted FWHM values are shown in Table~\ref{sample2}.

The application of our formalism to six MUSE galaxies is shown in Fig.~\ref{drift}. The labels in the plot indicate the fitted $f_{\rm r}$, together with an upper and a lower bound obtained by assuming that the photometric profiles are stretched/dilated by factors of 1.3 and 0.7 with respect to the real one: this corresponds to roughly twice the error in the similar photometric fit to the $N$-body data. In all cases the retrograde fractions, allowing the above uncertainty, are consistent with zero.

It is possible that our stellar velocity dispersion determinations are slightly underestimated if the FWHM of the observations has been overestimated. However, we have tested whether a negligible retrograde material function would also be found if we assume an underestimated FWHM. For this experiment we modelled the wavelength-varying FWHM of the instrumental line spread function provided by the MUSE pipeline calibration files. These files describe the profile of the lines in the pixel tables. When pixel tables are processed to build the datacubes this results in a small widening of the FWHM. The FWHM derived from the calibration files is $\sim0.2\,\AA$ narrower than that estimated by Eq.~\ref{bacon} in the blue side of the spectrum and $\sim0.1\,\AA$ narrower at the calcium triplet. Even when such an underestimated FWHM is used, only IC~217 has a mildly significant positive fraction of retrograde stars ($f_{\rm r}=0.070^{+0.026}_{-0.037}$).

We conclude that none of the six galaxies with a clear thick disc exhibit a significant retrograde thick-disc component.

\section{Discussion}

\label{discussion}

The kinematic data that we present show that retrograde material is not generally dominant at any height in the discs of the galaxies under study. For a couple of galaxies there is some evidence for strongly lagging -- non-rotating or counter-rotating -- material close to the centre, where a spheroid could be found. However, such evidence vanishes at large distances from the galaxy centre, where the rotation curves become flat.

With the numbers at hand (from this paper and from previous studies), only one thick disc in seventeen -- that in FGC~227 \citep{YOA05, YOA08} -- necessarily contains a large fraction of retrograde material.

Large amounts of retrograde thick-disc material would strongly support an external origin. The opposite, that a lack of such material excludes an accretion origin, is not necessarily true. The reasons for that are related to the way dynamical friction \citep{CHAN43} works, as shown below.

To first order, we can think of galaxies as rotating discs within a non-rotating or slowly rotating, pressure-supported, dark matter halo. At large distances from the galaxy centre, an accreted object is mostly influenced by the friction caused by the non-rotating halo and its behaviour does not depend on whether the accreted object is in a prograde or a retrograde orbit \citep[see, e.g.~the mass-loss rates as a function of time in][when the infalling galaxy is far away from the centre of the halo]{PEN02}. Differences between these two kinds of orbit arise when the interaction with the disc increases, as dynamical friction is more efficient for an accreted object with a velocity similar to that of the disc (a situation of near-resonance). Hence, the disc of a galaxy is more efficient at braking a prograde satellite. This has two effects. The most obvious one is that mergers are faster in the prograde case \citep[but only by a factor of at most two; ][]{PEN02, VILL08}. The second is that the accreted objects are strongly dragged towards the main galaxy disc plane in the case of a prograde merger with a low inclination \citep[$i<20\degr$;][]{PEN02, READ08}. Therefore, prograde mergers are the most likely to result in stars deposited into a thick-disc component (as opposed to being deposited into the stellar halo). However, a very-low-inclination retrograde merger would also result in a stellar distribution resembling that of a thick disc \citep{READ08}.

The fundamental question here is: if thick-disc stars were accreted in a small number of merger events, what would be the likelihood for a given galaxy to have a large fraction of retrograde material? One difficulty in answering this question is that many numerical studies on dynamical friction sample the prograde encounter parameter space much better than they do the retrograde one. However, satellite systems are pressure-supported -- with small net rotation -- so retrograde and prograde orbits are nearly equally likely \citep{SA07, READ08}.

In the discussion that follows, we only consider relatively major mergers as potential thick-disc progenitors. Small satellites take a long time to decay because dynamical friction scales with the mass of the infalling object to the second power. Furthermore, when small satellites are finally dissolved, they do so into structures that do not resemble a thick disc \citep{READ08}.

In \citet{PEN02} and \citet{VILL08}, a massive retrograde object would take at most two times longer than a prograde object to dissolve. Since the decay time for the massive satellites in those simulations is smaller than the Hubble-Lema\^itre time, both prograde and retrograde objects bound to the galaxy at its formation time are expected to have merged by now. However, among objects falling into the main galaxy potential well at a later time, not all have merged yet, and retrograde ones are the most likely not to have merged. Let us make a rough estimate of the number of prograde mergers versus that of retrograde mergers. For the models in \citet{PEN02}, a massive satellite with an initial apogalacticon of 55\,kpc and a moderate eccentricity $e=0.45$ would dissolve in $\sim5$\,Gyr if in a prograde orbit and in $\sim7$\,Gyr if in a polar orbit \citep[we assume that the decay time of a polar orbit is similar to that of a retrograde orbit, as shown for objects with larger eccentricities in][]{PEN02}. If we assume the dark matter halo merger rate per gigayear to go like $\propto(1+z)^2$ \citep[e.g.~][]{FAKH08} we obtain that the fraction of retrograde mergers is $\sim45\%$ for a galaxy of 10\,Gyr old. Here we have not accounted for the primary galaxy mass evolution -- which is likely to result in a smaller Coulomb logarithm at early times and hence into a longer decay time -- or for the possibility of nearly circular orbits with a decay time close to a Hubble-Lema\^itre time for a polar orbit \citep[on the other hand, satellites with an eccentricity larger than $e=0.45$ would decay faster than stated here;][]{PEN02}. Therefore, to an order of magnitude, we can estimate the ratio of retrograde-to-prograde mergers in a Hubble-Lema\^itre time to be about one third.

Among the prograde mergers, only the ones with $i<20\degr$ would be plane-dragged and dissolve into a thick disc, whereas the others would disperse into the halo. This can also be seen in \citet{QU11} where, using GalMer giant-dwarf merger simulations \citep{CHIL10} with $i=33\degr$ and $i=60\degr$ retograde encounters, the authors find a very small resulting fraction of retrograde stars at and below $z\sim1\,{\rm kpc}$ (their primary galaxy, however, is more massive than those studied here). It is to be expected that for retrograde encounters a smaller range of inclinations would result in a thick disc. In \citet{READ08} the only retrograde simulation ends up forming a thick disc and has an inclination of $i=10\degr$. Therefore, for the model galaxies in that set of simulations, this is a lower limit for the inclination of retrograde objects forming thick discs. Using that number, and assuming isotropic accretion, we obtain that a retrograde satellite has a chance that is ${\rm cos}\,70\degr/{\rm cos}\,80\degr\approx2$ smaller than that of a prograde one to end in an accreted thick disc.

If we combine the two factors -- that from the decay time and that from the in-plane drag -- we obtain that at the present time approximately one sixth of the objects accreted into a thick disc should be retrograde, and the others prograde. We warn that this number is only an order-of-magnitude estimate and certainly depends heavily on the mass distributions involved and on the orbital elements of the merging objects.

If we assume that all thick discs are accreted in a single merger event, a sixth of thick discs should have retrograde material. Considering a binomial distribution with $p=1/6$, there is a two sigma discrepancy between our prediction and the observed fraction of thick discs with retrograde material (1 in 17). If we assume that the thick disc is made in a few (two or three) accretion events, the tension worsens, since having any one of these coming from a retrograde orbit would likely leave an imprint detectable with our formalism, as described in Sect.~\ref{numerical}.

The above numbers, as well as the inferred origin for the Milky Way thick disc, are evidence against accretion of stars as the main thick-disc formation mechanism. There have been simulations indicating that only up to $\sim50\%$ of the masses of thick discs have been accreted \citep{A03, READ08}, but our spectroscopic data and those in the literature seem to indicate that the fraction is lower.

We note that a non-accretion origin does not necessarily imply that thick discs are not created by interactions. Indeed, merger events -- even those for which stars of a disrupted satellite end up in the halo -- heat the pre-existing thin disc and can participate in the formation of the thick disc \citep{KAZ08}.

\section{Summary and conclusions}

\label{conclusions}

In this paper we present new kinematical information for thick discs. Specifically, we observed eight nearby galaxies with the integral field unit MUSE at the VLT. Six of the galaxies have distinct thick discs according to \citet{CO18} and hence we can study the thick-disc kinematics. This increases the number of galaxies with known thick-disc kinematics to seventeen.

In Sect.~\ref{numerical} we present a formalism able to estimate the fraction of thick-disc retrograde material from the velocity and velocity dispersion maps. None of the thick discs in our sample show signs of retrograde star fractions larger than 10\%. Within the uncertainties, the retrograde star fraction is probably compatible with zero for the six galaxies.

Of the 17 thick discs with known kinematics only one shows clear evidence for retrograde material, which is a clear signature for an accretion origin. Using information from the simulations in \citet{PEN02} and \citet{READ08} we estimate that approximately one sixth of the satellites accreted into a thick disc had a retrograde orbit. This is in tension with the observed fraction of galaxies with retrograde thick-disc material as estimated from the rotation curves (1 in 17). We are planning new MUSE observations to confirm this result. Undoubtedly, additional new MUSE data for edge-on galaxies will be obtained by other groups and can also be used to increase the sample size.

Our findings, and the lack of evidence for large fractions of accreted material in the Milky Way thick disc, seem to discard accretion as the main thick-disc formation mechanism for most galaxies. The thick-disc origin is now narrowed down to either an internal origin or a scenario where a pre-existing thin disc is dynamically heated by encounters. Either way, the source of most of the thick-disc material is most likely to be internal.

\begin{acknowledgements}
 We thank the referee for in-depth comments that helped to quantify the fraction of retrograde material in the discs. We thank  Dr.~Lodovico Coccato for his help at fine-tuning the data reduction. We thank Dr.~Simon Conseil for his help with {\texttt ZAP} and Dr.~Peter Weilbacher for his suggestions on the line spread function handling. SC and HS acknowledge support from the Academy of Finland (grant No.~297738). HS, JHK, and RFP acknowledge financial support from the European Union’s Horizon 2020 research and innovation programme under Marie Sk{\l}odowska-Curie grant agreement No.~721463 to the SUNDIAL ITN network. JHK acknowledges additional support from the Spanish Ministry of Economy and Competitiveness (MINECO) under grant number AYA2016-76219-P, as well as from the Fundación BBVA under its 2017 programme of assistance to scientific research groups, for the project ``Using machine-learning techniques to drag galaxies from the noise in deep imaging'', and from the Leverhulme Trust through the award of a Visiting Professorship at LJMU. We acknowledge the usage of the HyperLeda database (http://leda.univ-lyon1.fr). This research has made use of the NASA/IPAC Extragalactic Database (NED) which is operated by the Jet Propulsion Laboratory, California Institute of Technology, under contract with the National Aeronautics and Space Administration.
\end{acknowledgements}

\bibliographystyle{aa}
\bibliography{rotation}

\begin{appendix}

\section{Individual galaxies map and plots}

\label{appen}

In this appendix we show the maps and plots derived for each of the galaxies in our sample. The plots are structured in two columns.

In the left column of plots we show the following.

\begin{enumerate}
 \item The white-light image obtained from the MUSE datacube.
 \item The stellar velocity map; the systemic velocity of the galaxy has been subtracted.
 \item The stellar velocity dispersion map.
 \item The gas velocity map as measured from emission lines; the systemic velocity of the galaxy has been subtracted.
\end{enumerate}

In the above four frames the diamond symbols indicate the galaxy centre and the north and the east are indicated by the ``backwards-L'' symbol at the top-right corner. The labels in the axes are in arcseconds. The ``NE'' and ``SW'' (``NW'' and ``SE'') labels indicate the northeastern and the southwestern (northwestern and southeastern) sides of the galaxy, respectively. The dividing line between the two halves is the midplane which is indicated by a thin continuous line. The two thick continuous line indicate the height above which the thick disc dominates the luminosity according to the fits in \citet{CO18}. These three lines, as well as the dashed lines, indicate the heights for which velocity curves have been constructed (see the plots in the right column).

The plots in the right column are structured as follows.
\begin{enumerate}
 \item Velocity curves for the heights indicated by the thin, the thick, and the dashed lines in the maps in the left column of plots. The curves for the NE (NW) half of the galaxy are shown at left and those for the SW (SE) half are shown at right. The vertical dashed lines indicate 1.3 times the position of the rotation curve flattening, $x_{\rm f}$. The colour coding indicates the height above the midplane to which a curve corresponds.
 \item Rotation velocity difference at a given height with respect to that of the midplane. The curves for the NE (NW) half of the galaxy are shown at left and those for the SW (SE) half are shown at right. The vertical dashed lines indicate 1.3 times the position of the rotation curve flattening according to our parametrisation in Eq.~\ref{equation}, $1.3\,x_{\rm f}$. The colour coding indicates the height above the midplane to which a curve corresponds.
 \item Average rotation velocity of the flatter part of the rotation curve ($x>1.3\,x_{\rm f}$) as a function of height for the red-shifted side of the galaxy (R) and the blue-shifted side of the galaxy (B). The vertical solid lines show the height where the thick disc starts to dominate, if applicable to the galaxy.
 \item Stellar and the gas rotation curves for selected heights. The circular symbols correspond to the stellar rotation curves and the lines to the gas rotation curves. Rotation curves for the midplanes are indicated by dark points and blue lines, and rotation curves for the largest height for which we constructed rotation curves are indicated by light points and red lines. The top panel is for the NE (NW) half of the galaxy and the lower panel is for the SW (SE) half. The vertical dashed lines indicate $x=1.3\,x_{\rm f}$. The dot lines indicate the zero axial distance to the centre and the zero velocity lines.
\end{enumerate}

\begin{figure*}
\begin{center}
\begin{tabular}{c c}
    \raisebox{-0.5\height}{\includegraphics[width=0.48\textwidth]{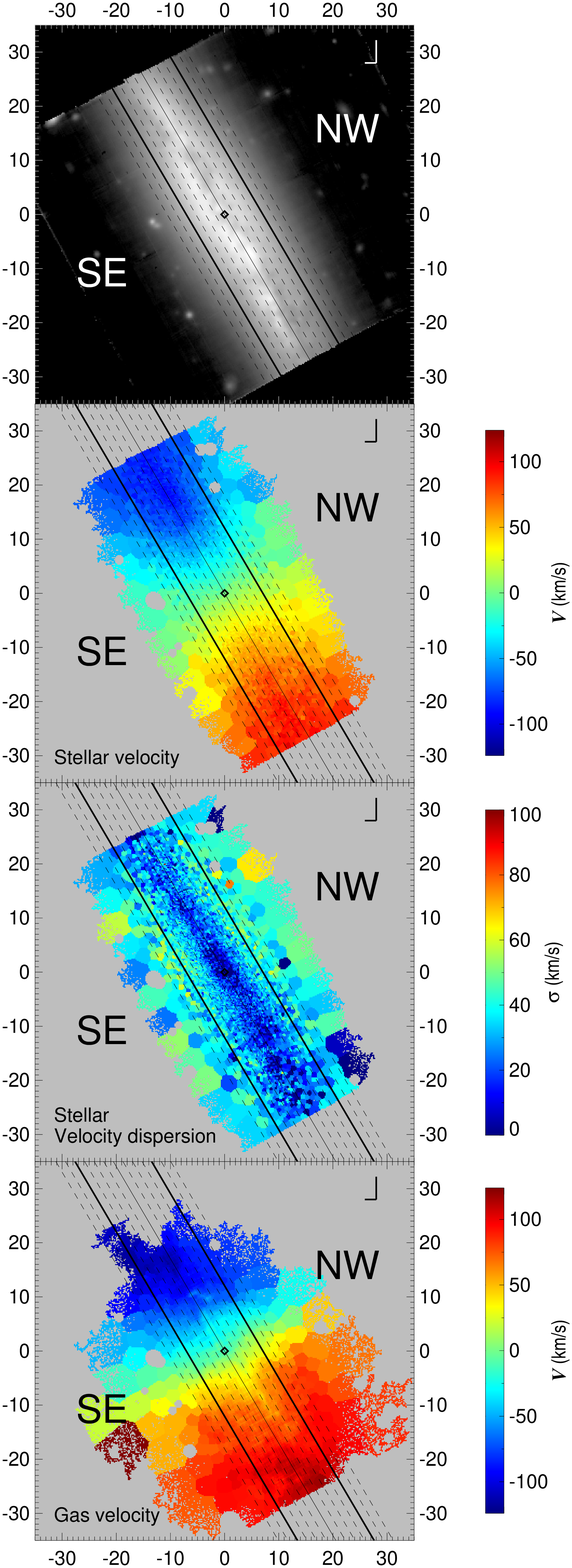}}&
  
  \begin{tabular}{c}
  {\Large{ESO~157-49}}\\
  \\
  \includegraphics[width=0.48\textwidth]{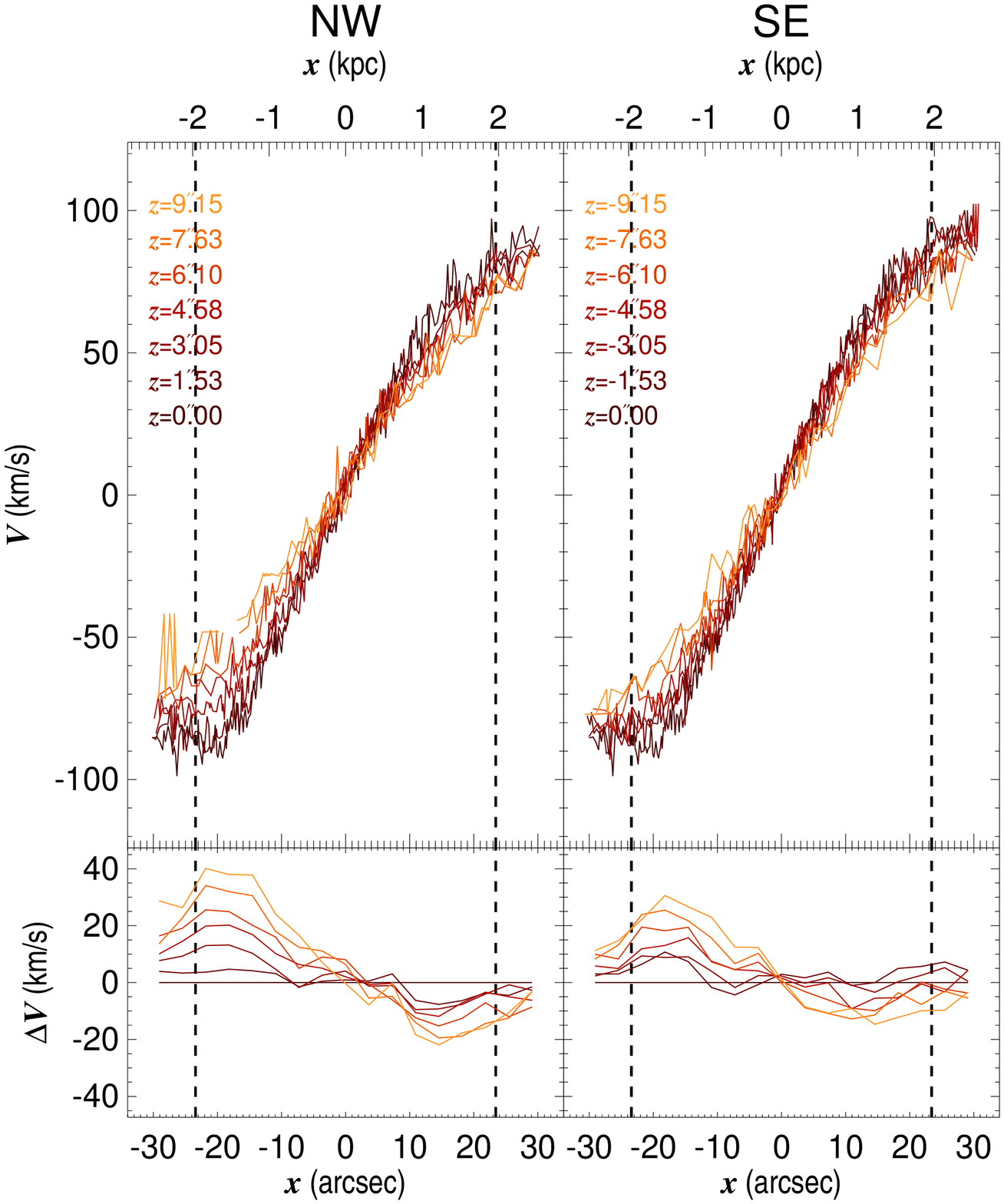}\\
  \includegraphics[width=0.48\textwidth]{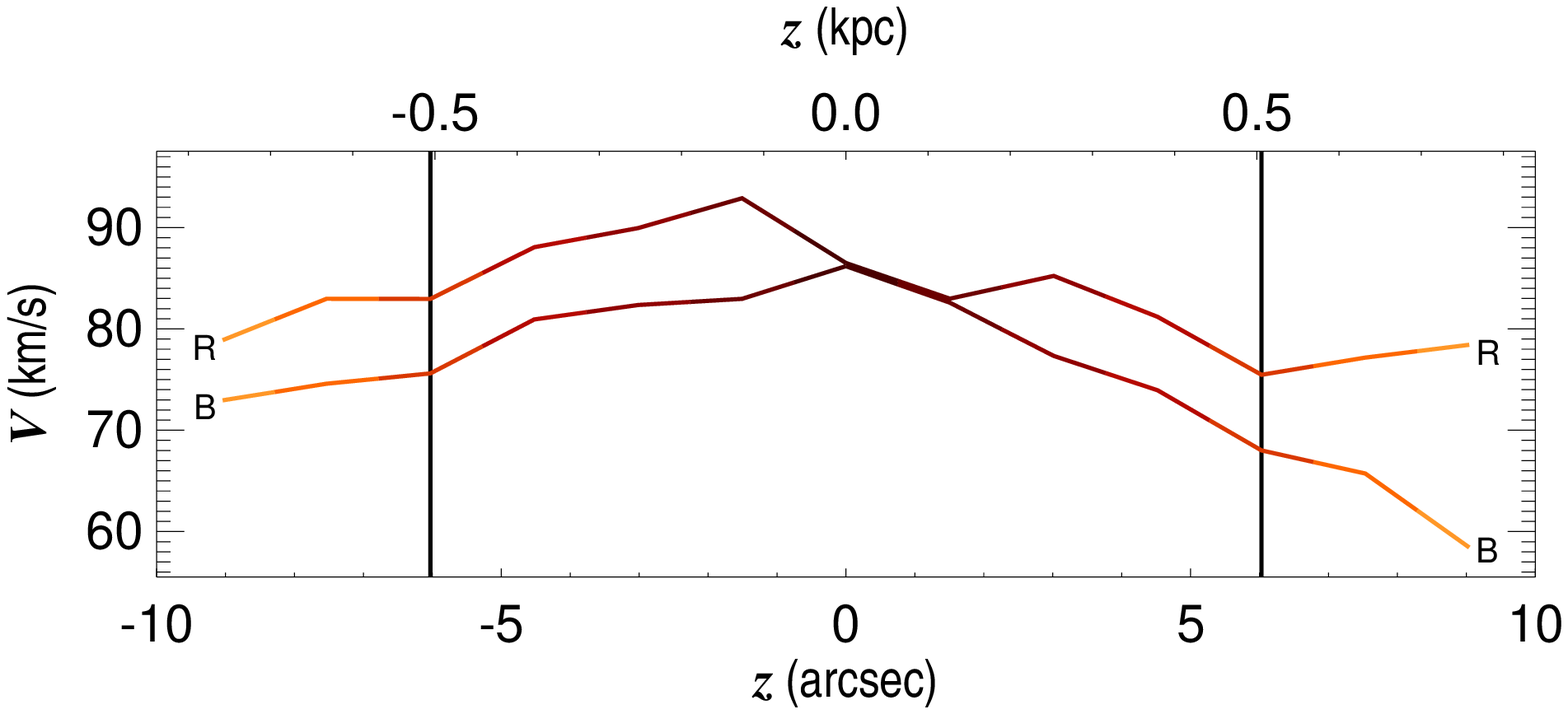}\\  
  \includegraphics[width=0.48\textwidth]{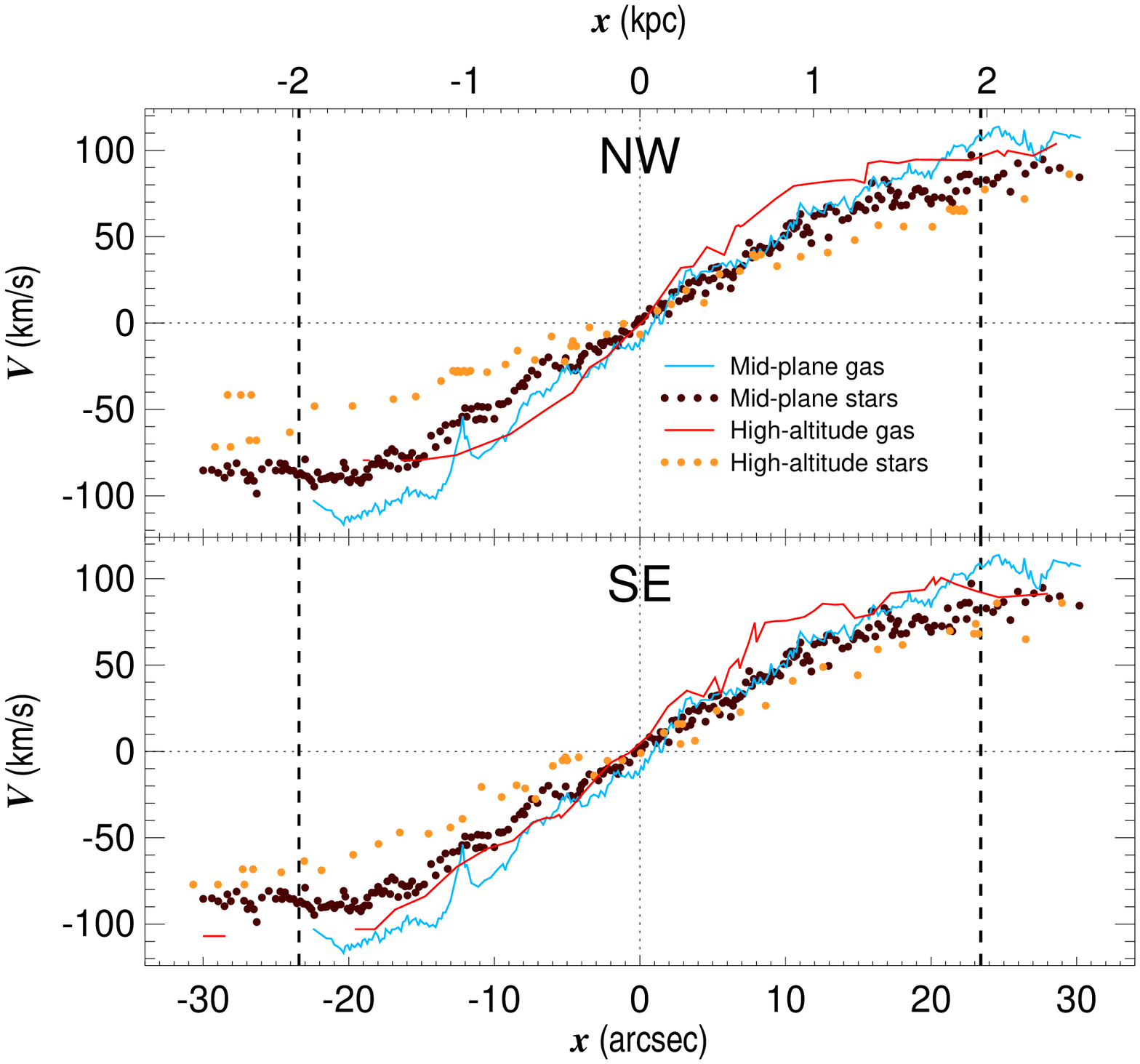}
  \end{tabular}
  
\end{tabular}

  \end{center}
  \caption{\label{ESO157-49} {Maps and plots corresponding to ESO~157-49.}}
\end{figure*}

\begin{figure*}
\begin{center}
 
\begin{tabular}{c c}
    \raisebox{-0.5\height}{\includegraphics[width=0.48\textwidth]{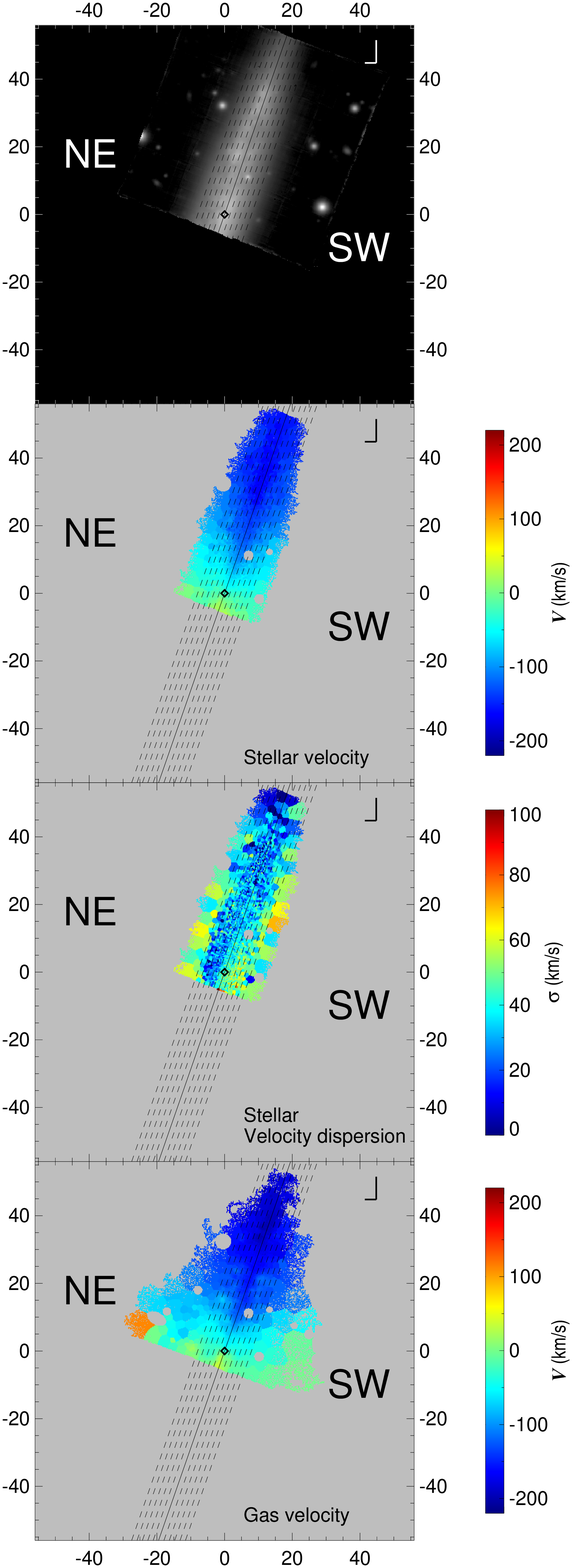}}&
  
  \begin{tabular}{c}
  {\Large{ESO~443-21}}\\
  \\
 
  \includegraphics[width=0.48\textwidth]{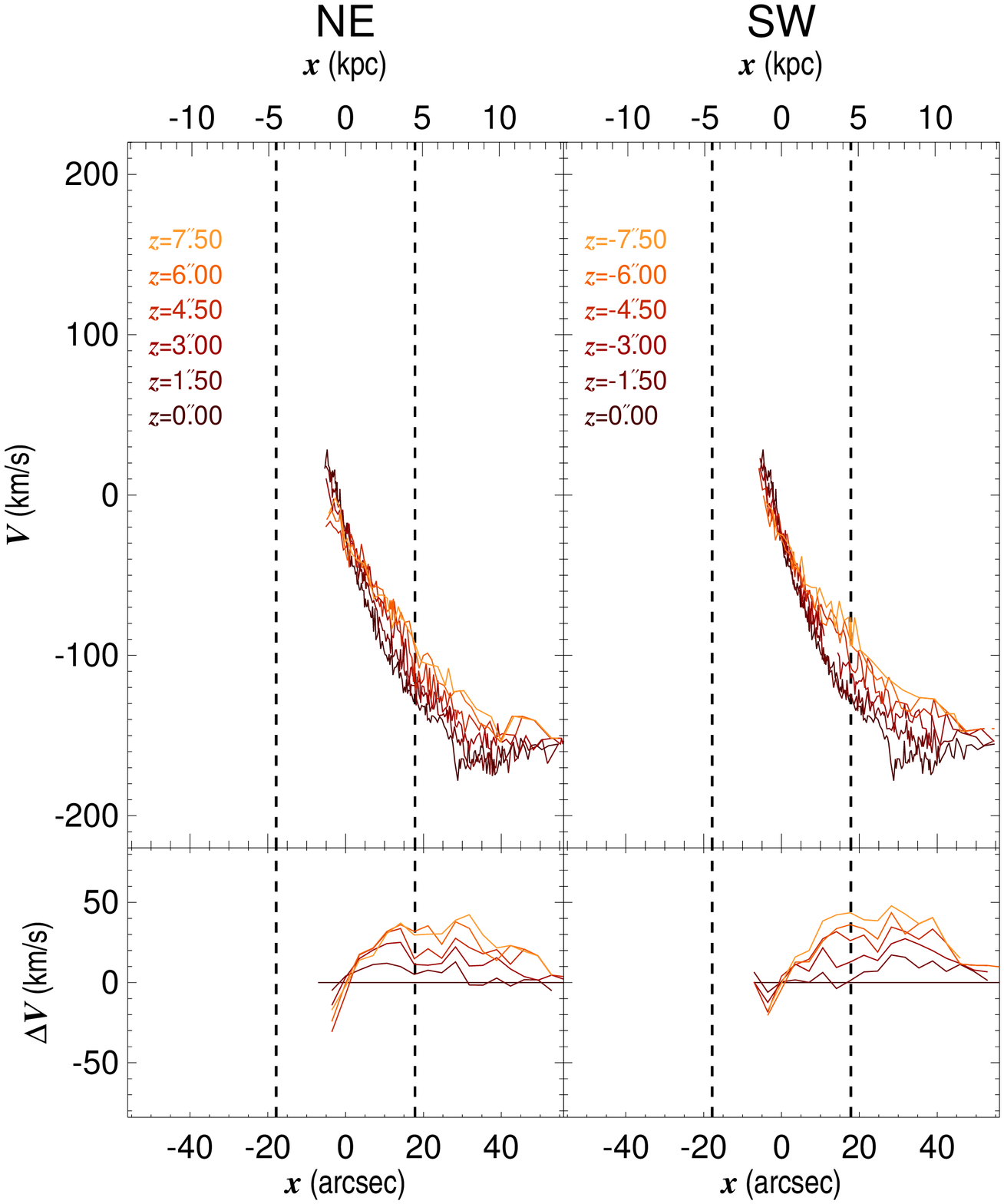}\\
  \includegraphics[width=0.48\textwidth]{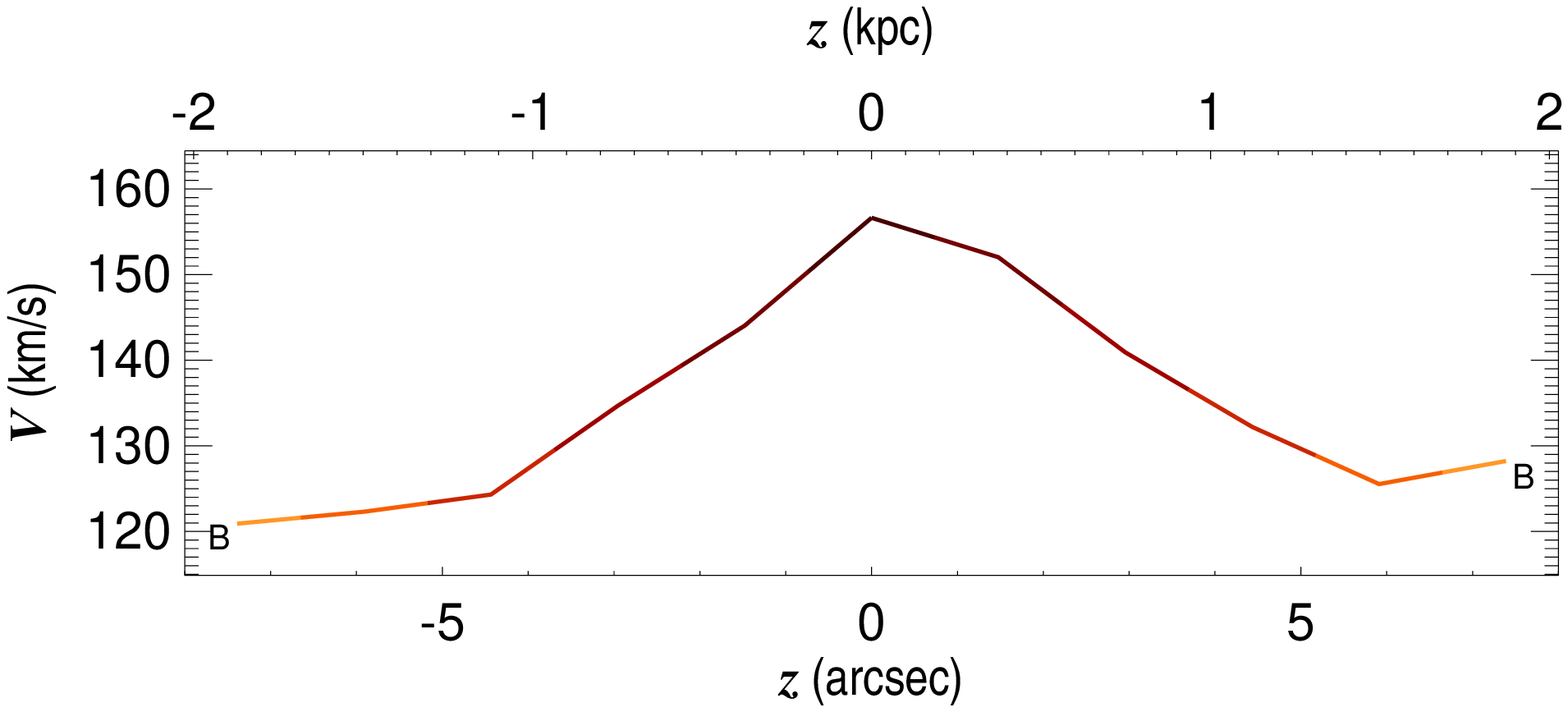}\\
  \includegraphics[width=0.48\textwidth]{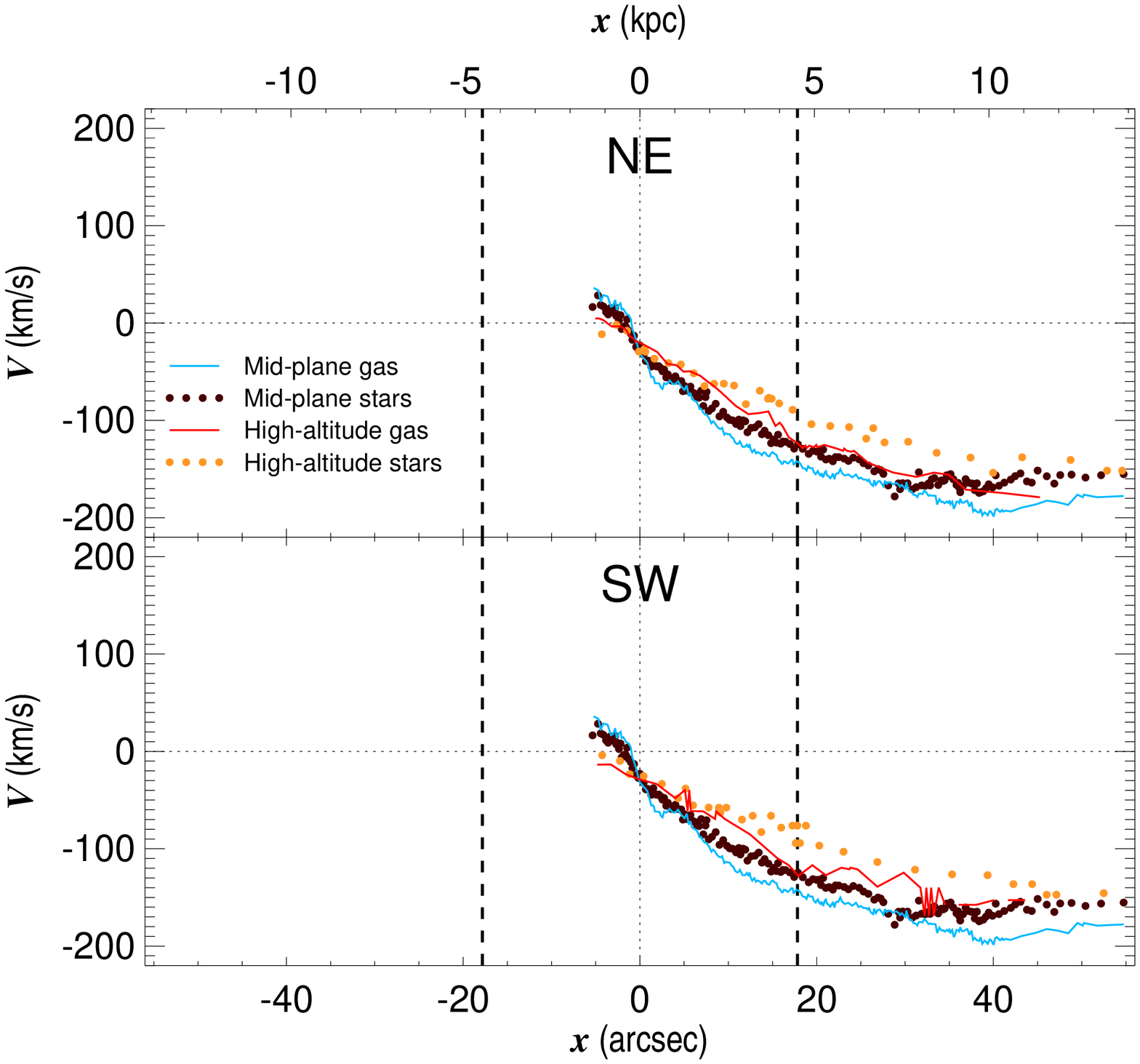}
  
  \end{tabular}
  
\end{tabular}

  \end{center}
  \caption{\label{ESO443-21} As in Fig.~\ref{ESO157-49}, but for ESO~443-21. \citet{CO18} found no distinct thin- and thick-disc structure for this galaxy.}
\end{figure*}

\begin{figure*}
\begin{center}
 
\begin{tabular}{c c}
    \raisebox{-0.5\height}{\includegraphics[width=0.48\textwidth]{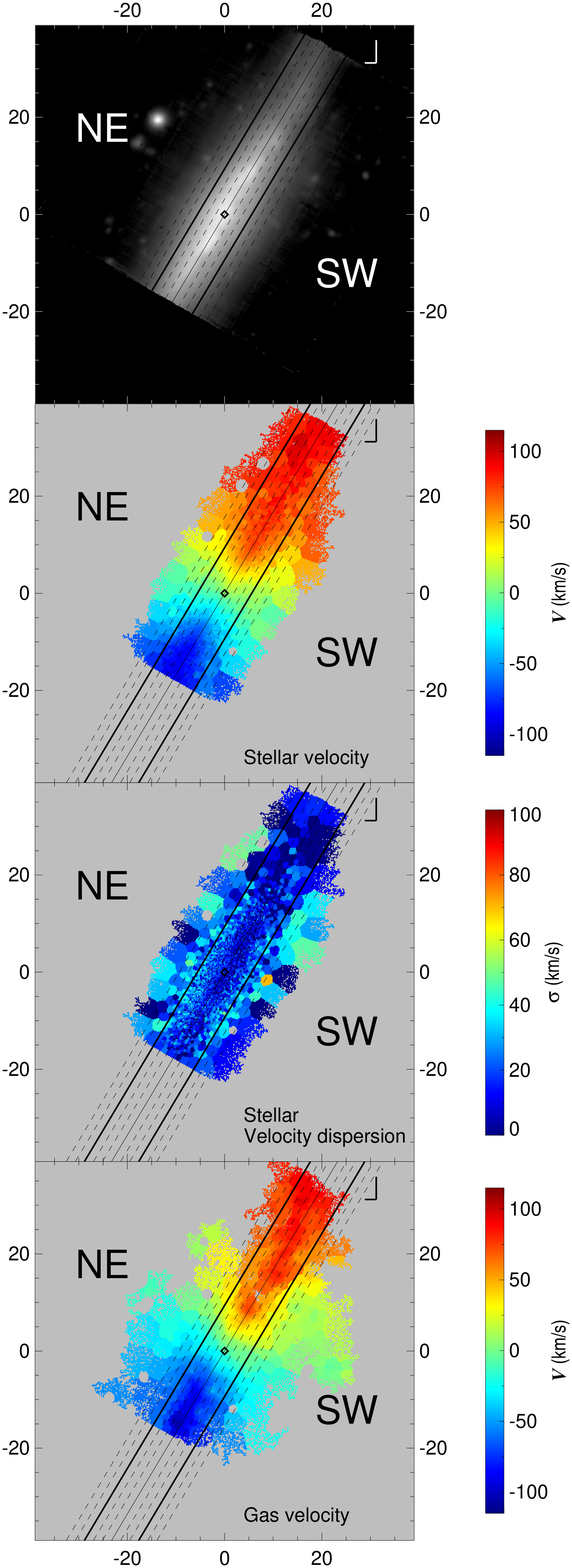}}&
  
  \begin{tabular}{c}
  {\Large{ESO~469-15}}\\
  \\
  \includegraphics[width=0.48\textwidth]{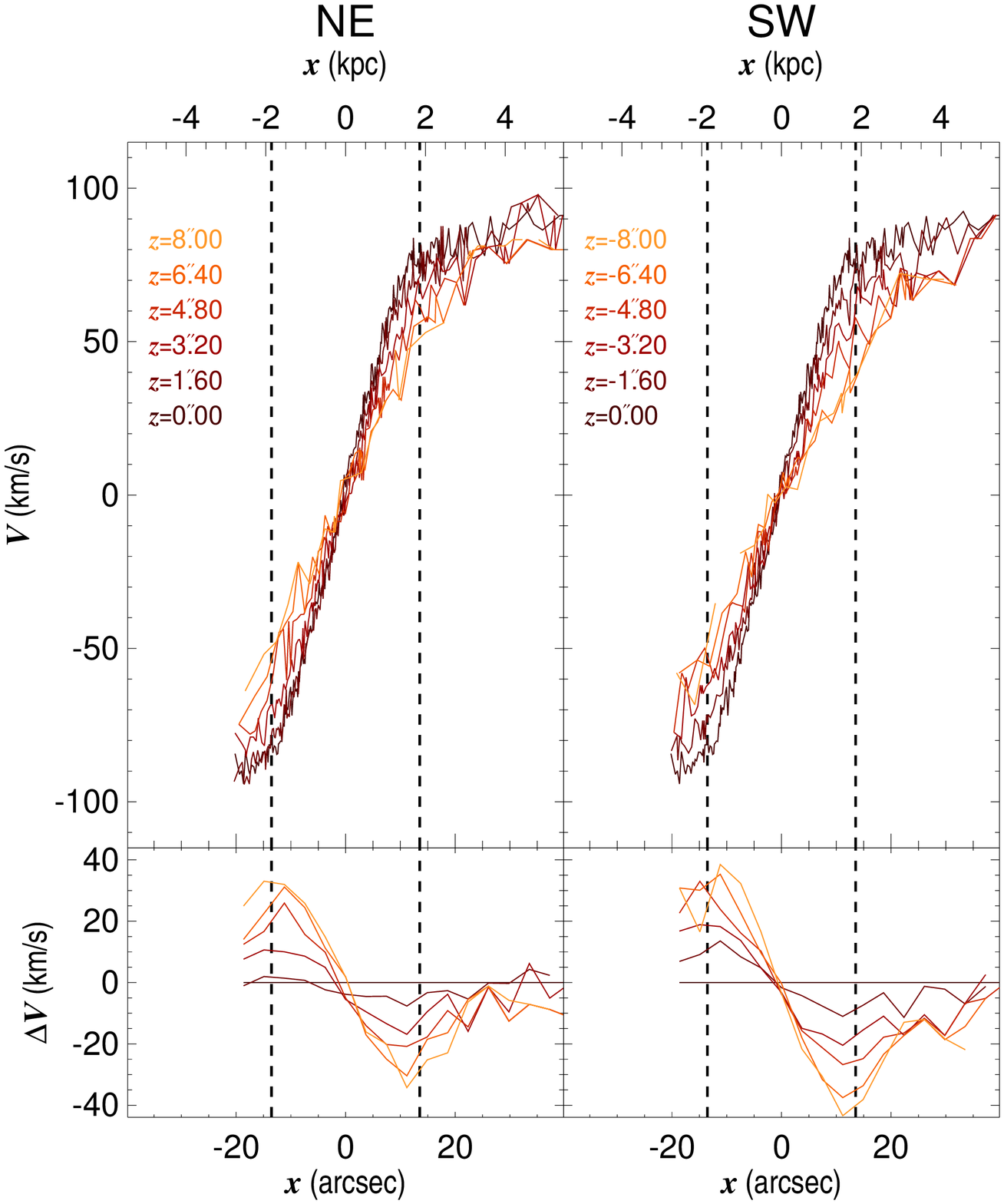}\\
  \includegraphics[width=0.48\textwidth]{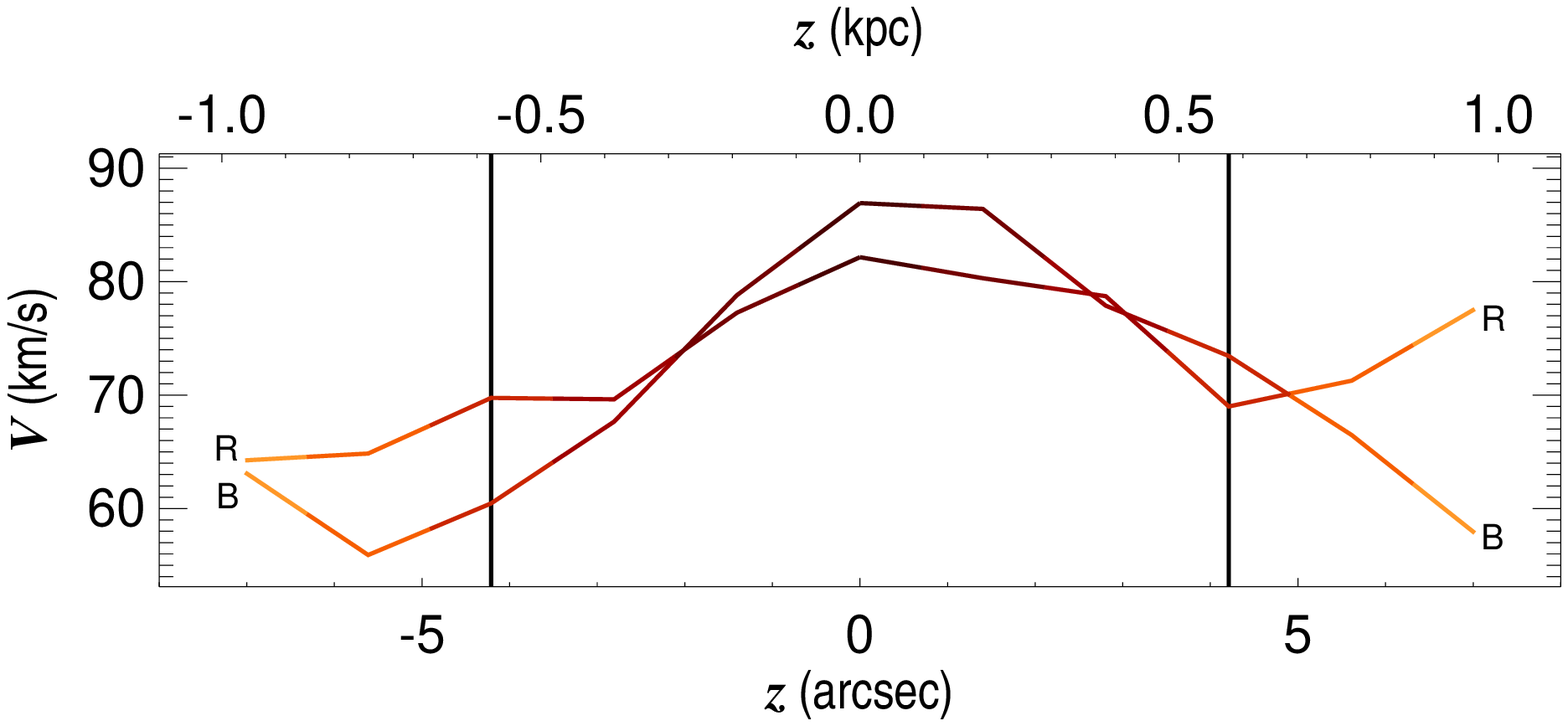}\\
  \includegraphics[width=0.48\textwidth]{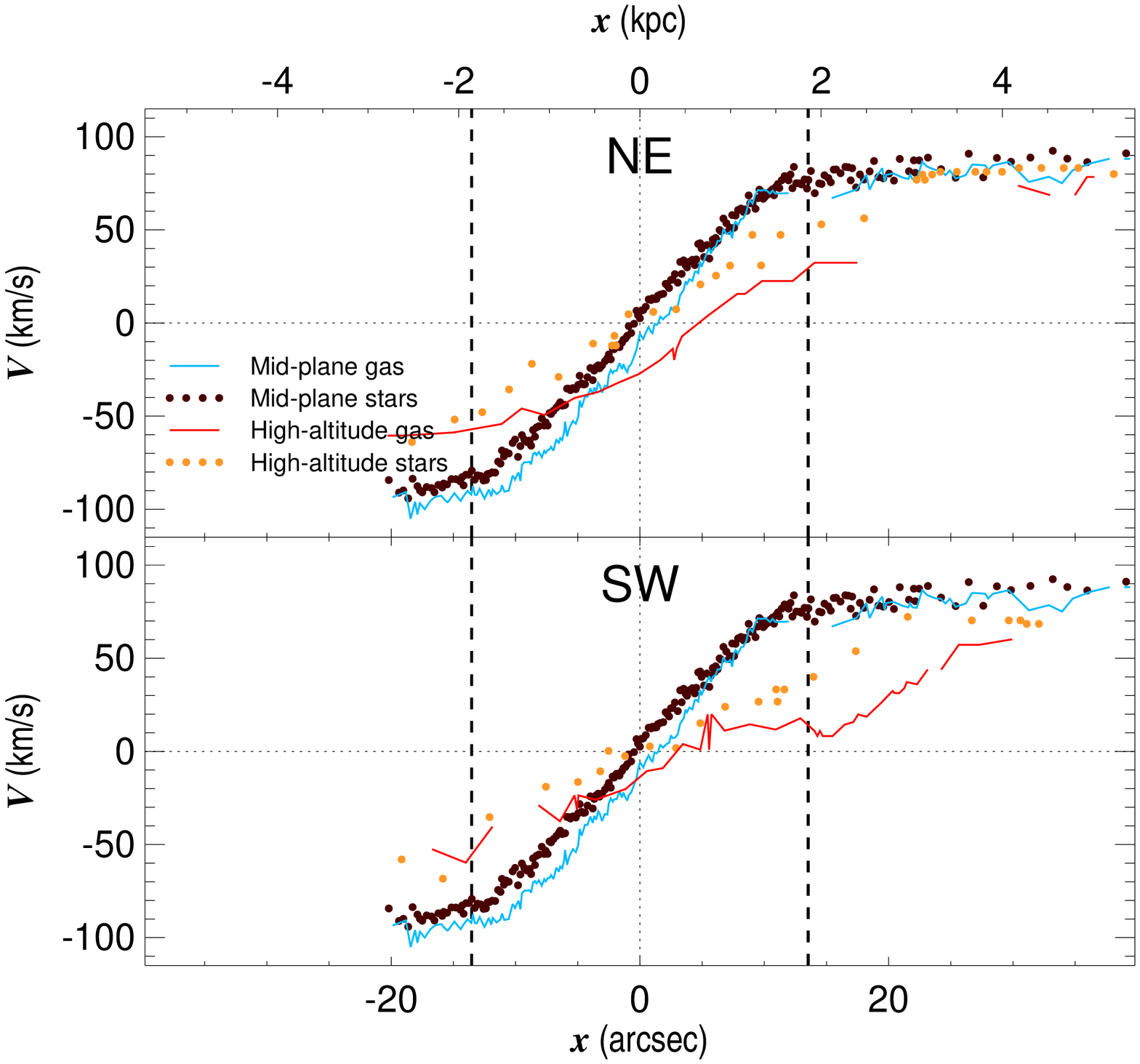}  
  \end{tabular}
  
\end{tabular}

  \end{center}
  \caption{\label{ESO469-15} As in Fig.~\ref{ESO157-49}, but for ESO~469-15.}
\end{figure*}

\begin{figure*}
\begin{center}
 
\begin{tabular}{c c}
    \raisebox{-0.5\height}{\includegraphics[width=0.48\textwidth]{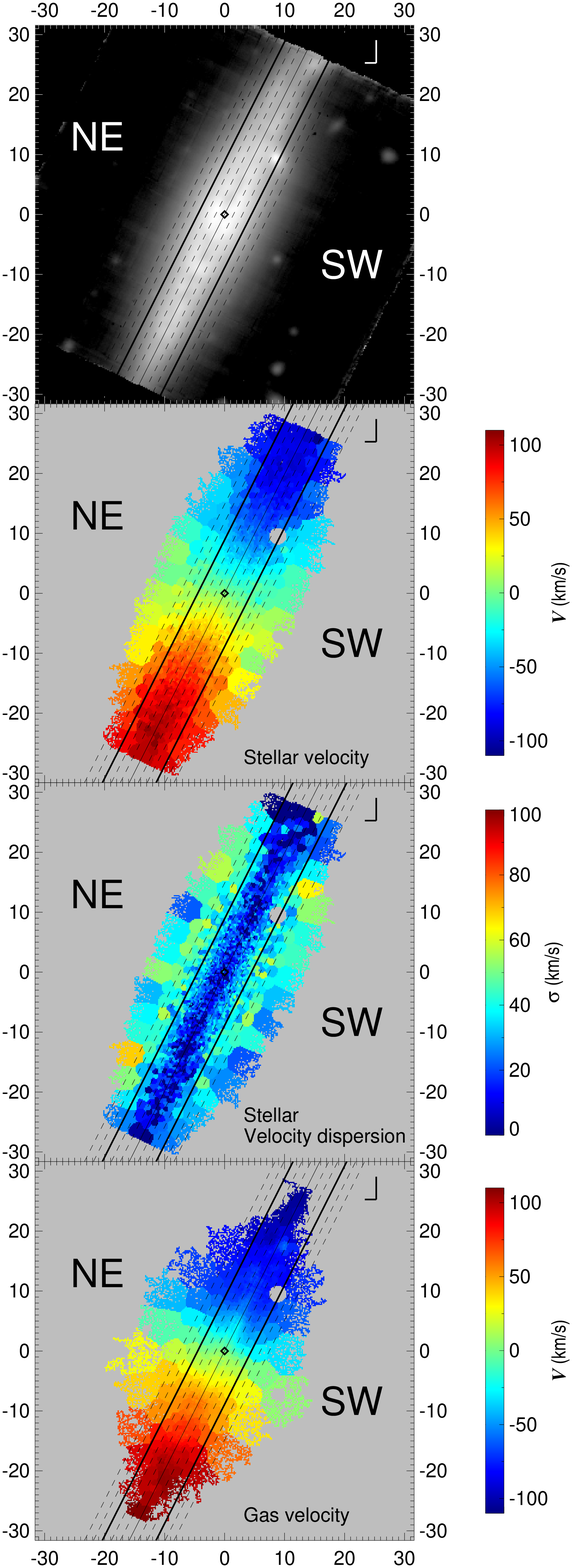}}&
  
  \begin{tabular}{c}
  {\Large{ESO~544-27}}\\
  \\
  \includegraphics[width=0.48\textwidth]{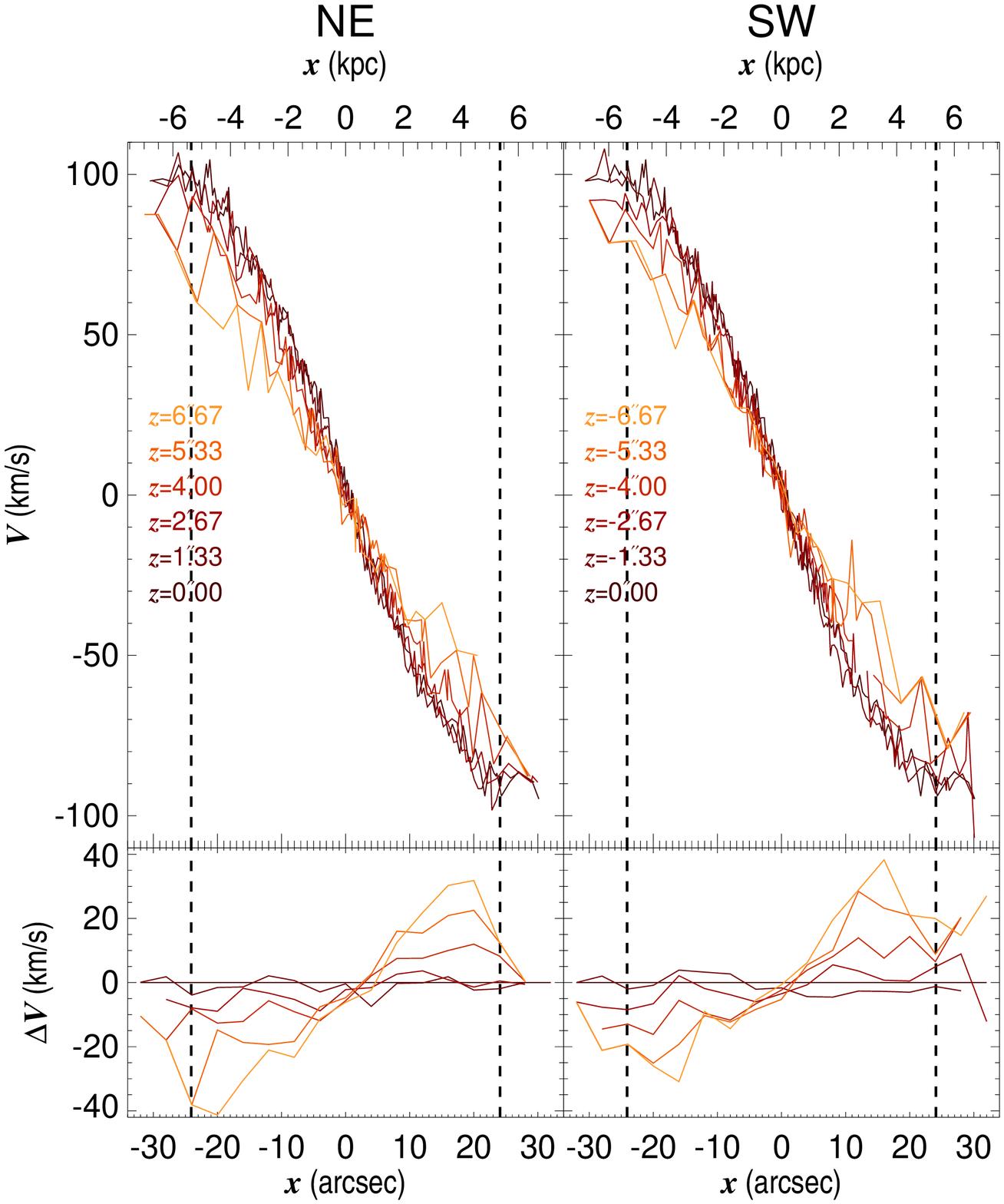} \\
  \includegraphics[width=0.48\textwidth]{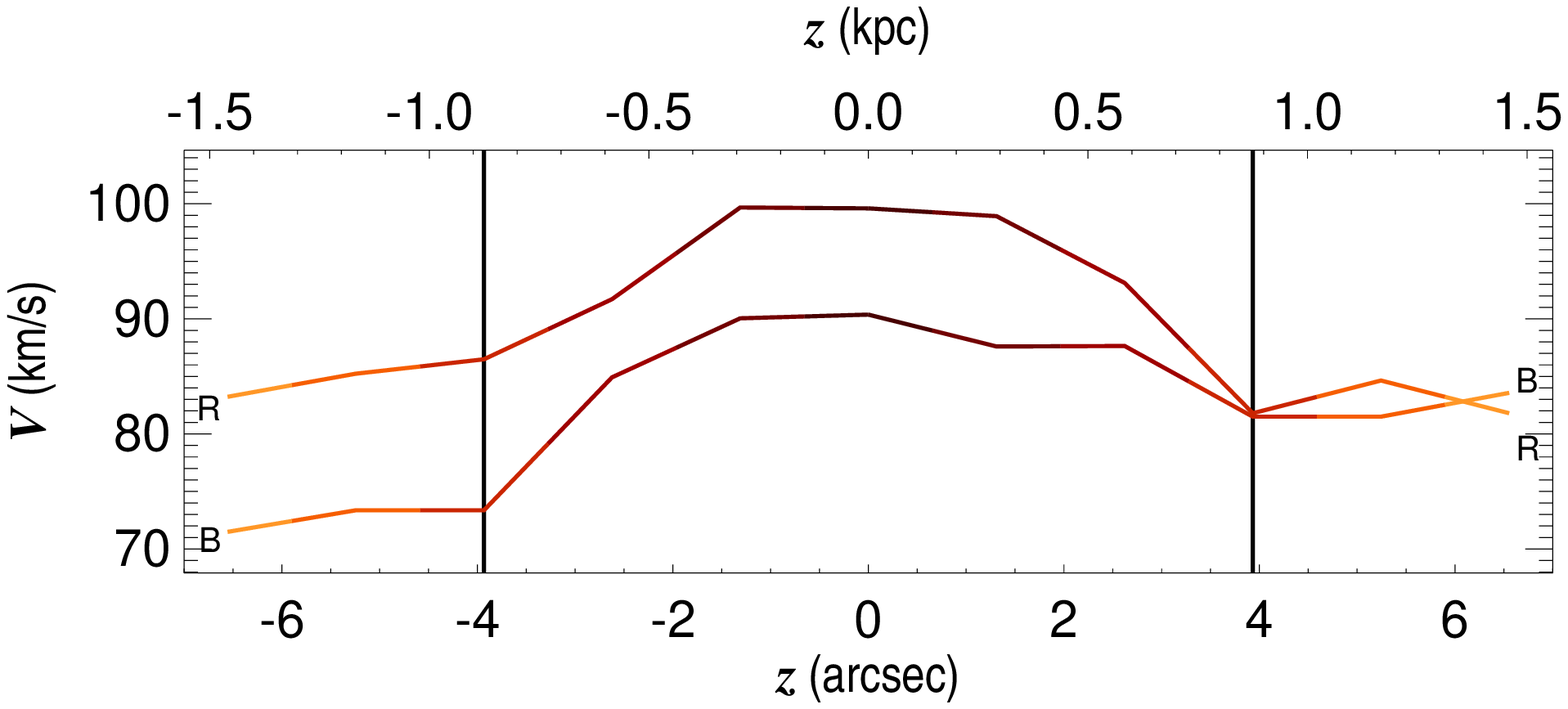}\\
  \includegraphics[width=0.48\textwidth]{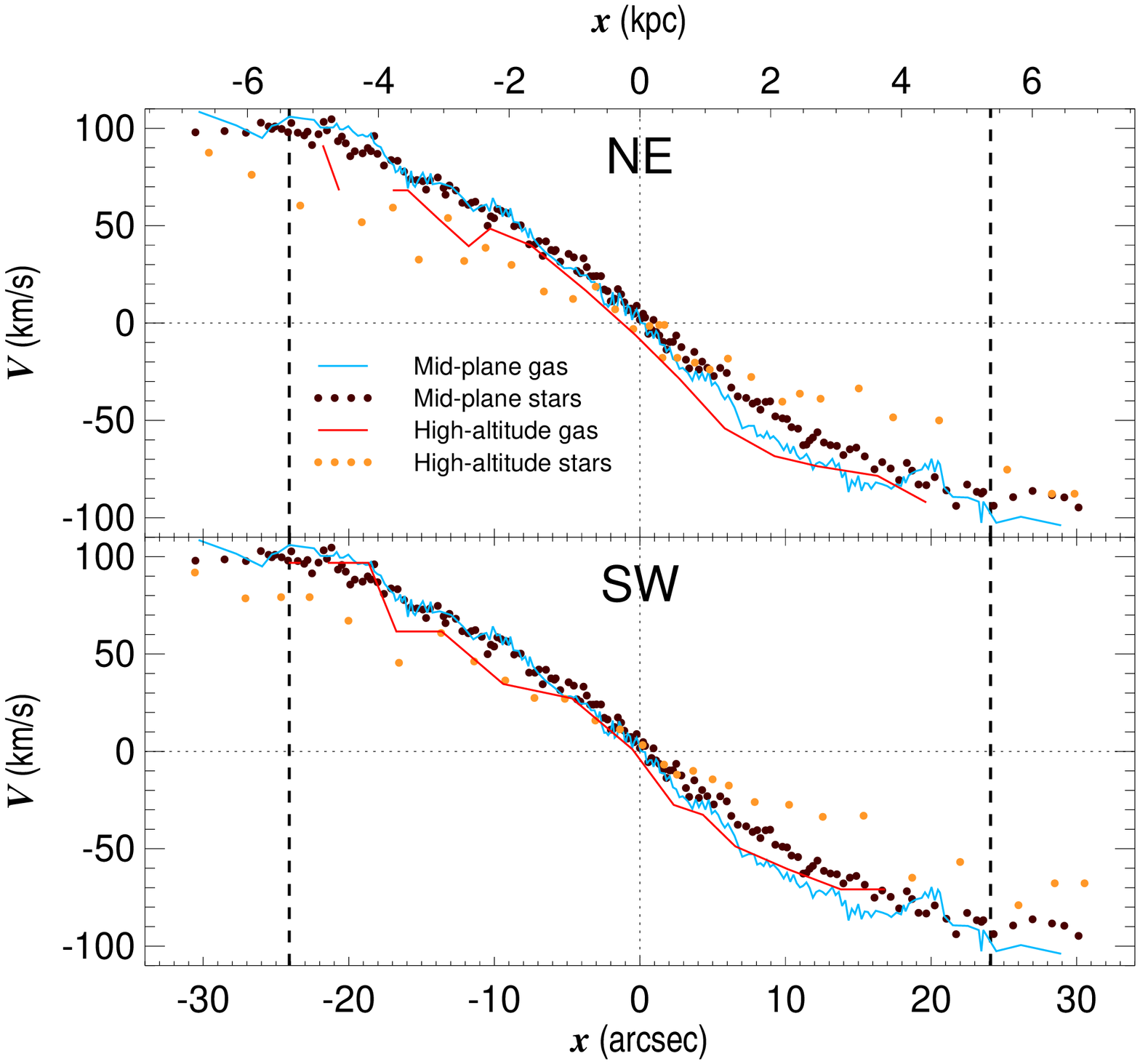}  
  \end{tabular}
  
\end{tabular}

  \end{center}
  \caption{\label{ESO544-27} As in Fig.~\ref{ESO157-49}, but for ESO~544-27.}
\end{figure*}

\begin{figure*}
\begin{center}
\begin{tabular}{c c}
    \raisebox{-0.5\height}{\includegraphics[width=0.48\textwidth]{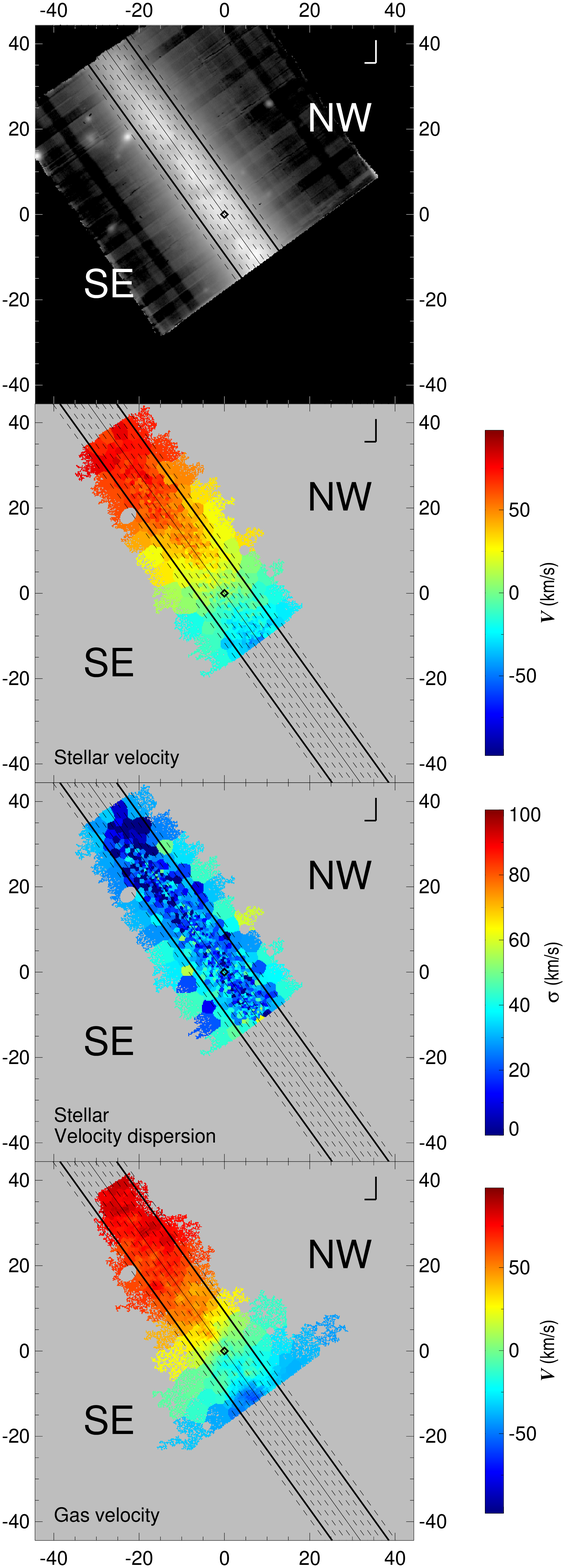}}&
  
  \begin{tabular}{c}
  {\Large{IC~217}}\\
  \\
  \includegraphics[width=0.48\textwidth]{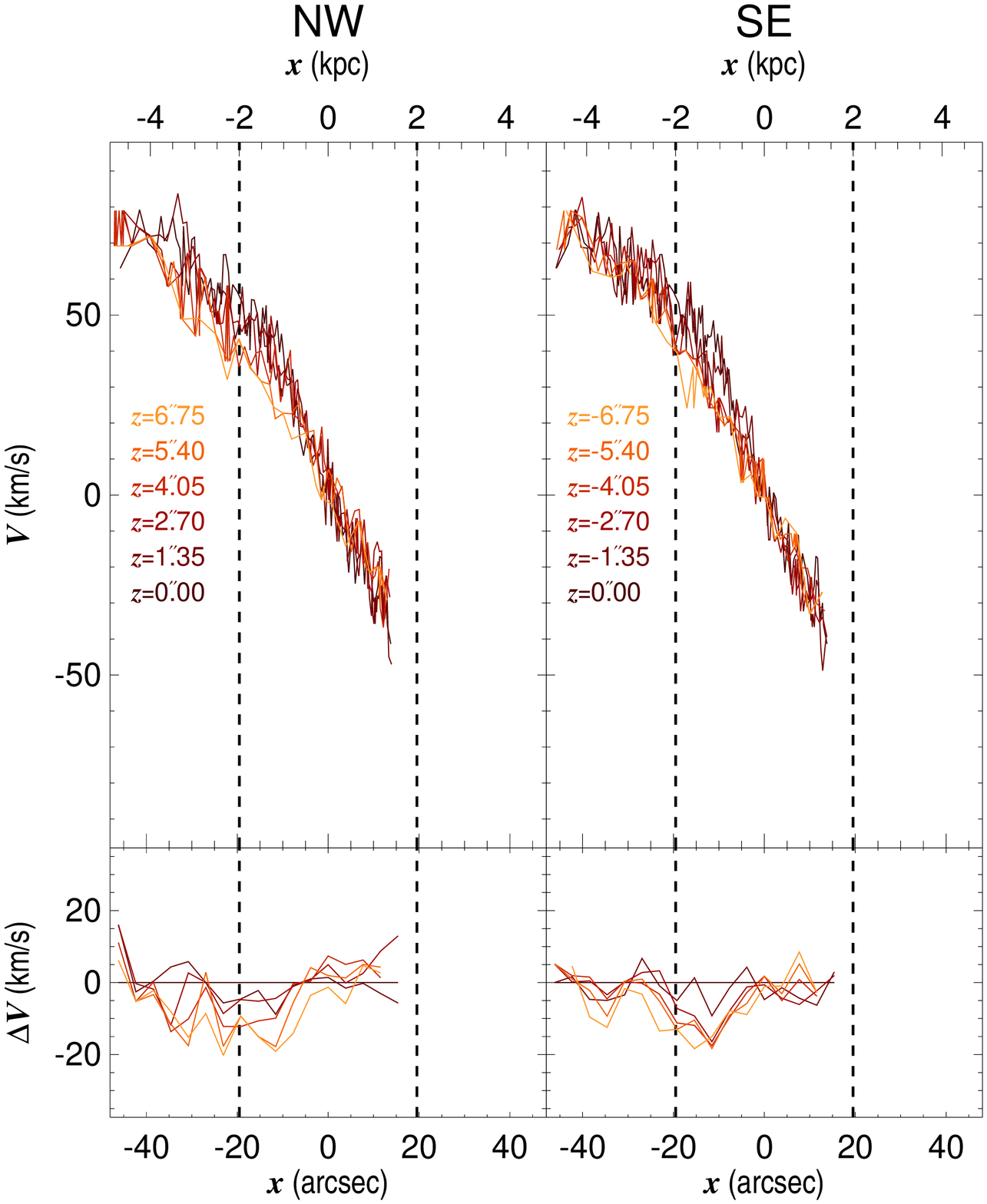}
  \\
  \includegraphics[width=0.48\textwidth]{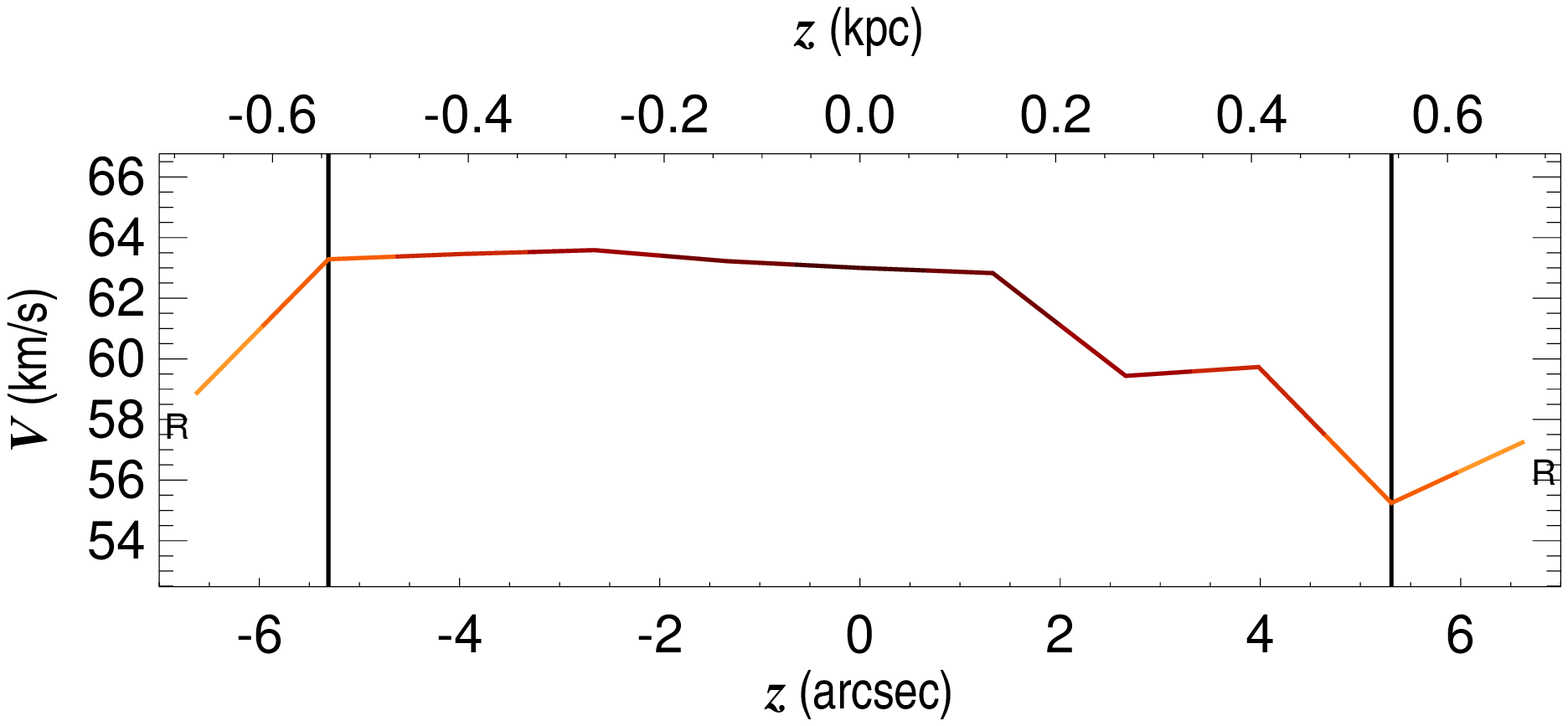}
  \\
  \includegraphics[width=0.48\textwidth]{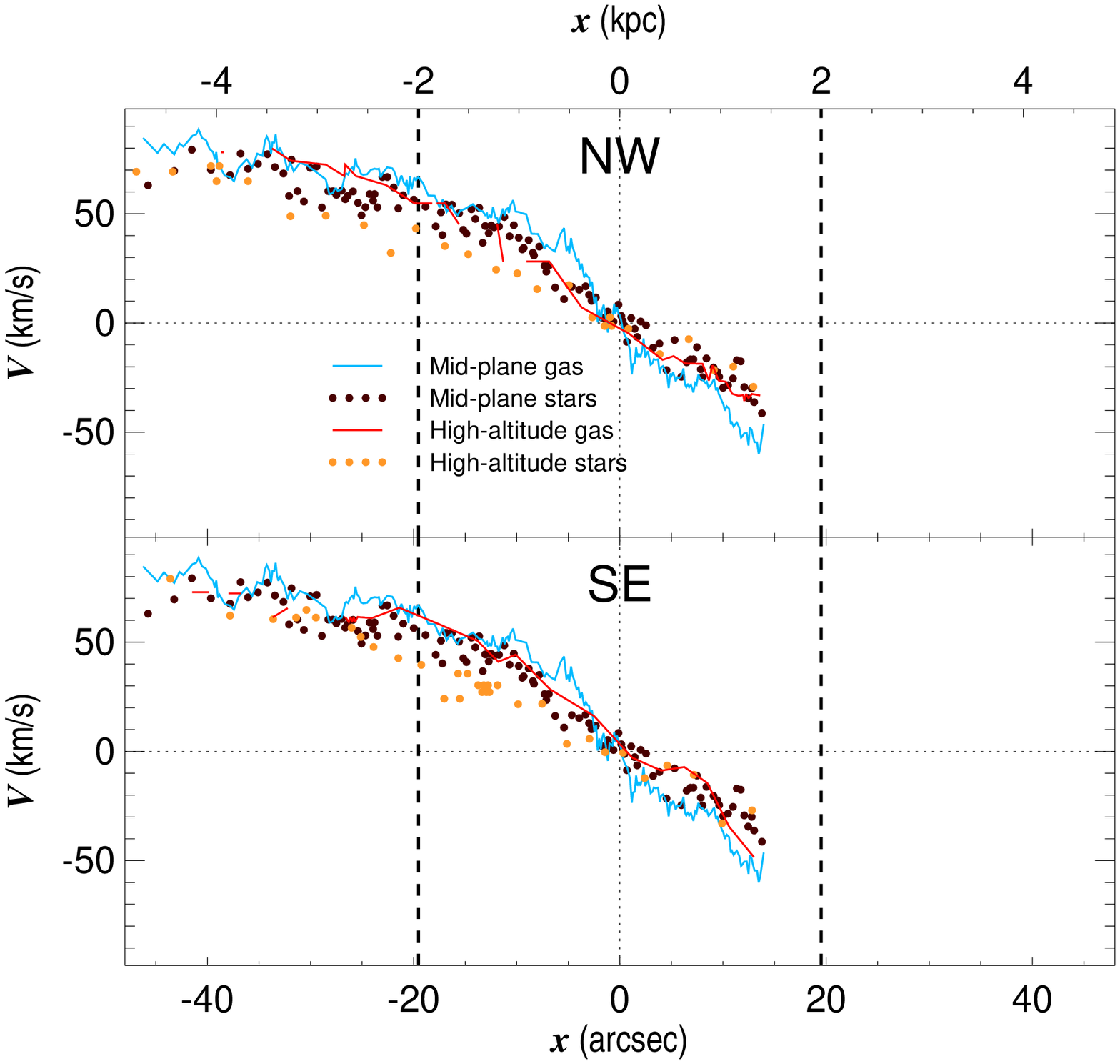}  
  
 \end{tabular}
  
\end{tabular}

  \end{center}
  \caption{\label{IC217} As in Fig.~\ref{ESO157-49}, but for IC~217.}
\end{figure*}

\begin{figure*}
\begin{center}
 
\begin{tabular}{c c}
    \raisebox{-0.5\height}{\includegraphics[width=0.48\textwidth]{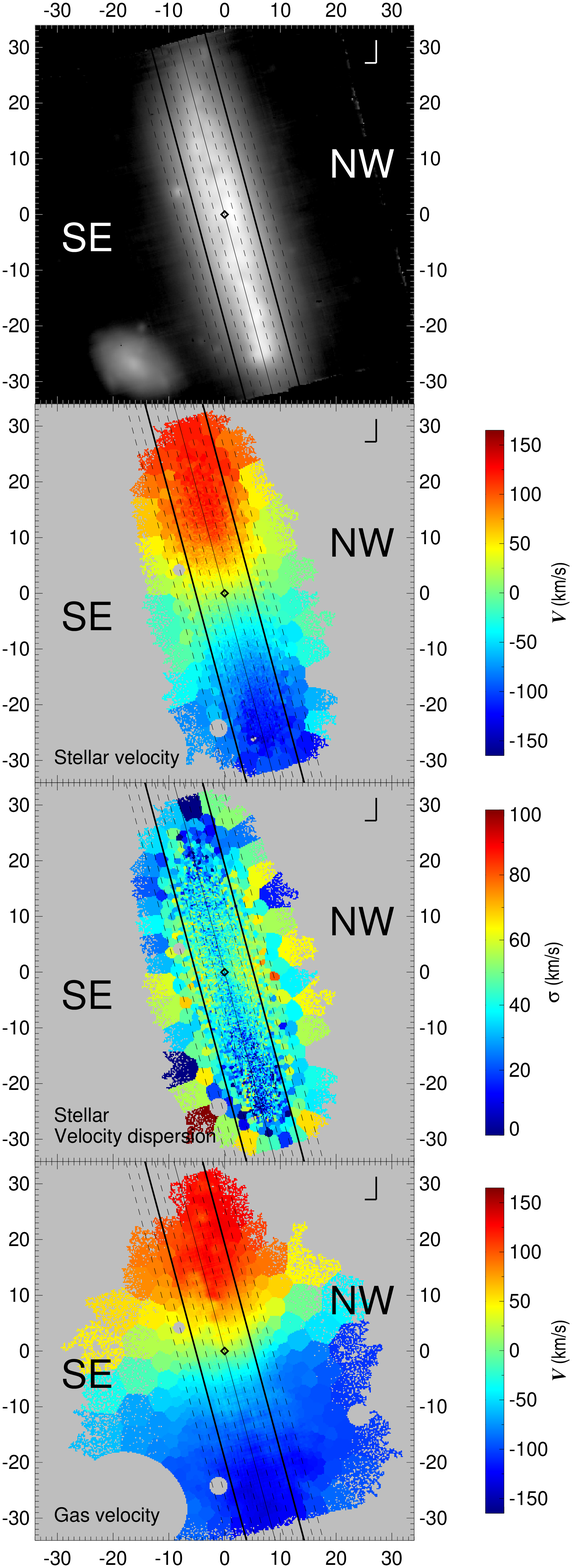}}&
  
  \begin{tabular}{c}
  {\Large{IC~1553}}\\
  \\
  \includegraphics[width=0.48\textwidth]{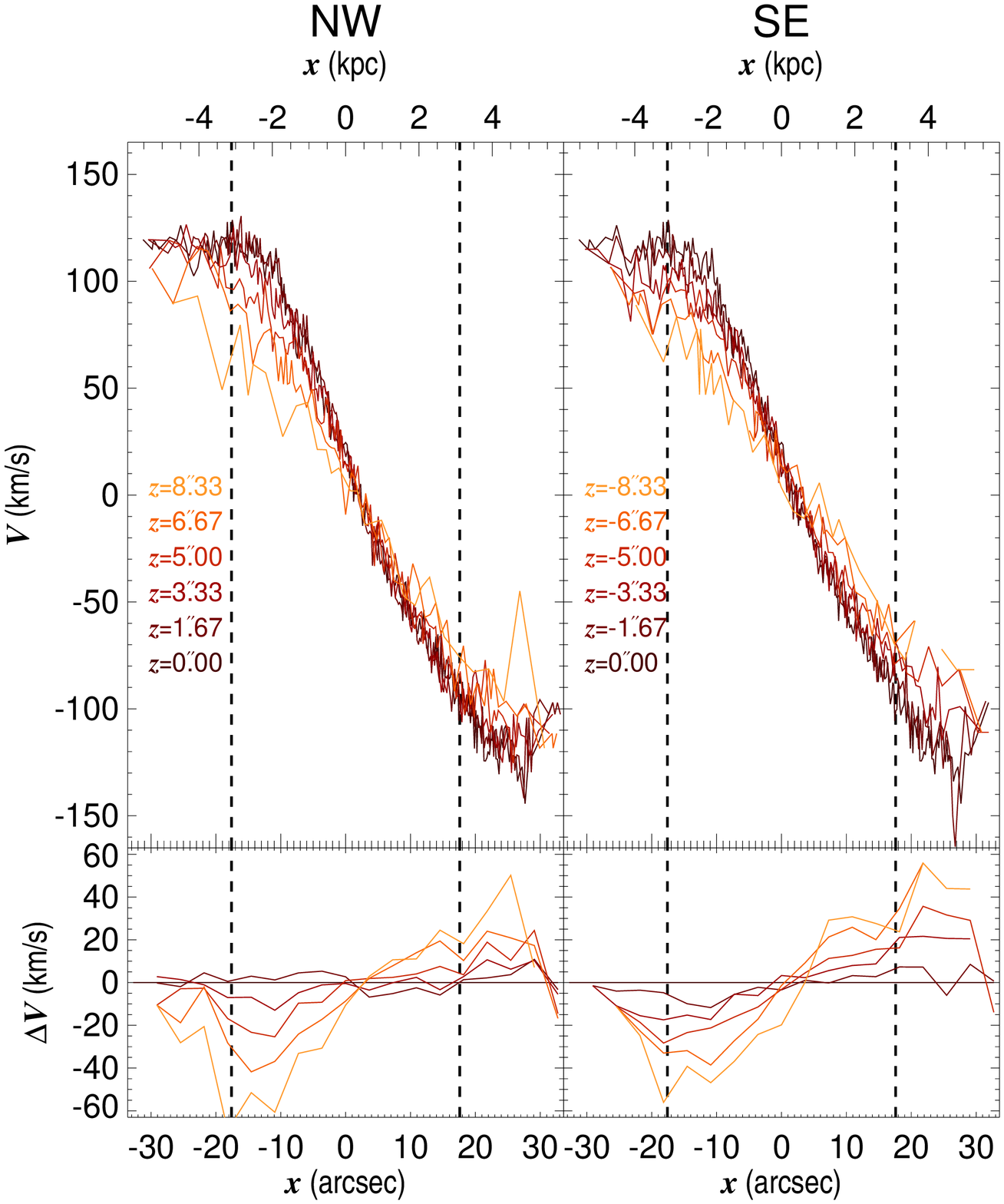}
  \\
  \includegraphics[width=0.48\textwidth]{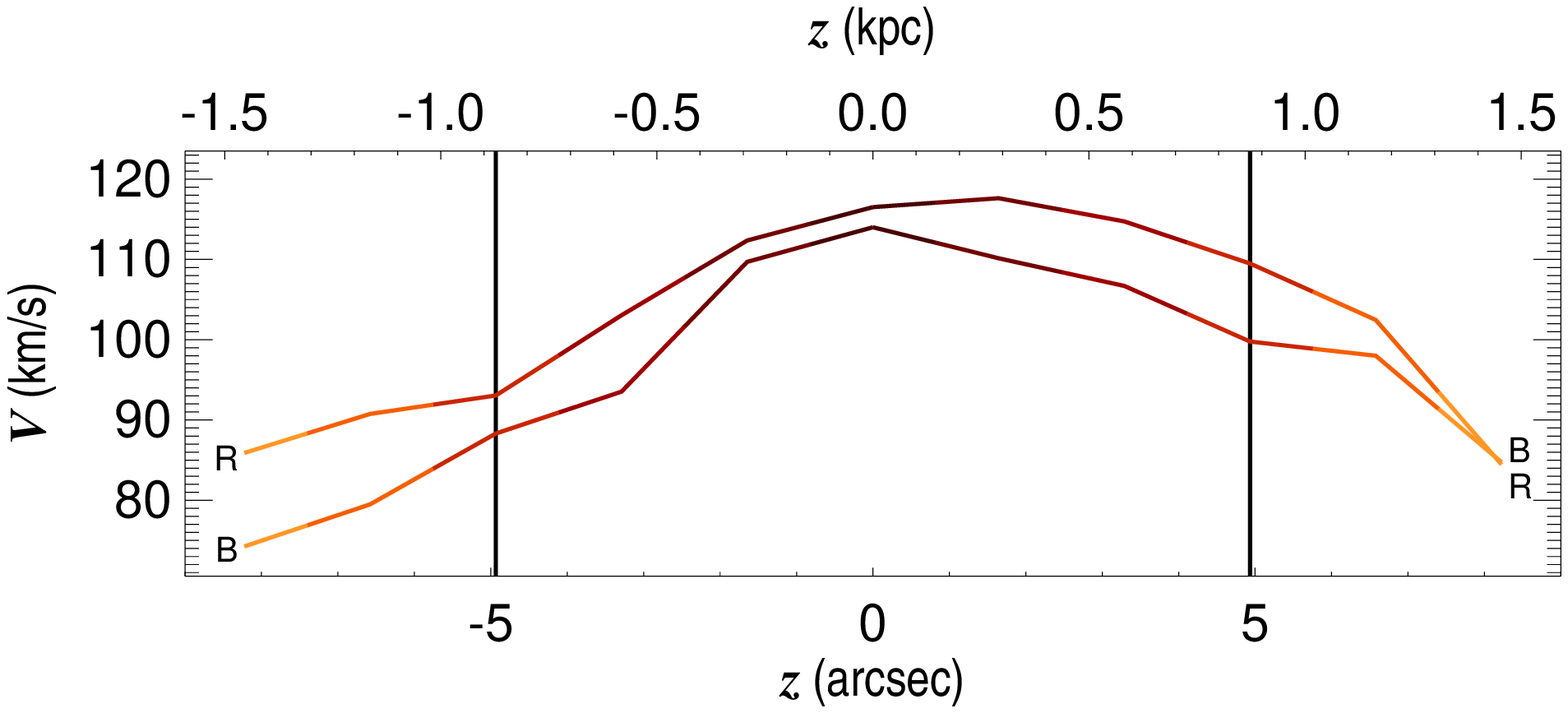}  
  \\
  \includegraphics[width=0.48\textwidth]{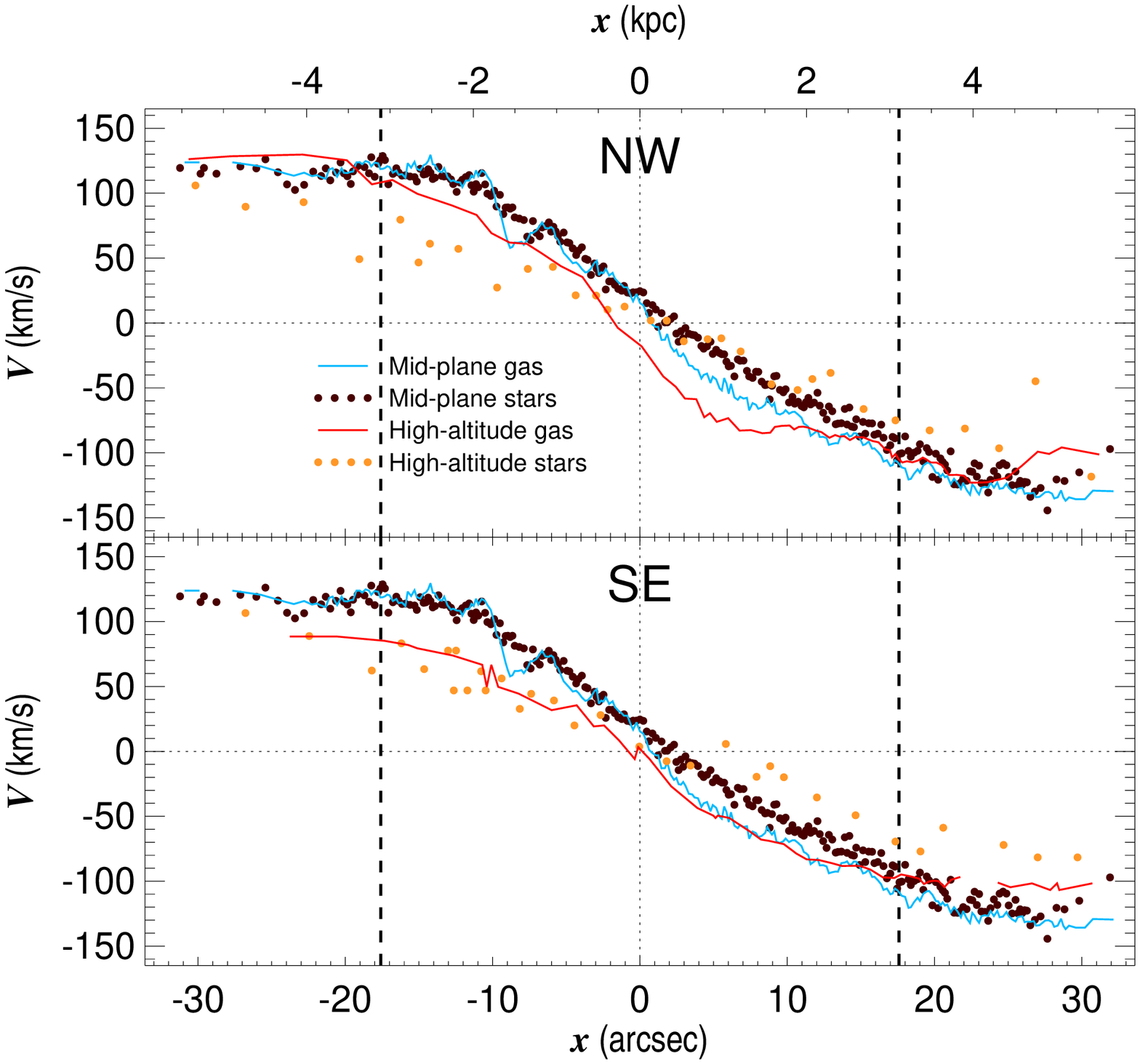}  
  \end{tabular}
  
\end{tabular}

  \end{center}
  \caption{\label{IC1553} As in Fig.~\ref{ESO157-49}, but for IC~1553. The small galaxy at the lower-left of the {\it top-left} panel is unrelated to IC~1553 and is found to have redshift $z=0.05$.}
\end{figure*}

\begin{figure*}
\begin{center}
 
\begin{tabular}{c c}
    \raisebox{-0.5\height}{\includegraphics[width=0.48\textwidth]{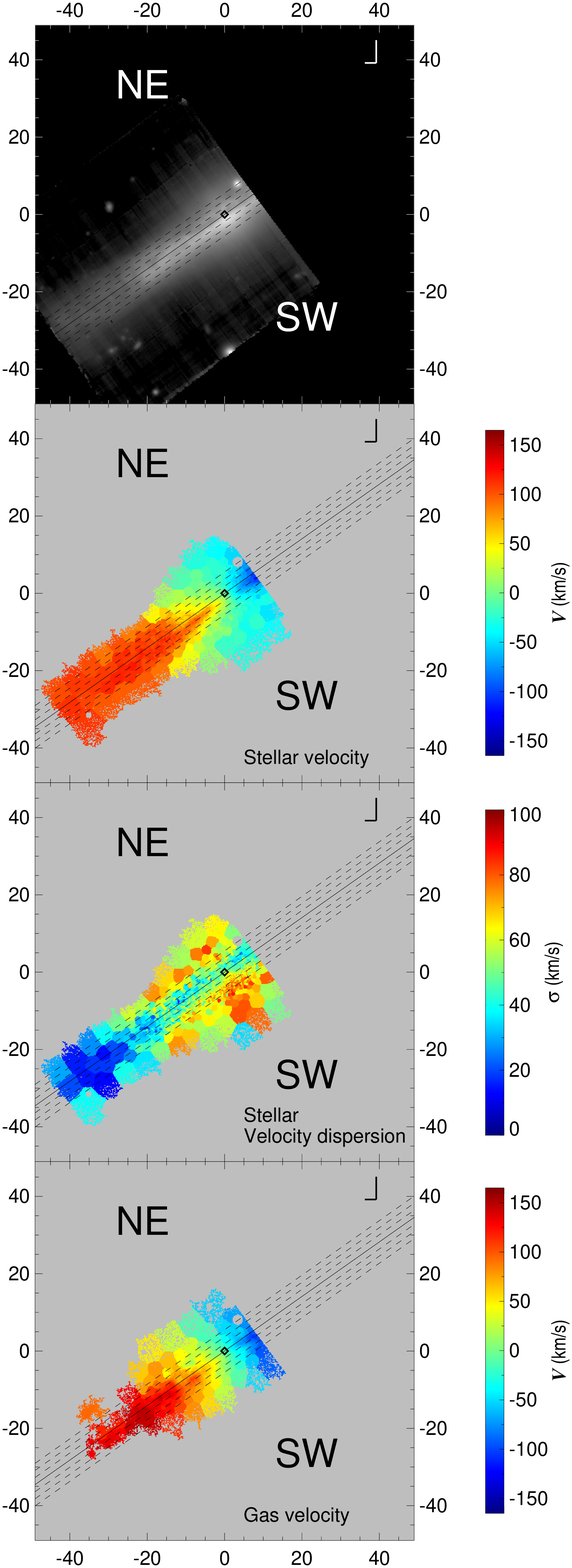}}&
  
  \begin{tabular}{c}
  {\Large{PGC~28308}}\\
  \\
  \includegraphics[width=0.48\textwidth]{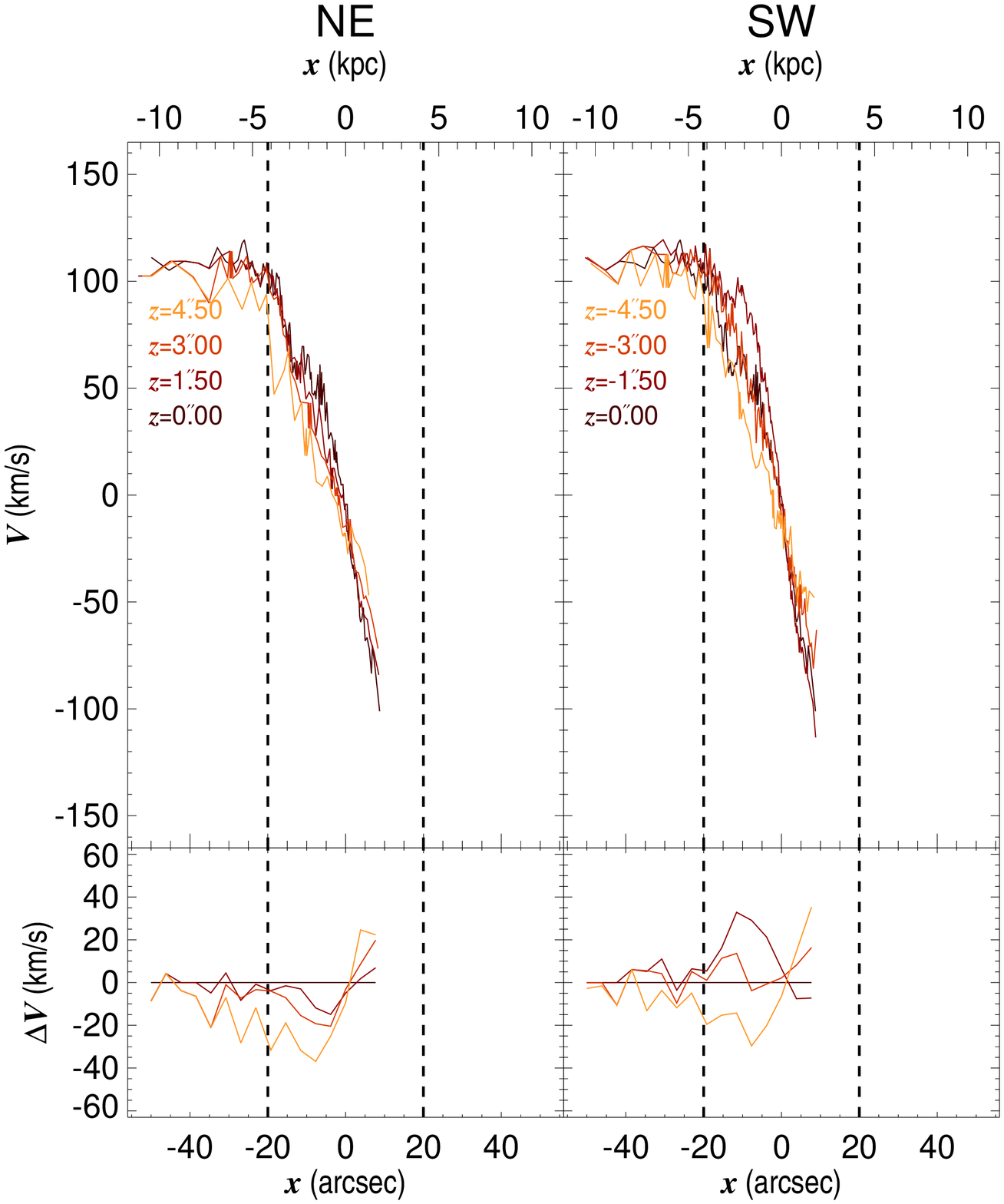}
  \\
  \includegraphics[width=0.48\textwidth]{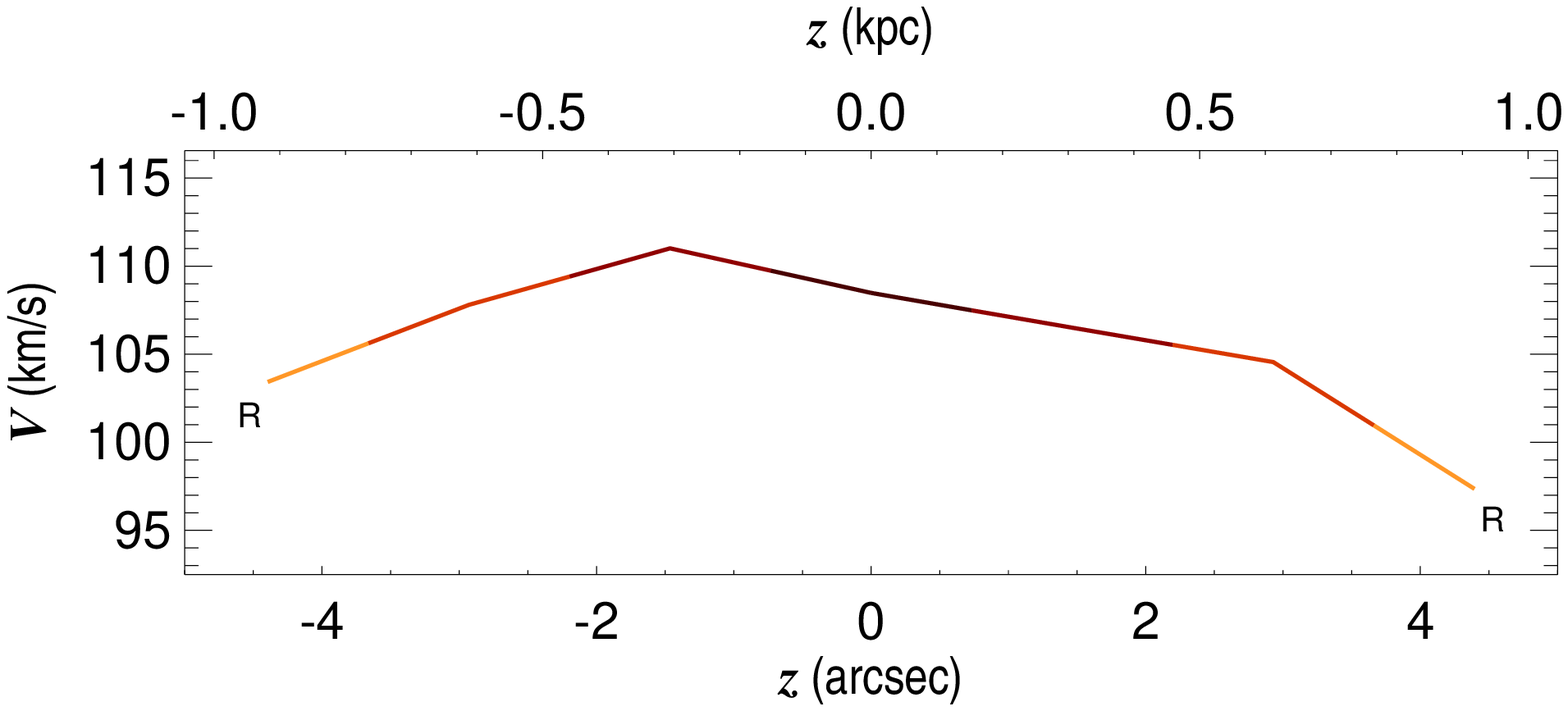}  
 \\
  \includegraphics[width=0.48\textwidth]{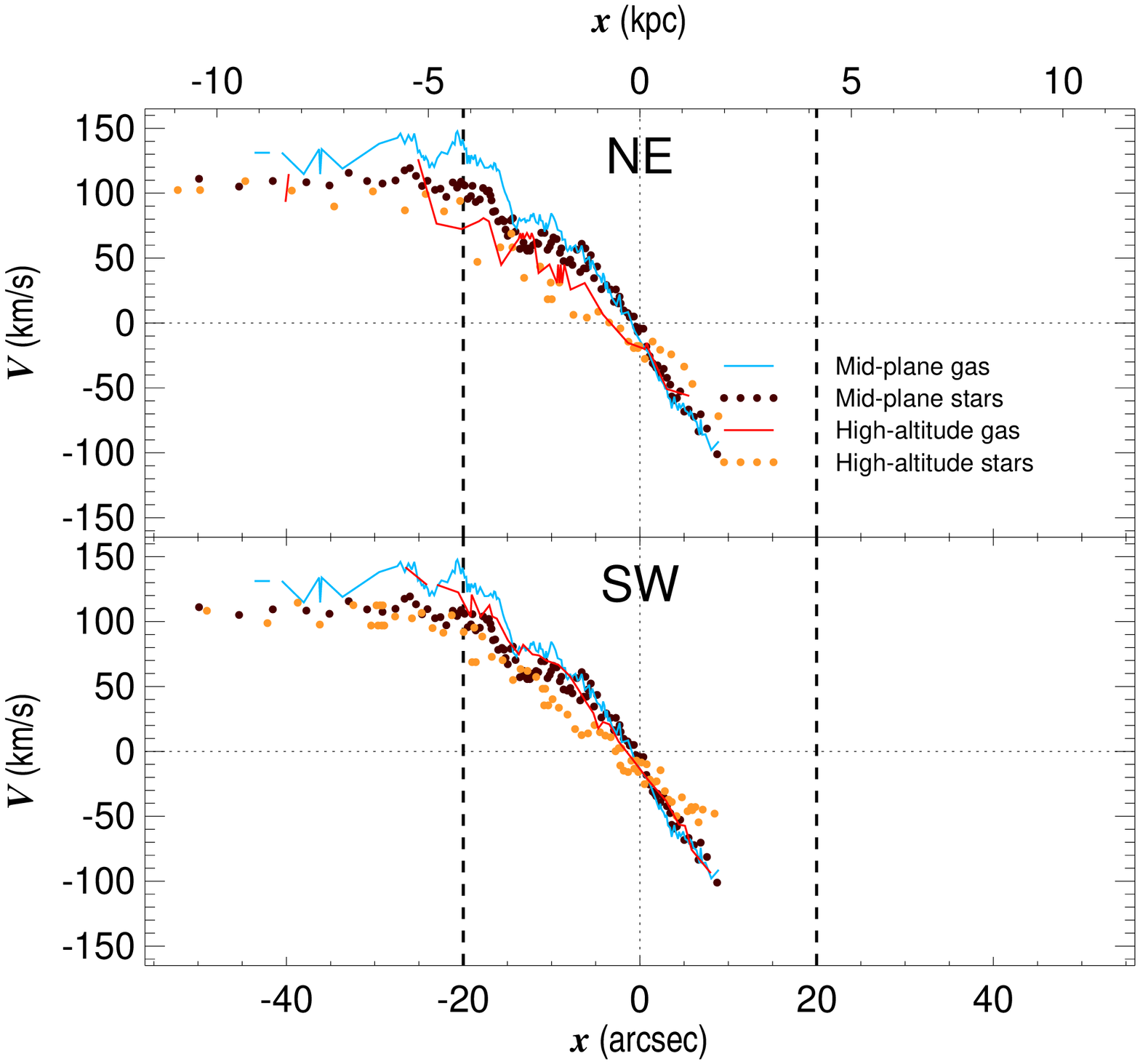}  
   \end{tabular}
  
\end{tabular}

  \end{center}
  \caption{\label{PGC28308} As in Fig.~\ref{ESO157-49}, but for PGC~28308. In this galaxy there is no height where the thin disc has a lower surface brightness than the thick disc once the PSF is accounted for \citep{CO18}.}
\end{figure*}

\begin{figure*}
\begin{center}
 
\begin{tabular}{c c}
    \raisebox{-0.5\height}{\includegraphics[width=0.48\textwidth]{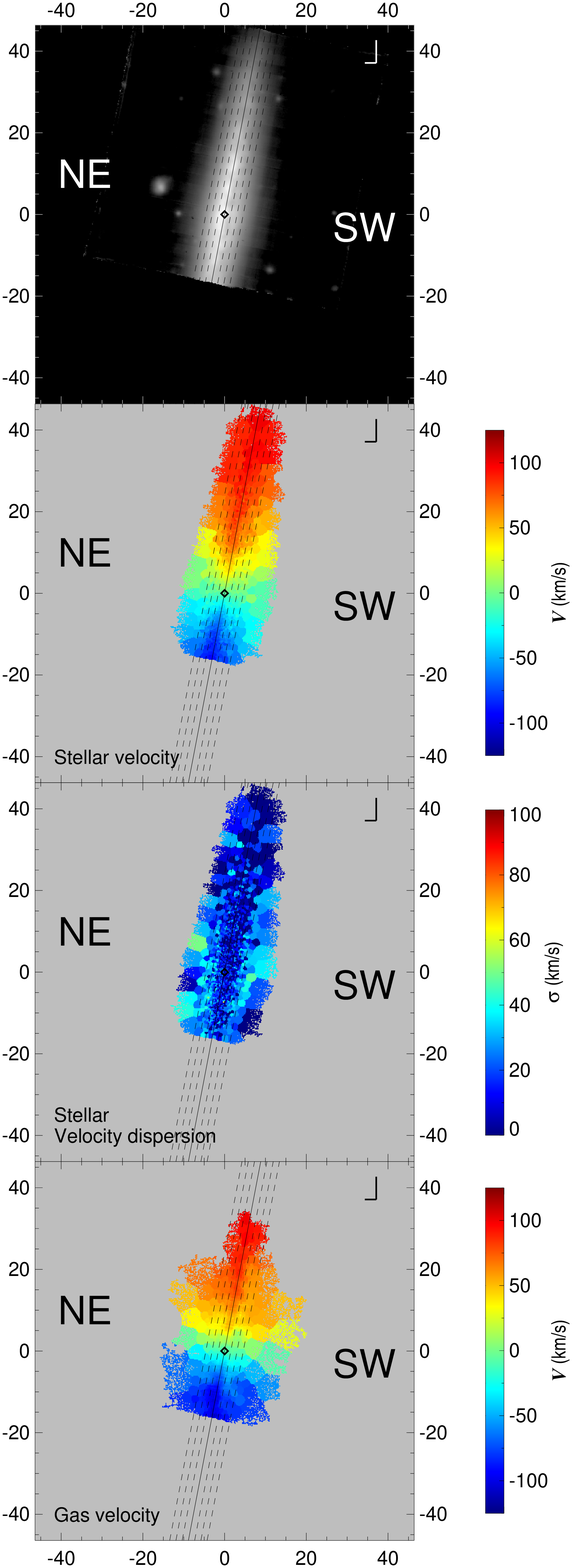}}&
  
  \begin{tabular}{c}
  {\Large{PGC~30591}}\\
  \\
  \includegraphics[width=0.48\textwidth]{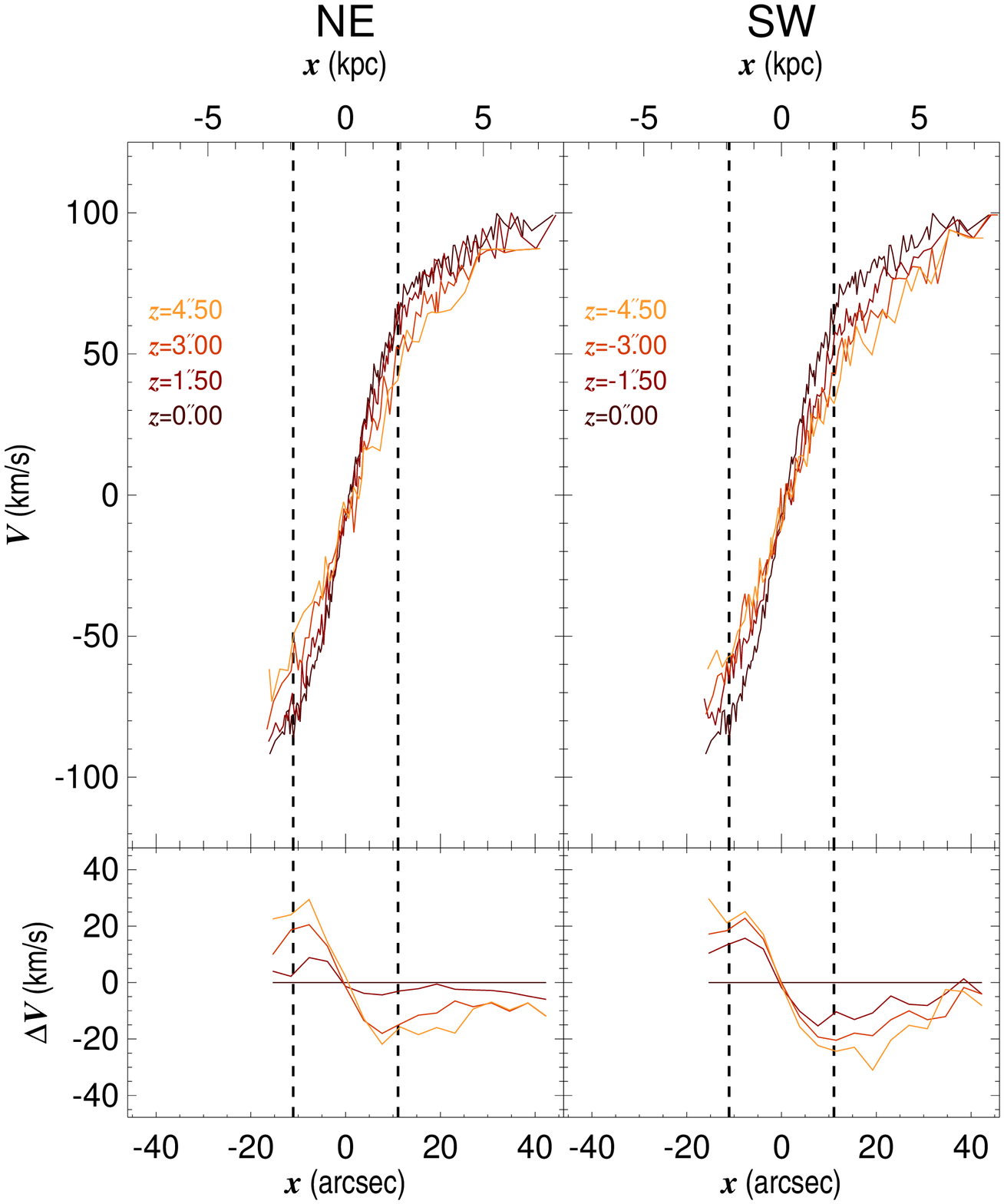}
  \\
  \includegraphics[width=0.48\textwidth]{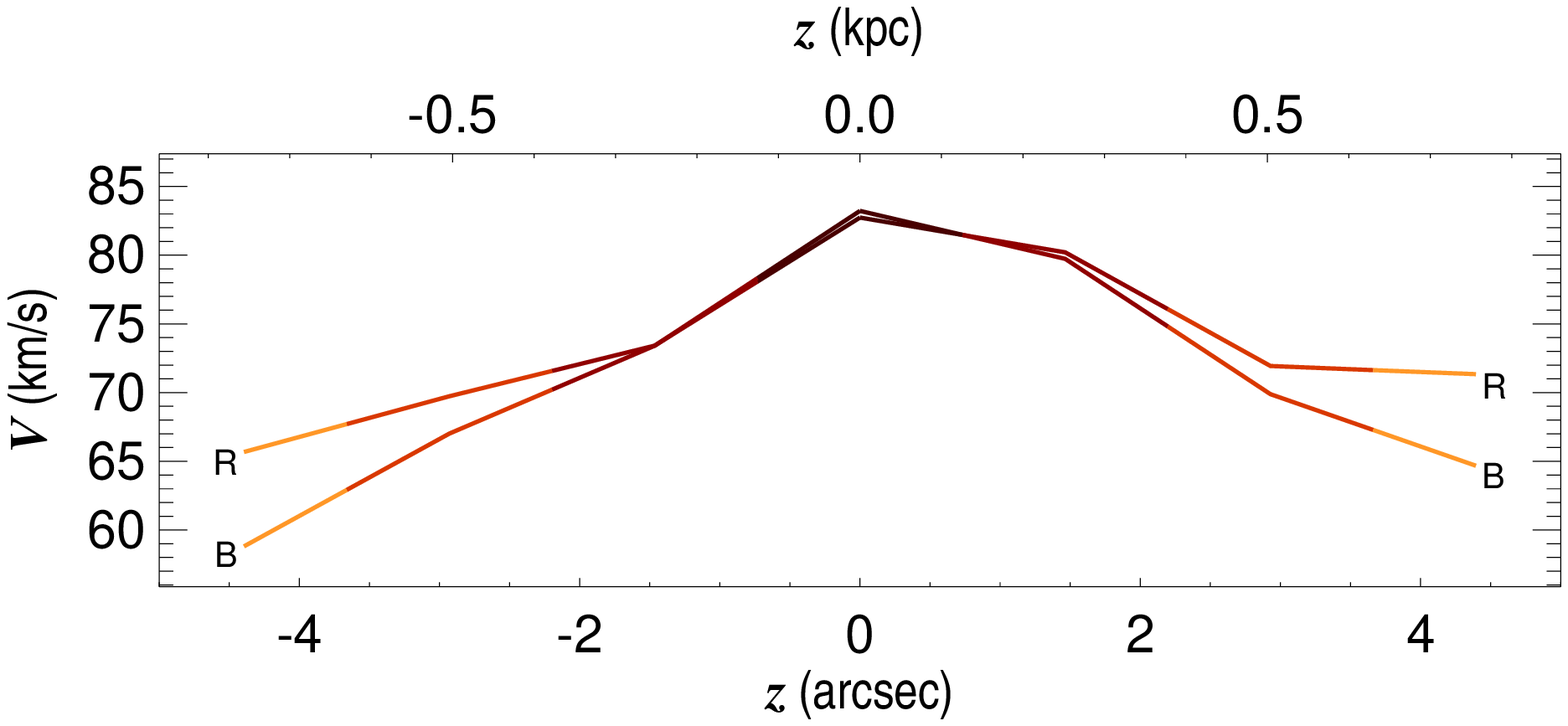}
 \\
  \includegraphics[width=0.48\textwidth]{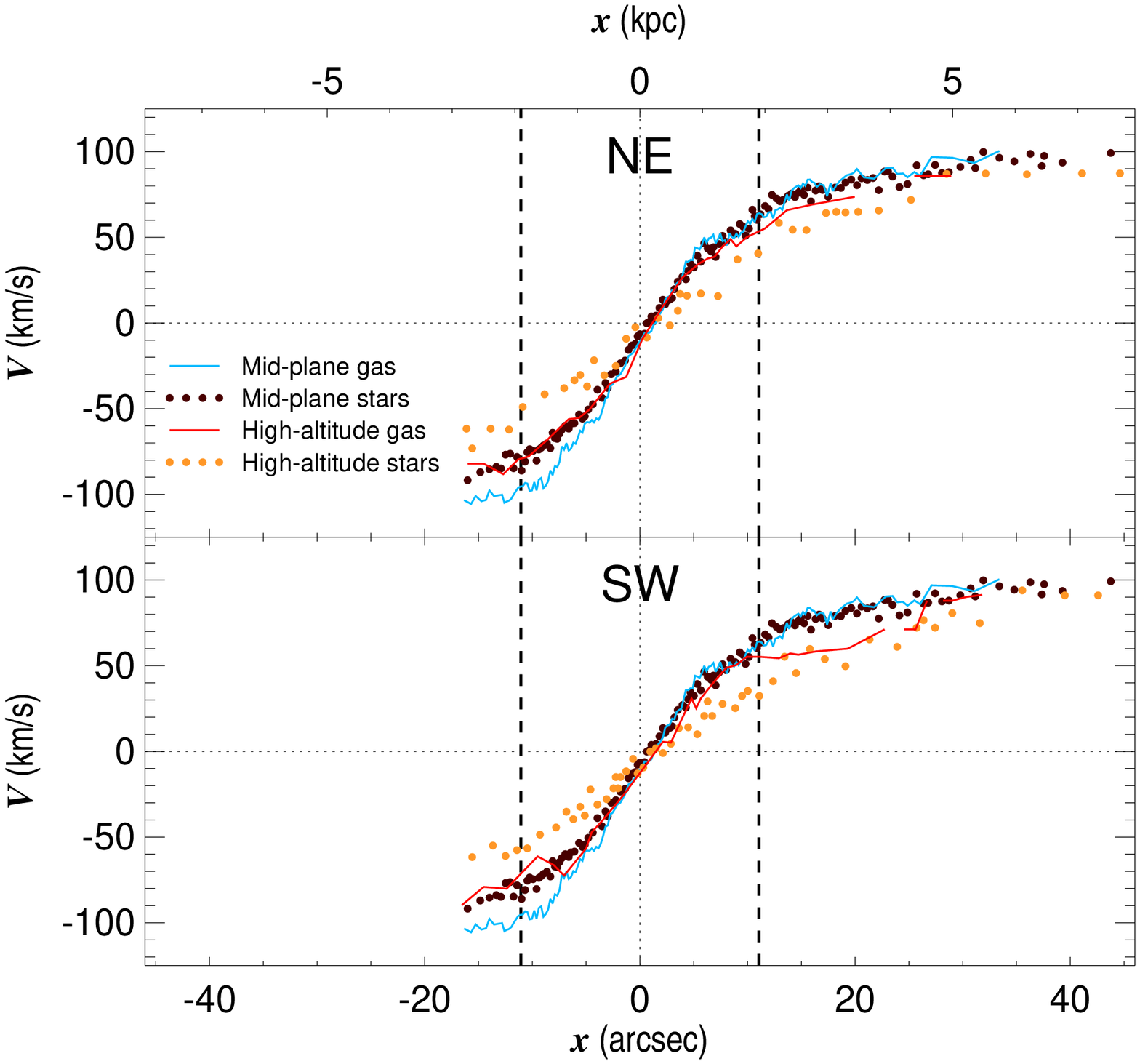}
  \end{tabular}
  
\end{tabular}

  \end{center}
  \caption{\label{PGC30591} As in Fig.~\ref{ESO157-49}, but for PGC~30591. \citet{CO18} found no distinct thin- and thick-disc structure for this galaxy.}
\end{figure*}

\end{appendix}

\end{document}